\documentclass[
aps,%
11pt,%
final,%
titlepage,%
oneside,%
nobibnotes,%
nofootinbib,%
superscriptaddress,%
showkeys,%
centertags]%
{revtex4}

\newcommand{\hurl}[1]{\href{#1}{#1}}

\usepackage{xpatch} 

\makeatletter

\patchcmd{\@ssect@ltx}
    {\addcontentsline{toc}{#1}{\protect\numberline{}#8}}
    {}
    {}
    {}
\makeatother

\renewcommand{\thesection}{\arabic{section}}\renewcommand{\thesubsection}{\arabic{section}.\arabic{subsection}}

\usepackage{amsmath}
\usepackage{graphicx}
\usepackage[colorlinks=true,linkcolor=blue,citecolor=blue,urlcolor=blue]{hyperref}

\usepackage{tabularx}

\usepackage{amssymb}

\usepackage{booktabs}
\usepackage{multirow}
\usepackage{soul}
\usepackage{microtype}
\usepackage{upgreek}
\graphicspath{{./Definitions/}}

\newcommand{\orcidicon}{\includegraphics[width=0.32cm]{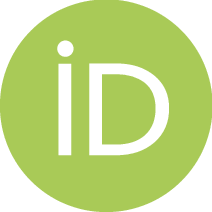}}

\newcommand{\be}{\begin{equation}}
\newcommand{\ee}{\end{equation}}

\newcommand{\orcidA}{\href{https://orcid.org/\orcidauthorA}{\orcidicon}}

\newcommand{\orcidauthorA}{0000-0002-2871-7711}

\usepackage{setspace}

\begin{document}


\singlespacing

\noindent {\it Universe} {\bf 2020}, {\it 6}(11), 212; \href{http://www.doi.org/10.3390/universe6110212}{doi:10.3390/universe6110212}

\vskip1cm

\title{\LARGE Einstein's Geometrical Versus  Feynman's Quantum-Field  Approaches to Gravity Physics: Testing by Modern Multimessenger Astronomy}

\author{\firstname{Yurij}~\surname{Baryshev}\orcidA}
\email{y.baryshev@spbu.ru; yubaryshev@mail.ru}
\affiliation{Astronomy Department, Mathematics \& Mechanics Faculty, Saint Petersburg State University,
28 Universitetskiy Prospekt, 198504 St. Petersburg, Russia}

\begin{abstract} 
Modern multimessenger astronomy delivers unique opportunity for performing crucial observations that allow for testing the physics of the gravitational interaction. These~tests include detection of gravitational waves by advanced LIGO-Virgo antennas, Event Horizon Telescope  observations of central relativistic compact objects (RCO) in active galactic nuclei (AGN), X-ray~spectroscopic observations of Fe $K_{\alpha}$ line in AGN, Galactic X-ray sources measurement of  masses and radiuses of neutron stars, quark stars, and other RCO. A very important task of observational cosmology is to perform large surveys of galactic distances independent on cosmological redshifts for testing the nature of the Hubble law and peculiar velocities. Forthcoming multimessenger astronomy, while using such facilities as advanced LIGO-Virgo, Event Horizon Telescope (EHT), ALMA, WALLABY, JWST, EUCLID, and THESEUS,
can elucidate the relation between Einstein's geometrical and Feynman's quantum-field approaches to gravity physics and deliver a new possibilities for unification of gravitation with other fundamental quantum physical interactions.
\end{abstract}

\keywords{gravitation; cosmology; multimessenger astronomy; quantum physics.}

\maketitle

\tableofcontents 


\newpage

\section{Introduction}
\label{intro}

Contemporary physics considers the whole observable Universe as the cosmic laboratory, where the fundamental physical laws must be tested in different astrophysical conditions and with increasing accuracy. Such basic theoretical assumptions as: constancy of the fundamental constants $\,c, \,G, \,h, \,m_p, \,m_e$, the~Lorentz invariance, the~equivalence principle, the~quantum principles of the gravity theory, validity of general relativity, and its modifications in strong gravity and at largest cosmological scales are now being investigated by modern theoretical physics 
and by contemporary astrophysical observations
(Cardoso \& Pani 2019~\cite{cardoso2019}, 
Ishak 2018~\cite{ishak2018}, 
De Rham~et~al.,~2017~\cite{derham2017},  
Giddings 2017~\cite{giddings2017}, 
\mbox{Debono \& Smoot 2016~\cite{debono2016},}
De Rham 2014~\cite{derham2014}, 
Clifton~et~al.,~2012~\cite{clifton2012},
Baryshev \& Teerikorpi 2012~\cite{baryshev-t12},
Uzan 2010~\cite{uzan2010},
Rubakov \& Tinyakov 2008~\cite{rubakov2008}, and
Uzan 2003~\cite{uzan2003}).

Since the first paper on Relativistic Astrophysics, published by
Hoyle~et~al.,~1964~\cite{hoyle64}, where~crucial role of relativistic gravity in studies of extremal astrophysical objects was discussed, more than fifty years have passed by.
Nowadays, relativistic astrophysics deals with high energy phenomena such as ultra-dense matter in  neutron and quark stars, strong gravity in black hole candidates of stellar and galactic masses,  gravitational radiation, and its detection, massive supernova explosions, gamma ray bursts, jets from active galactic nuclei and cosmological models of the Universe.
The common basis for all of these observed astrophysical phenomena is the theory of gravitation, for~which in modern theoretical physics one can separate two alternative approaches for description of gravity phenomena: Einstein's geometrical general relativity theory (GRT) and Feynman's non-metric quantum-field gravitation theory (QFGT).
Although classical relativistic gravity effects have the same values in both of the approaches, there are also dramatically different effects predicted by GRT and QFGT for relativistic astrophysics, which can be tested by multimessenger~astronomy.

The general relativity theory (GRT), which now achieves 100 years from its birthday (\mbox{Einstein 1915~\cite{einstein15}; Hilbert 1915~\cite{hilbert15}}) is the most developed geometrical description of the gravity phenomena (metric tensor $g_{ik}$ of Riemannian space).
The success of GRT in explanation of classical relativistic gravity effects and cosmological data is  generally recognized 
(Debono \& Smoot 2016~\cite{debono2016};  Will 2014~\cite{will14}); and, presented in many textbooks. 
However, there are some puzzling theoretical and observational problems, such as the problem of the energy-momentum pseudotensor and non-localizability of the gravitational field, information paradox of the event horizon, tension between Local and Global cosmological parameters of the standard model, which stimulate to search for alternative gravitation theories. 
\footnote{Debono \& Smoot review~\cite{debono2016}  contains list of 20 alternative  gravitation theories. A~comprehensive review of metric gravity theories by Clifton~et~al.~\cite{clifton2012} contains 13 metric theories and 1316 references.}


Especially, in~this review I present a comparison of general relativity predictions with main results of the Feynman's non-metric quantum-field approach to gravitation  theory (QFGT), which~he formulated in Caltech lectures during the 1962--1963 academic year
\cite{feynman71,feynman95}. Feynman's QFGT is based on  common principles with other fundamental physical interactions---gravity is described by the
symmetric tensor field $\psi_{ik}$ in Minkowski space---hence it gives a natural first step in the construction of the unification of gravitation with particle physics. 
In contrast to many claims against the feasibility of field approach, this review of its results demonstrate that  QFGT is consistent relativistic quantum gravity theory. In~particular, it predicts the same values for measured classical gravity effects (\mbox{Thirring 1961~\cite{thirring61}}, 
Baryshev 1990~\cite{baryshev90}, 2003~\cite{baryshev03}).

\subsection{ Key Discoveries of Modern Multimessenger~Astronomy}

In the beginning of the 21st century, several fundamental observational discoveries were done by  the multimessenger astronomy in electromagnetic, cosmic particles, neutrino, and gravitational wave channels. Among~them the following key new~results:
\begin{itemize}
\item[$\bullet$]{\emph{detection of the gravitational waves  from coalescent relativistic compact objects by LIGO-Virgo antennas;}}
\item[$\bullet$]{\emph{imaging of the supermassive black hole candidates  M87* and SgrA* by Event Horizon Telescope;}} and,
\item[$\bullet$]{\emph{establishing tensions between Local and Global cosmological parameters.} }
\end{itemize}

\subsubsection{Gravitational~Waves}

GW150914 was the first gravitational-wave signal detected by Advanced LIGO  interferometric antennas Abbott~et~al.,~2016a~\cite{abbott16a}.
Up to now, there are 68 GW detection during O1-O2-O3 observing runs. All LIGO-Virgo-Collaboration publications are presented in Abbott~et~al.,~2016c~\cite{abbott16c}. 
The first multimessenger observations of the gravitational waves and gamma-rays from a binary neutron star merger GW170817 and GRB 170817A is a breakthrough discovery that opened a new epoch in gravity physics Abbott~et~al.,~2016b~\cite{abbott16b}.

This means that the positive gravitational field energy carried by gravitational waves, was~localized by a GW detector, i.e.,~free gravitational field energy can be transformed to the kinetic energy of the moving LIGO mirrors. The~GW detection finished long-time old discussion about 
``DO GRAVITATIONAL WAVES CARRY ENERGY?''
(history review 
Cervantes-Cota~et~al.,~2016~\cite{cervantes2016}, Chen~et~al.,~2016~\cite{chen2016}).
Whether gravitational waves carry energy has been debated since the beginning of relativistic gravitation theory.  From~the work of Hilbert, Klein, and Noether, it was
established that there is no proper energy density for general relativity (and for any generally covariant theory of gravity). For~the density of gravitational energy-momentum Einstein had found a pseudotensor expression, i.e.,~not a proper
tensor, but rather an expression that was inherently reference frame dependent.
Einstein had used his pseudotensor and found that the transverse waves carry energy. However,~many objected
to using this pseudotensor to describe gravitational energy, including Levi-Civita, Bauer, and 
Schreodinger. Additionally, Eddington and Pauli argued that gravitational energy was non-localizable Chen~et~al.,~2016
~\cite{chen2016}.

Another interpretation of the GW detector length variations as a contracting  and stretching the ``space-time'' without energy taking from gravitational wave is possible in the frame of geometrical gravity theory. However, it introduces non-covariant
description of GW energy-momentum (\mbox{Maggiore 2008~\cite{maggiore08}}). It leads to some conceptual problems because of giving up the general covariance principle in geometrical description of the gravitational field~energy.

Indeed, according to Landau \& Lifshitz 1971~\cite{landau71} (\S101, p. 307):
``...it has no meaning to speak of a definite localization of the energy of the gravitational field in space...'' and ``so that it is meaningless to talk of wether or not there is gravitational energy at a given place''.
Also according to Misner, Thorne, Wheeler 1973
~\cite{misner73} (\S20.4, p. 467): ``...gravitational energy... is not localizable. The~equivalence principle forbids'', and~(\S35.7, p. 955):
``...the stress-energy carried by gravitational waves cannot be localized inside a wavelength'' and ``...one can say that a certain amount of stress-energy is contained in a given 'macroscopic' region of several wavelengths'~size''.

Note that, now, we have the observational fact of GW energy localization by the LIGO detector's mirror (1 m size) just well inside the GW wavelength (\mbox{$\lambda=c/\nu= 3000$ km for $\nu=100$ Hz}).
The~existence of positive localizable gravitational field energy is also consistent with firm observations of the energy loss via gravitational wave radiation from binary neutron star system PSR 1913+16
\footnote{R.A. Hulse and J.H. Taylor won in 1993 Nobel Prize in physics for this discovery. }
(\mbox{recently summarized in~\cite{weisberg16}}).
Accordingly, gravitational waves from collapsing relativistic cosmic objects, carry positive energy density which can be detected (localized) and analysed using modern multimessenger astronomy facilities.
This also confirms Feynman's words~\cite{feynman95} p. 220: “the situation is exactly analogous to electrodynamics - and in the quantum interpretation, every radiated graviton carries away an amount of energy $\hbar \omega$.”

\subsubsection{Imaging of Black Holes Candidates: Relativistic Jets and~Disks}

Breakthrough observations Akiyama~et~al.,~2019a~\cite{akiyama2019a} were performed by the  VLBI  Event Horizon Telescope (EHT see Doeleman~et~al.,~2009~\cite{doeleman09}) at the event-horizon-scale images (at wavelength 1.3 mm) of the supermassive black hole candidate in the center of the giant elliptical galaxy M87. The~image structure with sizes about $5.2\,R_{Sch}$, where $R_{Sch}=2GM/c^2$ is the 
Schwarzschild radius, is consistent with GRT prediction of the light ring around Black Hole, though~angular resolution (~$2R_{Sch}$) is not enough for distinction between disk radiation and light ring. The~problem of generation of powerful jet in GRMHD simulations still exists, because~the observed  optical/X-ray radiation of the M87 jet gives estimation its power about 
$10^{44} \div 10^{45}$ erg/s, while in considered ensemble of
GRMHD models the maximum achievable kinetic power 
$P_{jet} \sim 10^{43}$ erg/s (Akiyama~et~al.,~2019b~\cite{akiyama2019b}).

Schwarzschild radius-scale structures in the supermassive black hole candidates SgrA* and M87* can be directly achievable with future submm EHT observations  and this will give possibility to test relativistic and quantum gravity theories at the gravitational radius
(\mbox{Doeleman~et~al., 2008~\cite{doeleman08}}, Doeleman~et~al., 2009~\cite{doeleman09},
Doeleman~et~al., 2012~\cite{doeleman12},
Falcke \& Markoff 2013~\cite{falcke13}, \mbox{Johannsen~et~al., 2015~\cite{johannsen15}}).
The first results of EHT SgrA* observations at 1.3mm surprisingly indicate that, for the RCO in SgrA*, there are no expected for BH the light ring at radius $5.2R_{Sch}$ (Doeleman~et~al., 2008~\cite{doeleman08}: observed RCO size 
$\theta_{obs}= 37$ 
$\mu$as, while theoretical size of the ring
$\theta_{ring}= 53$ $\mu$as).

These observations have opened a new page in study of RCO. In~particular, EHT has been designed to answer the crucial questions: Does General Relativity hold in the strong field regime? Is~there an Event Horizon? Can we estimate Black Hole spin by resolving orbits near the Event Horizon? How do Black Holes accrete matter and create powerful jets? (Doeleman~et~al.,~2009~\cite{doeleman09}).

Recent important observational facts also come from studies of the black hole (BH) candidates at the centers of luminous Active Galactic Nuclei and stellar mass Black Hole Candidates. The analysis of the iron $K_{\alpha}$ line profiles and luminosity variability gave amazing result: the estimated radius of the inner edge ($R_{in}$) of the accretion disk around central relativistic compact objects (RCO) is about $(1.2 - 1.4) R_g$, where $R_g = GM/c^2 = R_{Sch}/2$, i.e.,
less than the Schwarzschild radius $R_{Sch}$ of corresponding central mass
(Fabian 2015~\cite{fabian15}, Wilkins \& Gallo 2015~\cite{wilkins15},
King~et~al.,~2013~\cite{king13}).
Existing observational data~\cite{king13} demonstrate that, in the nature, the Schwarzschild radius is not a limiting size of relativistic
compact objects (RCO).
For example, in~the case of Seyfert 1 galaxy Mrk335
$R_{in} \approx 0.615\, R_{Sch} = 1.23 R_g$,
which means that BH should be a Kerr BH rotating with linear velocity about 0.998c. Also,~the~emissivity profile sharply increases to smaller radius of the \mbox{disk (Wilkins 2015~\cite{wilkins15}).}

\subsubsection{Tensions between Local and Global Cosmological~Parameters}
Modern cosmological observations are well described by the standard cosmological LCDM model based on Friedmann's solutions of GRT field equations
(Debono \& Smoot 2016 review~\cite{debono2016}). 

However,
in the last years, there has been growing evidence for a number of “tensions” between the derived parameters of the early Global Universe and measured parameters of the late Local Universe. Both the Cosmic
Microwave Background (CMB) data and the Local Universe (LU) observations have revealed an underlying discrepancies that cannot be ignored 
Verde, Treu \& Riess 2020~\cite{verde2019},
Di~Valentino, Melchiorri \& Silk 2020a, 2020b
~\cite{divalentino2020a,divalentino2020b}.

It comes from comparison of the measured Hubble Constant
for the early and late Universe, so-called “the $H_0$ tension”
Riess~et~al.,~2020~\cite{riess2020}; 
Verde~et~al.,~2019~\cite{verde2019};
Lin~et~al.,~2019~\cite{lin2019}; and, from uncertainty in curvature density parameter---“the curvature tension”
Handley 2019~\cite{handley2019}.
Additionally,~problems in the galaxy formation theory have raised the fundamental question on necessity for consideration a more general initial conditions in the cosmological
N-body simulations 
Peebles 2020~\cite{peebles2020}, 
Benhaiem, Sylos Labini \& Joyce 2019~\cite{benhaiem2019}.

Further evidence on modern “crisis” in cosmology was found from recent combined analysis of the Planck CMB power spectra anisotropy and large-scale structure data (Di Valentino~et~al.,~2020a, 2020b~\cite{divalentino2020a,divalentino2020b})   . 
Their results cast doubt on the standard values of basic
LCDM parameters for inflation, non-baryonic dark matter and dark energy. Instead of the flat universe with constant cosmological term, they suggest to consider the Phantom Dark Energy Closed (PhDEC) cosmological~model.

The conclusion of the recent works~\cite{divalentino2020a,divalentino2020b,handley2019,riess2020,verde2019,lin2019} 
is that either LCDM needs to be replaced by a
drastically different model or~else there are significant, but still undetected, systematics. 
The new theoretical suggestions call for new
observations and stimulate the investigation of alternative theoretical models and~solutions.

In addition, there are number of observational and conceptual difficulties of LCDM scenario, which~discussed in the contemporary literature. Among~them, the~discovery of strong inhomogeneous galaxy distribution and galaxy flows  in the Local Universe, which include such super-large structures as Great Wall, Sloan Wall,  
South Pole Wall, and others, with sizes up to 400 Mpc 
Pomarede~et~al.,~2020~\cite{pomarede2020},
Tully~et~al.,~2019~\cite{tully2019},
Hoffman~et~al.,~2017~\cite{hoffman2017}, and
Tully~et~al.,~2014~\cite{tully14}.
The size 400 Mpc is much more than the 120 Mpc of the baryon acoustic oscillations, where galaxy correlation function of the standard LCDM model must crosses zero level. Complex walls-filament-void  structures was discovered by means of distance measurements  independent on redshifts in real space for CosmicFlows-2 catalog
(Courtois~et~al.,~2013~\cite{courtois2013})
and power-law correlation function in redshift space for 2MRS catalog (Tekhanovich \& Baryshev 2016
~\cite{tekhanovich16}).

Additionally, such problems are discussed: the cold dark matter crisis on galactic and sub-galactic scales
(Kroupa 2012~\cite{kroupa12}); the LCDM crisis at super-large scales
(Sylos Labini 2011~\cite{sylos11},
Clowes~et~al.,~2013~\cite{clowes13},
Horvath~et~al.,~2015~\cite{horvath15};
Shirokov~et~al.,~2016~\cite{shirokov16}),
existence at high redshifts the old galaxies and the supermassive quasars
~\cite{dolgov2018, dolgov2019,yang2020}.

The general approach to the observational testing of the standard and possible alternative cosmological models was formulated as the ``practical cosmology''
by Allan Sandage 1997~\cite{sandage97}. Modern state of the practical cosmology
is presented in~\cite{baryshev-t12} and applied in
~\cite{shirokov20a, shirokov20b}.

\subsubsection{Conceptual Problems of the Gravity~Physics}

Direct detection (localization) of the gravitational waves by LIGO-Virgo antennas raises the conceptual problem of the GW energy localization in geometrical gravity theory. For~the GRT, it is a consequence of the equivalence principle in the metric gravity theories: ``...This corresponds completely to the fact that by a suitable choice of coordinates, we can 'annihilate' the gravitational field in a given volume element, in~which case, from~what has been said, the~pseudotensor $t^{ik}$ also vanishes in this volume element'' (\cite{landau71}, \S101, p. 307). Note that there is no
such problem in electrodynamics and Feynman's quantum-field gravitation theory, where the field energy-momentum tensor exists and GW energy density $T^{00}_g =\epsilon_g (t, \vec{r})$ 
in classical limit is well-defined in each point of the Minkowski~space. 

Conceptual obstacles of GRT, which are directly related to these observations, include well-known ``energy-momentum pseudo-tensor'' and ``horizon'' problems. The~energy localization problem is that, within GRT, there is no tensor characteristics of the energy-momentum for the gravity \mbox{field
\cite{trautman66,landau71,misner73,ehlers07,maggiore08,baryshev08a,strauman13}}.
Landau \& Lifshitz 1971~\cite{landau71} called this quantity pseudo-tensor of energy-momentum and noted that covariant divergence of the total energy-momentum
tensor (right side of the Einstein's field equations Equation~(\ref{efeq1}))
does not express the energy-momentum conservation for matter plus gravity field.
The ``pseudo-tensor''(meaning non-tensor) character of the gravitational energy-momentum in GRT has been discussed from time-to-time for a century (\mbox{see a review Baryshev 2008a~\cite{baryshev08a}}), causing surprises for each new generation of physicists. However, rejecting the Minkowski space inevitably leads (according to Noether theorem) to deep difficulties with the definition and conservation of the energy-momentum for the gravitational field (\mbox{see Sections~\ref{sec1.3} and \ref{sec2.5}}).

There are several paradoxes that are related to the concept of black hole horizon, which were emphasized by Einstein 1939~\cite{einstein39}. The~information paradox was recently discussed by Hawking 2014~\cite{hawking14}, 2015~\cite{hawking15}, 't Hooft 2015~\cite{thooft15}, and~the incompatibility of classical and quantum concepts of the BH horizon was considered by Chowdhury \& Krauss 2014~\cite{chowdhury14}.
The infinite time formation of the classical BH event horizon (in the distant observer's coordinates) and finite time of BH quantum evaporation means that a BH should evaporate before its formation \cite{chowdhury14}.
Stephen Hawking claimed in~\cite{hawking14} that
``There would be no event horizons and no firewalls.  The~absence of event horizons mean that there are no black holes - in the sense of regimes from which light can't escape to infinity''.
Although there is no escape from a black hole in classical theory, but,~in quantum theory, energy and information can escape from a black hole. It means that an explanation of the gravity physics requires a theory that successfully merges gravity with the quantum fields of other fundamental forces of nature
(actually this is the goal of the field gravitation theory, as~we discussed below).

The theory of internal structure of the neutron stars is based on the Tolman--Oppenheimer--Volkov GRT equation of hydrostatic equilibrium for the high-density equation of state.
Importantly, first~radius-mass measurements of several neutron stars,  discovered that  there is no reasonable equations of state, which is able to describe the observed parameters of the pulsars, e.g.,~millisecond pulsar PSR J0030+0451 observed by NICER (Miller~et~al.,~2019~\cite{miller2019}). Additionally, difficulties arise for coalescence of two neutron stars in the event GW170817/GRB170817A giving the resulting mass (\mbox{2.73--3.07})$M_{Sun}$ Abbott~et~al.,~2020~\cite{abbott2020}.  

Additional conceptual problem of the BH physics arises when one considers the process of BH formation from coalesces of two neutron stars or two black holes. The~point is that the any binary system of relativistic compact objects has finite binding energy, equals the work against the gravity force needed for destruction of the binary system. However, after the coalescence to one black hole, we get at the classical level the ''infinite binding energy'', because~of impossibility to destroy the~BH. 

For cosmological solutions of GRT equations, there are several conceptual questions discussed in the literature: the Newtonian character of the exact Friedmann equation
(Baryshev 2008c~\cite{baryshev08c}, 2015~\cite{baryshev15});
violation of the energy-momentum conservation within any comoving local volume
(Harrison 1995~\cite{harrison95}, Baryshev 2008c~\cite{baryshev08c},
2015~\cite{baryshev15});
violation of the velocity of light by space expansion velocity for galaxies
observed at high redshifts (Harrison 1993~\cite{harrison93},  2000~\cite{harrison00},
Baryshev~\&~Teerikorpi 2012~\cite{baryshev-t12}
Baryshev 2015~\cite{baryshev15}).

So all, mentioned above, recently discussed the observational and theoretical problems of gravitation theory and cosmology point to need for reanalysis of alternative possibilities for construction of the theory of gravitational~interaction.

\subsection{The Quest for Unification of of the Gravity with Other Fundamental~Forces}\label{sec1.2}

The success of the Standard Model of electromagnetic, weak, and strong interactions was achieved on the way of unification of the fundamental physical forces in the frame of the quantum field theory (QFT). Now, it has reached a respectable status as an accurate and well-studied description of sub-atomic forces and particles, though~difficult conceptual and technical problems remain to be solved
(Bogolubov \& Shirkov 1993~\cite{bogolubov93};
Wilczek 1999~\cite{wilczek99}, 2015a~\cite{wilczek15a},
2015b~\cite{wilczek15b};
Blagojevich 1999~\cite{blago99};
Pavsic 2002~\cite{pavsic02};
't Hooft 2004~\cite{thooft04};
Maggiore 2005~\cite{maggiore05}; and,
\emph{``Approaches to Fundamental Physics''} 2007~~\cite{approaches07}).

Especially important that the unification of gravitational interaction with general quantum physics includes all fundamental principles of modern quantum theory, such as particle-wave duality, quantum uncertainty, amplitude probability for exchange by energy quanta, and, in~particular, recently studied irreversibility, non-locality, and quantum entanglement (Kadomtsev 2003~\cite{kadomtsev2003},
Rauch~et~al.,~2018~\cite{rauch2018},
Erhard, Krenn \& Zeilinger 2020~\cite{erhard2020}).

\subsubsection{Future Unified~Theory}
It is expected that future ``Core Theory'' of physics will unify all fundamental forces (electromagnetic, weak, strong and gravitation) and also deliver
unification of forces (bosons) and substances (fermions) via transformations of supersymmetry (Wilczek 2012, 2015a~\cite{wilczek12,wilczek15a}).

There is an important obstacle for the unification of  fundamental forces with the geometrical gravitation theory (general relativity theory---GRT):  the conceptual basis of GRT is principally different from the Standard Model
(Ehlers 2007~\cite{ehlers07}; Approaches to Fundamental Physics 2007~\cite{approaches07}).
Gravity in the frame of GRT is not a force
(de Sitter 1916~\cite{desitter16a}) and has no generally covariant EMT, so quantization is applied to the curved  Riemannian space-time
(Rovelli 2004~\cite{rovelli04,approaches07}). However the concept of gravitation energy quanta cannot be properly (tensorial) defined in a theory,
 where the energy of gravitational field is not localized
 (Ehlers 2007~\cite{ehlers07}).

The QFT reconciled Quantum Mechanics with the Relativistic Field Theory by construction of interacting substances via material fields that does obey the laws of Lorentz invariance, gauge invariance, and causality.
The concept of a field energy has crucial meaning in the QFT, because
the energy in a quantized field comes in quantized energy packages, which, in all respects, behave like elementary particles.
The association of forces (or, more generally, interactions) with exchange of particles is a general feature of quantum field theory~\cite{approaches07}.
Electric and magnetic forces between charged particles are
explained as due to one particle acting as a source for electric and magnetic fields, which then influence others. With~the correspondence of fields and particles, as~it arises in quantum field theory, Maxwell's ED corresponds to the existence of photons, and~the generation of forces by intermediary fields via the exchange of real and virtual~photons.

\subsubsection{Quantum Electrodynamics as the Paradigmatic~Theory}
\label{sec-1-2-2}
The first step for constructing quantum electrodynamics (QED) is to develop
the classical electrodynamics (ED)---the
relativistic classical vector field ($A^i(x^k)$).
In this paper, the ED theory will be used as a primary example for preparation of the classical part of the QFT. Accordingly, below, I emphasize the crucial points of ED (Landau \& Lifshitz 1971~\cite{landau71}), 
\footnote{We use main definitions and notations similar to Landau \& Lifshitz~\cite{landau71},
so the Minkowski metric $\eta_{ik}$ has signature $(+,-,-,-)$, 4-dimensional tensor indices are denoted by Latin letters $i,k,l...$ which take on the values 0, 1, 2, 3, and~Greek letters $\alpha,\beta,\mu,\nu...$  take the values 1,2,3.}
which will be compared with geometrical and field gravitation theories. The~second step is to unite the relativistic classical field with quantum mechanical~principles.

\vspace{10pt}
\textit{Classical electrodynamics.}
As the basic principles of ED one may consider following~items:
\begin{itemize}
\item[$\bullet$]{\emph{the inertial reference frames;}}
\item[$\bullet$]{\emph{the flat Minkowski space-time;}}
\item[$\bullet$]{\emph{the relativistic vector field $A^i(t,\vec{x})$;}}
\item[$\bullet$]{\emph{the Least (Stationary)  Action Principle;}}
\item[$\bullet$]{\emph{the conservation of charges;}}
\item[$\bullet$]{\emph{the gauge invariance principle;}}
\item[$\bullet$]{\emph{the localizable positive energy density ($T^{00}_{(em)}$) of the field.}}
\end{itemize}

The action $S$ for the system, containing  electromagnetic field with charged particles, must consist of three parts:
\begin{equation}
 \label{act-em}
  S = S_{(\mathrm{f})} + S_{(\mathrm{int})} + S_{(\mathrm{m})} = -\frac{1}{c}
  \int \left(\frac{1}{16\pi}F_{ik}F^{ik} + \frac{1}{c}A_i j^i +
  \eta_{ik}T_{(p)}^{ik}
  \right) d\Omega\,.
\end{equation}
The notations (f), (int), and (m)
refer to the actions for the electromagnetic field, the~interaction, and~the particles. The~physical dimension of each part of the action is
$$[S]= [energy\,\, density]\times [volume]\times[time],$$
meaning that the  \textbf{definition of energy density of the field must exist} within the conceptual bases of the principle of stationary action,
$j^{\,i}$-four-current,
$A^i$-four-potential, and~$F^{ik}$-electromagnetic field tensor
\begin{equation}
\label{4-pot-em}
A^i = (\varphi, \vec{A})\,,\,\,\,\,\,F_{ik}=A_{k,i} - A_{i,k}
\end{equation}

From the Least Action Principle ($\delta S =0$) by means of the variation of four-potentials $A^{\,i}$  for the case of \textbf{ fixed sources $j^{\,i}$ } we get field
equations with conserved sources
\begin{equation}
\label{4-field-eq-em}
(A^{k,i} - A^{i,k})_{,\,k} = - \frac{4\pi}{c} j^{\,i}\,\,\,\,\,\,
where \,\,\,\,\,\,\, j^{\,i}_{\,\,,i} =0 \,.
\end{equation}
Following Schwinger's ''source theory''
\cite{schwinger73,schwinger76}
in ED the
electromagnetic field source is 4-current  $j^{\,i} (x^k) = (c\rho_e, \vec{j}\,)$,
which together with the Lorentz invariant law of charge conservation (scalar~restriction
$j^{\,i}_{\,\,,i} = 0$) excludes the scalar source of the four-vector field, i.e.,~the scalar photons. In~fact the logic of spin 0 particle exclusion is following:
\begin{equation}
\label{source-exclusion}
current \, conservation \Rightarrow  scalar\, source\,
exclusion \Rightarrow
gauge \, invariance \Rightarrow constraint\, field
\end{equation}
The left side of the field equations Equation~(\ref{4-field-eq-em}) allows for the gauge invariance in the form:
\begin{equation}
\label{gauge-em}
A^i \rightarrow A^i + \xi^{,i}
\end{equation}
which allows for it to use the Lorentz gauge condition
\begin{equation} \label{lorentz-gauge}
A^i_{,\,i}=0 \,\,\,\,\,\,\textrm {i.e.,}\,\,\,\,\,\,
\frac{1}{c}\frac{\partial \varphi}{\partial t} +
\rm{div}  \vec{A} = 0
\end{equation}
and the final field equations has ordinary wave equation form
\begin{equation} \label{4-f2-eq-em}
\left(\triangle - \frac{1}{c^2}
\frac{\partial^2}{\partial t^2} \right) A^i = - \frac{4\pi}{c} j^{\,i}\,\,\,.
\end{equation}

The gauge invariance Equation~(\ref{gauge-em}) is consistent with the conservation of
the source of the field Equation~(\ref{4-field-eq-em}) and with the deleting of the ``scalar'' photons.
Indeed, the~four-vector field (four components) can be decomposed  into (3 + 1) components~\cite{fronsdal58,barnes65}.
Four-potential $A^i$ has four independent components, which correspond to
one spin-1 ($2s+1=3$) and one spin-0 ($2s+1=1$) particles, then current conservation law allows for excluding the source of the spin 0 particles, so only photon with spin 1 is real:
\begin{equation}
\label{4-comp-em}
  \{A^{i}\} = \{1\} \oplus \{0\} \,\,\,\Rightarrow\,\,
\,\,current\,\, conservation \,\, \Rightarrow \,\,\, \{A^{i}\} = \{1\} \,.
\end{equation}

The canonical energy-momentum tensor (EMT) of the electromagnetic field, after~symmetrization, has the form:
\begin{equation} 
\label{EMT-em}
T^{ik}_{(em)} = \frac{1}{4\pi}(-F^{il}F^k_l +
\frac{1}{4}\eta^{ik}F_{lm}F^{lm}) \,\,\,,
\end{equation}
which has following important~features:
\begin{itemize}
\item[$\bullet$]{\emph{$T^{ik}_{(em)}=
T^{ki}_{(em)}$ -symmetry condition;   }}
\item[$\bullet$]{\emph{$T^{00}_{(em)} = (E^2 + H^2)/8\pi > 0$ - localizable field energy density, positive for both static and wave field, corresponding to the positive photon energy $E_{photon} = h\nu$;}}
\item[$\bullet$]{\emph{$T_{(em)}= \eta_{ik}T^{ik}_{(em)} =0 $ -trace of the EMT is zero;
for mass-less particles (photons);}}
\item[$\bullet$]{\emph{the EMT from action $S$ is defined not uniquely; and,}}
\item[$\bullet$]{\emph{the EMT is gauge invariant.}}
\end{itemize}

The localization and positiveness  of the energy density of the electromagnetic field means that the 00-component 
of the energy-momentum tensor 
$T^{00}_{(em)}(\vec{r}, t) >0$
is defined for any point $(\vec{r}, t)$ of the Minkowski space-time
and it can be transformed (localized) in the
kinetic energy of charged particles (e.g.,~detection of an electromagnetic
wave).

Considering in action $S$
variation the trajectory of a moving charge particle for the case of the
\textbf{fixed four-potential}
gives the four-equations of motion for charged particle:
\begin{equation}
\label{eq-motion-em}
mc\frac{du_i}{ds} = \frac{e}{c}F_{ik}u^k \,\,\,,
\end{equation}
or in 3-d form it gives the Lorentz force ($i=\alpha$) and its work ($i=0$):
\begin{equation}
\label{eq-3d-motion-em}
\frac{d\vec{p}}{dt} = e\vec{E} + \frac{e}{c}[\vec{v}\times\vec{H}] \,\,\,,
\end{equation}
and
\begin{equation}
\label{eq-kin-energy-em}
\frac{dE_{kin}}{dt} = e\vec{E}\cdot \vec{v} \,\,\,.
\end{equation}
Thus, in~ED, the~fundamental role plays the \textbf{concept of force, work produced by force, positive energy density of the field, and its localization}.

\textit{Quantum extension of electrodynamics.}
Adding to ED the quantum physical requirements---\mbox{the uncertainty} principle, probability amplitudes,
the principle of superposition, quanta of the field energy as mediators of force, and~others, the~quantum electrodynamics (QED) was constructed in the frame of QFT and~then unified into
electro-weak theory and grand unified theory. The~canonical quantization and Feynman's functional integral quantum field theory are presented in  general textbooks on QFT (e.g.,~Bogolubov \& Shirkov 1982~\cite{ bogolubov93},
Ryder  1984~\cite{ryder1984},
\mbox{Sadovskii  2019~\cite{sadovskii2019}}).

The QED  uses the concept of  ``force-mediating quantum particles''  for describing the electromagnetic force.
The virtual-particle description of static force explains  the inverse-square behavior of the Coulomb law and predict the repulsive force between the charges with the same~signs.

In the path integral formulation of the QED, from~the Lagrangians in action Equation~(\ref{act-em}), while taking~into account the field
Equation~(\ref{4-f2-eq-em}), together with the Lorentz invariant law of current conservation (scalar restriction $j^{\,i}_{\,\,,i} = 0$), 
one can get the amplitude of particles exchange in the form 
$-j'^m\,(\eta_{mn}/k^2)\,j^n$, where the photon propagator is 
$\,\,D_{mn}(k)=\eta_{mn}/k^2$.

The  quantum current-current interaction amplitude, in~the case of exchange of one particle having  four-momentum $k^m =(\omega,\,\kappa,\,0,\,0)$,  can be written in the form (Feynman~et~al.,~1995~\cite{feynman95}):
\begin{equation}
- j'_m \left(\frac{1}{k^2} 
\right)\,j^m\,\, =\,\, 
\frac{j'_0\,j_0}{\kappa^2} + \frac{1}{\omega^2 - \kappa^2}
\left(j'_2 \,j_2 + j'_3 \,j_3
\right)\,,
\label{q-ampl-em}
\end{equation}
where the frequency dependent part of the amplitude Equation~(\ref{q-ampl-em}) gives electromagnetic~radiation.

The instantaneous part of the amplitude Equation~(\ref{q-ampl-em}) corresponds to the Poisson equation for electrostatic potential
\begin{equation}
\Delta \varphi = -4\pi\,\rho_e
\label{coulomb-e} \,\, ,   
\end{equation}
which, now, has the quantum nature. 
Thus, the Coulomb law for the electrostatic force is explained through the probability amplitude, which gives the energy of interaction
\begin{equation}
U(r) = \frac{e^{-}_1\,e^{-}_2}{r}
= - \frac{e^{-}_1\,e^{+}_2}{r}\,\,,
\end{equation}
and corresponds to the repulsion for similar signs and to the attraction of opposite sings of~charges.

Note that, in the case of massive photon (Proca mass term), the propagator can be written in the form
$\,\,D_{mn}(k)=\eta_{mn}/(k^2-m^2$).
As a result, the~exchange amplitude between two conserved sources is the same in the limit  $m \rightarrow 0$,
no matter whether the vector field $A^i$ is intrinsically massive (propagates~three~degrees of freedom) or if it is massless 
(propagates two degrees of freedom). 
Therefore, it is impossible to probe the difference between an exactly
massive vector field and a massive one with arbitrarily small mass
~\cite{derham2014}.

\subsection{ Einstein's Geometrical and Feynman's Quantum-Field Gravitation~Physics}\label{sec1.3}

\subsubsection{Two Ways in Gravity~Physics}
\label{sec-1-3}
Since the beginning of the 20th century  two really alternative approaches were put forward  for the description of the gravitational interaction
in theoretical~physics.

\emph{The first approach} is the geometrical Einstein's \emph{general relativity theory}
(GRT), which is based on the geometrical concept of curved Riemannian
(actually pseudo-Riemannian) space and rejects  the ordinary physical concept of force in application to gravitation.
GRT was founded by Einstein 1915~\cite{einstein15}; 1916a~\cite{einstein16a}; and, Hilbert 1915~\cite{hilbert15},
and it gives an example of geometrical way in construction of gravity theory.
GRT operates with such concepts as  metric tensor $g_{ik}$, geodesics, curvature, equivalence of the free fall to the inertial motion.
Wheeler termed this approach \emph{geometrodynamics}, underlining
the fact that geometry is not a passive background but
becomes a dynamical physical entity that may be deformed, stretched
and even spread in the form of gravitational waves. Geometrical gravity
treats the gravitational interaction as the curvature of space and it has a singular
position among other physical interactions, which are based on the physical concept of the force caused by the exchange of the field quanta
in Minkowski~space.

During one hundred years, GRT was developed and successfully applied to many gravity phenomena in the Solar System, galactic, and extragalactic astronomy
(Debono \& Smoot 2016~\cite{debono2016},
Will~2014~\cite{will14}; Straumann 2013~\cite{strauman13};
Kopeikin, Efroimsky, Kaplan 2011~\cite{kopeikin11};
Schuts 2009~\cite{schuts2009};
Brumberg 1991~\cite{brumberg91};
Misner, Thorne \& Wheeler 1973~\cite{misner73};
Weinberg 1972~\cite{weinberg72}, 2008~\cite{weinberg08};
Landau \& Lifshitz 1971~\cite{landau71}; and,
Zeldovich \& Novikov 1984~\cite{zeldovich84}
).
However, general relativity is not a quantum theory and many attempts to construct geometrical quantum gravity theory have not yet brought generally accepted convincing solution of the ``Quantum theory's last challenge'' (Amelino-Camelia 2000~\cite{amelino00}; Approaches to Fundamental Physics 2007~\cite{approaches07};
Wilczek 2015a,b~\cite{wilczek15a,wilczek15b}).

\emph{The second  approach} for alternative understanding gravity was already suggested by Poincar\'{e}, who~considered gravitation as a fundamental force in
relativistic space-time. As~early as 1905, Poincar\'{e} in his work
``On the dynamics of the electron''
put forward an idea about relativistic theory for all physical
interactions, including gravity, in~flat 4-d space-time
(now called Minkowski space). He~pointed out that analogously to
electrodynamics, gravitation should propagate with the velocity
of light, and~there should exist mediators of the interaction---gravitational waves, \emph{l' onde gravifique}, as~he called them
(Poincar\'{e} 1905~\cite{poincare05}; 1906~\cite{poincare06}).
A few years later in his lecture on ``New concepts
of matter'' Poincar\'{e} wrote about inclusion Planck's discovery of the
quantum nature of electromagnetic radiation into the framework of future
physics for all fundamental interactions.
Poincar\'{e}, thus, could be rightfully regarded as the visionary of that
approach to gravity that describes gravitation as the relativistic
quantum field in Minkowski~space.

According to Feynman's \emph{Lectures on Gravitation}
(1971~\cite{feynman71}, 1995~\cite{feynman95})
(Caltech lectures in 1962--1963)
the \emph{field gravitation theory} (FGT) must be relativistic and quantum, which is described by symmetric second rank tensor field $\psi_{ik}$ in Minkowski spacetime. Accordingly, as~in the theory of electromagnetic interaction we have electrodynamics (ED) and quantum electrodynamics (QED), in~the case of FGT we should consider ``\emph{gravidynamics}'' (GD) and
\emph{quantum-field gravitation theory} (QFGT)---``\emph{quantum gravidynamics}'' (QGD).
Within QFGT, as~in QED, general concepts of force and localizable positive field energy density naturally exist, and~QFGT should be included  in the list of the field theories of fundamental physical~interactions.

Because of the great success of general relativity in explanation of existing experimental and observational facts in gravity physics, the~field gravitation theory up to now has been outside general attention. However
the field approach to gravitation was partly developed by number of famous physicists, among~them Birkhoff 1944~\cite{birkhoff44}; Moshinsky 1950~\cite{moshinsky50}; Thirring 1961~\cite{thirring61}; and,
Kalman 1961~\cite{kalman61}. Attempts for a field-theoretical description
of the gravitational field quantization were made by Bronstein 1936~\cite{bronstein36};  Fierz \& Pauli 1939~\cite{fierz39}; , Ivanenko \& Sokolov 1947~\cite{ivanenko47};  Gupta 1952a,b~\cite{gupta52a,gupta52b}; Feynman 1963~\cite{feynman63}; 1971~\cite{feynman71};  Weinberg 1965~\cite{weinberg65};  Zakharov 1965~\cite{zakharov65};  Ogievetsky \& Polubarinov 1965~\cite{ogievetsky65}, Blagojevich 1999~\cite{blago99},
Maggiore 2008~\cite{maggiore08}; and,~others.

It is important to note that in Feynman's \emph{Lectures on Gravitation} the
gravitational field is initially described as the \textbf{reducible} symmetric second rang tensor $\psi^{ik}$, which can be presented as a direct sum of three irreducible representations of the Poincare group:
four-tensor (traceless), four-vector, and four-scalar (5 + 4 + 1 = 10 components).
Gauge invariance (and corresponding EMT conservation) only excludes four components (four-vector) and, hence, leaves direct sum of \textbf{two irreducible} representations:
\textbf{spin-2  and spin-0} parts (i.e.,~six independent components).
The irreducible spin-2  representation
\textbf{$\phi^{ik} = \psi^{ik} - (1/4)\eta_{ik}\psi^{ik}$}
describes \textbf{the attractive force} and Feynman (as many other
authors of the spin-2 derivations of GRT equations) tried to construct
gravitation theory based on the spin-2 field only, so excluding spin-0~field.

The fundamental role of the 4-scalar spin-0 component
(the trace \textbf{$\psi = \eta_{ik}\psi^{ik}$}),
which is the second irreducible part of the gauged  total
reducible symmetric tensor potential $\psi^{ik}$,
was found and developed by Sokolov and Baryshev
\cite{sokolov80,baryshev-soc97,sokolov19,sokolov15,baryshev86,baryshev88,baryshev90,baryshev03,baryshev08a}.
Intriguingly this irreducible \textbf{intrinsic 4-scalar} field
corresponds to \textbf{the repulsive dynamical field}, which
in the sum with the pure spin-2 field gives the Newtonian
gravity force
and also all classical relativistic gravity effects.
As a result, a~consistent field gravity theory (FGT) has been developed,
where the central role
belongs to the inertial frames, Minkowski space and localizable positive
energy of the gravitational field, \textbf{including its scalar part}.
Although many important questions are waiting for further~work.

The relation between GRT and FGT was discussed in the literature with a very wide spectrum of opinions. There is a statement about
the identity of the field gravity to the
general relativity, so they are ``just different languages'' leading to the same  experimental predictions,
and after ``repairing'' the spin-2 approach becomes GRT
(Misner, Thorne, Wheeler 1973~\cite{misner73}).
Additionally, there is the claim, that the metric gravity theory is the only possible way to construct the correct gravitation theory (Misner,~Thorne,~Wheeler 1973~\cite{misner73}; Straumann 2013~\cite{strauman13}).
Let us consider the real state of art of the~problem.

\subsubsection{Special Features of the Geometrical~Approach}
Within a geometrical approach, the gravitational interaction is described
as a curvature of space-time with the metric $g_{ik}$. A~deep analysis of the GRT basic principles was done by Ehlers 2007~\cite{ehlers07} and Straumann 2013
~\cite{strauman13}. Here, I emphasize that GRT is a  non-quantum relativistic
theory of the gravitational interaction and based on the following
fundamental concepts:

{\small
\begin{itemize}
\item[$\bullet$]{\emph{the non-inertial reference frames;}}
\item[$\bullet$]{\emph{the equivalence principle and geometrization
    of gravity;}}
\item[$\bullet$]{\emph{the curved Riemannian space-time with metric $g_{ik}$;}}
\item[$\bullet$]{\emph{the geodesic motion of matter and light;}}
\item[$\bullet$]{\emph{the general covariance; and,}}
\item[$\bullet$]{\emph{the geometrical extension of Stationary Action Principle.}}
\end{itemize}
}
Note that the equivalence principle (EP) played an important role in the history of the general relativity formulation. EP has many forms---from non-relativistic to philosophical, which are not equivalent and difficult to test. Actually in experiment one tests the \emph{universality of the free fall}, which is expected to be independent on the structure and motion of a test body
(also known as the \emph{effacing principle} \cite{kopeikin11,strauman13}).
Another form of the EP is the \emph{geometrization} principle, i.e.,~the metric representation of gravitational potentials and geodesic motion in Riemannian space, which now is considered to be the primary initial assumption of the geometrical approach (Ehlers 2007~\cite{ehlers07}).
However, the geometrical quantum gravity approach predicts violation of the EP and Lorentz invariance (Amelino-Camelia~et~al.,~2005~\cite{amelino05};  Bertolami~et~al.,~2006~\cite{bertolami06}).

On the bases of its initial principles, general relativity was developed and successfully applied  to the number of experiments and observations in the weak gravity conditions (\cite{will14,strauman13}).
Strong~gravity~GR~predictions, like gravitational collapse to singularity,
black hole existence, and~global space expansion, may only be observed within astrophysical conditions, where the interpretation of the data allows for different possibilities due to specific passive character of the astronomical observations and the dominance of
distortion and selection effects which influence real astronomical data. So, in~spite of many claims about proved existence of black holes and space expansion,  up~to now there is no direct \textbf{experimental/observational proof} of the GRT strong gravity effects,
which are still hypothetical models for the observed astrophysical phenomena (see discussion in Sections~\ref{sec4} and \ref{sec5}).

In conditions of the weak gravity
general relativity is a well verified theory.
It has passed all available tests in the Solar System and binary pulsars.
Nevertheless, more accurate and conceptually new
tests in the weak-field regime are still needed, as well as tests of strong-gravity effects
(Fabian 2015~\cite{fabian15};
 Wilkins 2015~\cite{wilkins15}; Baryshev 2015~\cite{baryshev15};
 Sokolov 2015~\cite{sokolov15};
 Will 2014~\cite{will14};  Doeleman 2009~\cite{doeleman09};
 Baryshev 2008b~\cite{baryshev08b}; and,
 Bertolami~et~al.,~2006~\cite{bertolami06}).

\subsubsection{Problem of the Gravitational Field Energy-Momentum  in~GRT}

The most puzzling feature of general relativity is
the absence of the tensor character of the ``energy-momentum tensor''
for the gravity ``field''.
This was clearly exposed already by
Einstein 1918~\cite{einstein18a,einstein18b};
Schr\"{o}dinger 1918~\cite{schrodinger18}; Bauer 1918~\cite{bauer18},
and more recently discussed by
Landau \& Lifshitz 1971~\cite{landau71};
Misner, Thorne \& Wheeler 1973~\cite{misner73};
Logunov \& Folomeshkin 1977~\cite{logunov77};
Strauman 2000~\cite{strauman00}, 2013~\cite{strauman13};
Pitts \& Schieve 2001~\cite{pitts01}; Xulu 2003~\cite{xulu03};
Ehlers 2007~\cite{ehlers07}; and,
Baryshev 2008a~\cite{baryshev08a}.

The problem of the energy of the gravity field in general relativity has a long history, it was, in~fact, born together with Einstein's equations. Hilbert 1917~\cite{hilbert17}   was the first who noted that ``I contended ... in general
relativity ... no equations of energy ... corresponding to those in orthogonally invariant theories''. Here, ``orthogonal invariance'' refers to theories in the flat Minkowski space.
Emmy~Noether~1918~\cite{noether18}, a~pupil of Hilbert, proved that the symmetry of Minkowski space is the cause of the conservation of the energy-momentum tensor of all physical fields. Many results of modern relativistic quantum field theory are based on this theorem. Accordingly, the \textbf{``prior geometry'' of the
Minkowski space } in the field theories has the advantage of
\textbf{guarantee the tensor character of the energy-momentum, its localization and its conservation} for the fields. However, in GRT, there is no global Minkowski space, so there is no EMT of the gravitation field and its~conservation.

In fact, Einstein \& Grossmann 1913~\cite{einstein13} came close to Noether's result when they wrote: ``remarkably the conservation laws allow one to give a physical definition of the straight line, though, in our theory, there is no object or process modeling the straight line, like a light beam in ordinary relativity theory''. In~other words, they stated that the existence of conservation laws implies the flat Minkowski geometry. In~the same article, Einstein \& Grossmann also emphasized that the gravity field must have an energy-momentum tensor as all other physical fields. However, in~the final version of general relativity, Einstein rejected this requirement in order to have a generally covariant gravity theory with no prior Minkowski~geometry.

Schrodinger 1918~\cite{schrodinger18} showed that the mathematical object
$t^{ik}$
suggested by Einstein in his final general relativity for describing the energy-momentum of the gravity field may be made vanish by a coordinate transformation for the Schwarzschild solution if that solution is transformed to Cartesian coordinates. Bauer 1918~\cite{bauer18} pointed out that Einstein's energy-momentum object, when calculated for a flat space-time, but in a curvilinear system of coordinates, leads to a nonzero result. In~other words,
$t^{ik}$ can be zero when it should not be, and~it can be nonzero when it should
(this also emphasized by Landau \& Lifshitz 1971~\cite{landau71} \S 101 p. 307).

Einstein 1918a~\cite{einstein18a} replied that already Nordstrom informed him about this problem with $t^{ik}$. Einstein noted that in his theory $t^{ik}$ is not a tensor and also it is not symmetric. He also withdrew his previous demand of the necessity to have an energy-momentum tensor: ``there may very well be gravitational fields without stress and energy~density''.

The ``pseudo-tensor'' (meaning ``non-tensor'') character of the gravity field in GR has simple mathematical cause. As~emphasized by Landau \& Lifshitz 1971
(\cite{landau71} \S 101 p. 304) due to Bianchi identity the \emph{covariant divergence} of the right part of Einstein's equation (which is the EMT of matter $T^{ik}_{(m)}$) is equal to zero, i.e.,~$T^{ik}_{(m)\,\,;k} = 0$.
However, for conserved quantity one should have the ordinary partial divergence:
$((\sqrt{-g})T^{ik}_{(m)})_{\,\,,k} = 0$.
Accordingly, Landau \& Lifshitz suggested to consider
pseudo-tensor (non-tensor)  of energy-momentum of gravitational field which should be added
to the EMT of matter and allow to fulfill the needed equation
$((\sqrt{-g})(T^{ik}_{(m)} + t^{ik}))_{\,\,,k} = 0$.

There are many different expressions for pseudo-tensors, but the problem of
coordinate dependent (non-physical) definition of the
gravity energy-momentum still exists at fundamental level---gravitational field is not a matter within GRT.
This also demonstrates that
rejecting the Minkowski space inevitably leads to deep difficulties with the definition and conservation of the energy-momentum for the gravity~field.

\subsubsection{Attempts to Resolve the Gravitational
Energy-Momentum Problem in Geometrical~Approach} 

The main question of the gravity physics is the role of
the global Minkowski space in the gravitation theory.
Within the geometrical approach Minkowski space is a tangent
space at each point of curved space (the local Lorentz invariance).
The field
approach utilizes the global Minkowski space to describe all
four fundamental physical interactions as material fields
in~space.

According to Noether's theorem in the relativistic field theory, the conservation of the energy-momentum relates to the flat global Minkowski space.
However, in general relativity there are no conservation laws
for the energy-momentum of the matter plus gravity field,
just because of the absence of the global Minkowski space.
The energy problem has deep roots in  the geometrical approach, which uses
curved space and non-inertial reference frames, while the field approach based on the Minkowski space with inertial reference frames naturally contains local EMT for the gravity~field.

Note that, in GRT, there is a suggestion for consideration of observable gravity effects without using the general covariant concept of the gravitational energy
(Strauman 2013~\cite{strauman13}; Maggiore 2008~\cite{maggiore08}).
Also in GRT there are attempts to construct of a ``quasi-local'' energy-momentum and angular momentum  to save the physical concept of the energy for the gravity ``field'' (Szabados~2009~\cite{szabados09}).
There are also several suggestions how to overcome the energy-momentum problem by a modification of general relativity, or~by postulating additional constraints on the metric of the Riemannian space, or~by introducing Minkowski and Riemannian metrics together.
This  has led to different ``field-geometrical''  gravity theories
(which actually belong to metric gravity theories)
having different equations and predictions
( e.g.,~Logunov 2002~\cite{logunov02};
Logunov \& Mestvirishvili 1989~\cite{logunov89}, 2001~\cite{logunov01};
Yilmaz 1992~\cite{yilmaz92};
Babak \& Grishchuk 2000~\cite{babak00};
Pitts \& Schieve 2001~\cite{pitts01}; and, Xulu 2003~\cite{xulu03}).

All these theories are geometrical (they use \textbf{geometrization principle}) and they predict some differences with GRT only in the case of the strong gravity field, which is not directly observed yet.
However, as~we shall show below, the~results of the consistent Poincare--Feynman
field approach has led to predictions that differ from GRT, even in the weak field conditions, which, in principle, can be tested by experiments in the Earth laboratories and by observations while using terrestrial and space~observatories.

\subsubsection{Special Features of the Feynman~Approach}

 Feynman discussed the strategy of the QFGT and suggested constructing
``theory of gravitation as the 31st field to be discovered''
(\cite{feynman95} p. 15).
He analyzed basic principles of the QFGT  and emphasized, that ``geometrical interpretation is not really necessary or essential for physics''(\cite{feynman95} p. 113).

Accordingly, the natural relativistic quantum field approach to gravitational interaction should be developed on the way where
other fundamental interactions
already have been constructed. Feynman emphasized that the
``world cannot be one-half quantum and one-half classical'' and
``it should be impossible to destroy the quantum nature of fields''
(\cite{feynman95} p. 12).

Modern physics deals with four presently known fundamental interactions:
the electromagnetic, the~weak, the~strong, and the gravitational.
The first three interactions are described by using
Lagrangian formalism of the relativistic quantum field theory in Minkowski space. The~QFGT theory also should be based on the same Lagrangian concepts, also including specific scalar-tensor character of the gravitational
field:
{\small
\begin{itemize}
\item[$\bullet$]{\emph{the inertial reference frames;}}
\item[$\bullet$]{\emph{the flat Minkowski space-time wih metric $\eta^{ik}$;}}
\item[$\bullet$]{\emph{the reducible symmetric tensor potentials $\psi^{ik}(x^m)$
with trace $\psi(x^m) = \psi^{ik}\eta_{ik} $;}}
\item[$\bullet$]{\emph{the universality of gravitational interaction;}}
\item[$\bullet$]{\emph{the Stationary  Action Principle (Lagrangian formalism);}}
\item[$\bullet$]{\emph{the conservation law of energy-momentum tensor;}}
\item[$\bullet$]{\emph{the gauge invariance principle;}}
\item[$\bullet$]{\emph{the localizable positive energy density of the gravitational field;}}
\item[$\bullet$]{\emph{the gravitational field energy quanta as mediators of the gravity force; and}}
\item[$\bullet$]{\emph{the uncertainty principle and other quantum postulates.}}
\end{itemize}
}
In Section~\ref{sec-3}, we discuss how to construct the consistent Poincare--Feynman field gravity theory based on these initial principles.
The energy of the gravitational field should play the central role in a reasonable theory of gravitational interaction. Feynman's notorious words in a letter to his wife ``Remind me not to come to any more gravity conferences''
(\cite{feynman95} Foreword p. xxvii)
are related to this very issue, he did not wish to discuss the question of whether there is energy of the gravitational field. For~him, gravitons were particles carrying the energy-momentum of the field: ``the situation is exactly analogous to electrodynamics---and in the quantum interpretation, every radiated graviton carries away an amount of energy
$\hbar \omega$'' (\cite{feynman95} p. 220).

Nowadays, when the Nobel Prize in Physics (1993) was given for the discovery of the binary pulsar PSR 1913+16, which is emitting positive energy of gravitational radiation, and the~Advanced LIGO gw-antennas have detected the gravitational waves (i.e.,~have localized positive gw-energy), it is clear that Feynman was right when insisting on the necessity to have proper concept of the energy density of the gravitational~field.

\subsubsection{Why Is QFGT  Principally Different from GRT?}

The history of the field gravity approach is characterized by many
controversial opinions and misleading claims (a review in~\cite{baryshev96}).
From time to time at a gravity conference, a physicist appeared  who announced that he ultimately had just derived the full non-linear Einstein's equations from the spin 2 field approach and he will demonstrate it at the next conference.
However, at the next conference the situation was~repeated.

The incompatibility of geometrical (GRT) and quantum field (FGT) approaches
exists on the level of the adopted initial conceptual principles
(Ehlers 2007~\cite{ehlers07}), which we have
considered above. The~most important difference is the geometrization principle
in GRT (gravitational potentials are described by the metric tensor $g^{ik}$ of the Riemannian  space), while in FGT gravitational potential $\psi^{ik}$ is the material field in Minkowski space with metric $\eta^{ik}$. Accoridngly, the gauge transformation in FGT is related to potentials in a fixed inertial frame,
while in GRT the gauge transformation is the change of coordinates.
In the field gravity approach there is usual localizable energy-momentum tensor (EMT) of the gravitation field, while, in the geometrical approach, there is no
tensor quantity for the gravitational energy-momentum (problem of pseudo-tensor).

In the frame of QFGT, the symmetric tensor potential $\psi_{ik}$ actually
corresponds to
the  reducible representation of the Lorentz group, which can be decomposed
to the direct sum of three irreducible representations:
traceless four-tensor, four-vector, and four-scalar (5 + 4 + 1 = 10 components). 
\footnote{ Corresponding projection operators can be found in
Barnes 1965~\cite{barnes65},
Maggiore 2005~\cite{maggiore05}.}
After the four gauge conditions,
one excludes four-vector field (spin-1 and spin-0 particles: four components),
so the initial reducible tensor field will only contain two irreducible
representations corresponding to
spin-2 and spin-0 particles ($\{2\}+\{0'\} \Rightarrow  5 + 1 =6$
components).
The gauge freedom is also consistent with the
four conditions from conservation of the field source, so that
the two types of particles have corresponding parts of the source,
and the final field equations describe two real dynamical~fields.

Accordingly, according to Wigner's theorem~\cite{wigner1939}, the~QFGT is the \textbf{scalar-tensor} field gravitation theory.
Note that the \textbf{intrinsic} scalar part of the symmetric
tensor field (trace $\psi(x^m) = \psi^{ik}\eta_{ik}$)
is an observable part of the
classical gravity experiments (see Section~\ref{sec-3}).
The most radical difference of QFGT from GRT is that the field approach
works with the two parts of
the gravity physics - the traceless spin-2 \textbf{attraction} field and the
intrinsic scalar spin-0 \textbf{repulsion} field (the trace of the tensor potential).
This fact demonstrates the principal incompatibility of FGT and GRT, though~there is common region of applicability of geometrical and field approaches (coincident predictions for classical relativistic gravity effects in the weak field regime).

However, up~to now,
there are attempts to ``prove of identity'' of GRT and FGT approaches
by using two opposite ways, so called ``top-down'' and ``down-top''~argumentations.

The \emph{top-down} approach starts from the ``top'' full non-linear Einstein equations and goes to the ``down''---linear weak field approximation, where the metric tensor
$g_{ik}$ of the Riemannian space only slightly differs from the
metric tensor $\eta_{ik}$ of the Minkowski space.
Accordingly, metric tensor $g_{ik}$ is defined by the relation (first suggested by Einstein 1916b~\cite{einstein16b}):
\begin{equation}
g_{ik} = \eta_{ik} + h_{ik}
\label{n-h}
\end{equation}
where  the quantities $\mid h^{ik}\mid \ll 1$, and~the
\textbf{rigorous identities must be fulfilled} for the metric tensor
of any Riemannian space:
\begin{equation}
g^i_k \equiv \delta^i_k = diag(1,1,1,1) \,\,\,\,\,\,
{\rm{and}} \,\,\,\,\,\,
Trace(g_{ik}) = g_{ik} g^{ik} \equiv 4
\label{n-h-conditions}
\end{equation}

In this weak field approximation, the Einstein's field equations  are equivalent to the field equations of the relativistic symmetric second rank tensor field
$h_{ik}$ in Minkowski space with metric $\eta_{ik}$.
Working with the linear approximation of the Einstein's equations, one usually uses \textbf{the convention} that
indices are rased and lowered by the flat metric $\eta^{ik}$. However strictly speaking in the frame
of the geometrical approach such procedure is ``illegal'', because it violates the general covariance principle ($\eta^{ik}$ and $h^{ik}$ are not tensors of the initial curved space). For~the rasing and lowering indices, one should use the sum (Equation (\ref{n-h})), which must obey the strict identities (Equation (\ref{n-h-conditions})). As~Schuts 2009~\cite{schuts2009} (p. 199) emphasized: ``Thus 
$\eta^{im} \eta^{nk}h_{mn} = h^{ik}$---is not the deviation of $g^{ik}$ from~flatness''.  

Subsequently, in~fact  using the field-theoretical approach in Minkowski space,  one can calculate the retarded potentials and emission of gravitational waves, which  corresponds to the field quanta---spin 2 massless particles, together with additional condition $h = 0$ (TT-gauge):
(Bronstein 1936~\cite{bronstein36}; \mbox{Fierz \& Pauli 1939~\cite{fierz39}}; Ivanenko \& Sokolov 1947~\cite{ivanenko47}; Gupta 1952a,b~\cite{gupta52a,gupta52b}; Feynman 1963~\cite{feynman63}; Zakharov 1965~\cite{zakharov65}; Maggiore 2008~\cite{maggiore08}; and, Straumann 2013~\cite{strauman13}).

It is clear that such ``derivation'' does not prove an ``identity'' of GRT and FGT.
The Equation~(\ref{n-h}) means that the \emph{geometrization principle} is given up because quantities $\eta^{ik}$ and $h^{ik}$ are not tensors of the Riemannian  space, but~they are tensors only for the Minkowski space.
Here one meets the point where the initial principles of general relativity are replaced by the initial principles of the field gravitation~theory.

The \emph{down-top} approach starts from the ``down'' linear equations for material symmetric tensor potentials $\mid\psi^{ik}\mid \ll 1$ in Minkowski space and
goes to a derivation of the ``top'' nonlinear Einstein's equations for the
``effective metric tensor'' $f^{ik} = \eta^{ik} + \psi^{ik}$ of the Riemannian  space
\mbox{(Weinberg 1965~\cite{weinberg65})},
Ogievetsky \& Polubarinov 1965~\cite{ogievetsky65};
Deser 1970~\cite{deser70};
Feynman 1971~\cite{feynman71}; 1995~\cite{feynman95};
Misner, Thorne, Wheeler 1973~\cite{misner73} ).
However, there are fundamental obstacles of such transformation of material tensor field $\psi^{ik}$ of Minkowski space into non-material metric
$g^{ik} = \eta^{ik} + h^{ik}$.

First, the~strict properties of the metric tensor of the Riemannian space
Equation~(\ref{n-h-conditions})
and the general tensor rules for physical quantities in Minkowski space
demand that for the sum of two quantities
$\eta^{ik} + h^{ik}$  and  $\eta^{ik} + \psi^{ik}$,
one gets  the following expressions
(correct to the first order of $h^{ik}$ and~$\psi^{ik}$) :
$$\textbf{Geometrical\,\,\, approach}\;\;\;\;\;\;\;\;\;\; \;\;\;\;\;\; \;\;\;\;\;\;
\textbf{Field\,\,\, approach} \;\;\;\;\;\;\;\;\;\;\;\;\;  $$
\begin{eqnarray}
\label{weak-gr-fg}
  g_{ik}(\vec{r},t) = \eta_{ik} + h_{ik}(\vec{r},t)\;\;\;\;\;\;\;\;\;\;
  {\rm{and}} \;\;\;\;\;\;\;\;\;\;
  f_{ik}(\vec{r},t) = \eta_{ik} + \psi_{ik}(\vec{r},t) \;\;\;\; \;\;\;\;
  \nonumber\\
  g^{ik}(\vec{r},t) = \eta^{ik} - h^{ik}(\vec{r},t) \;\;\;\;\;\;\;
  {\rm{while}} \;\;\;\;\;\;\;\;\;\;
  f^{ik}(\vec{r},t) = \eta^{ik} + \psi^{ik}(\vec{r},t) \;\;\;\;\;\;\;
  \nonumber\\
  g^i_k \:= \delta^i_k \;\;\;\;\;\;\;\;\;\;\;\;\;\;\;\;\;
  {\rm{while}} \;\;\;\;\;\;\;\;\;
  f^i_k(\vec{r},t) \:= \delta^i_k + \psi^i_k(\vec{r},t)\;\;\;\;\;\;\;\; \;\;\;
  \nonumber\\
  g_{ik}\cdot g^{ik} = 4\;\;\;\;\;\;\;\;\;\; \;\;\;\;
  {\rm{while}} \;\;\;\;\;\;\;\;
  f_{ik}\cdot f^{ik} = 4 + 2\psi(\vec{r},t) \;\;\;\;\;\;\;\;\;\;\;\;
\end{eqnarray}
As we see from Equation~(\ref{weak-gr-fg}), there is essential difference between
the geometrical approach and the field approach.
The consistent field approach demands that the sum of
two tensors must be a tensor of Minkowski space.
Indeed the trace of the ``effective metric'' $f^{ik}$ is
a function of space-time $f^{ik}f_{ik} = 4 + 2\psi + O(\psi_{ik}^2)$,
due to the trace of the gravitational potentials $\psi^{ik}(\vec{r},t)$,
i.e.,~$Tr(\psi^{ik}) = \psi(\vec{r},t) = \eta_{ik}\psi^{ik}(\vec{r},t)$.
Hence, tensor 
$\hat{\textbf{f}}=\{f^{ik}\}=\{f_{ik}\}$
can not be the metric tensor of a Riemannian space
and, in the geometrical approach, the scalar part of the symmetric tensor
field is~lost.

From Equation~(\ref{weak-gr-fg}), we see that
a tensor of Riemannian space $g^{ik}$ is presented by the sum
of two non-tensor quantities $\eta ^{ik}$ and $|h_{ik}|<<1$.
For example,  in~the third identity of Equation~(\ref{weak-gr-fg}) the components
$h^i_k$ strictly speaking must be zero $h^i_k\equiv 0$.
The different signs of the quantities $\hat{\textbf{h}}$ for covariant and contravariant components of the metric tensor $\hat{\textbf{g}}$
are caused by the exact identity $g_{ik}\cdot g^{ik} \equiv 4$
valid for the
\emph{trace of the metric tensor} of any Riemannian four-space.
In the frame of  FGT, the tensor $f_{ik} = \eta_{ik} + \psi_{ik}$ cannot be a metric tensor of a Riemannian space (the $f_{ik}f^{ik} = 4 + 2\psi \neq 4$),
so the field approach
cannot be identical to a metric gravity~theory.

Feynman, in his lectures
on gravitation, also tried to derive the full nonlinear Einstein's Lagrangian
by iterating the Lagrangian of the spin 2 field.
Misner, Thorne \& Wheeler~\cite{misner73} (Chapter 7, p. 178)
wrote that ``tensor theory in flat spacetime is internally inconsistent;
when repaired, it becomes general relativity''.
They refereed to papers by Feynman~\cite{feynman63}, Weinberg~\cite{weinberg65}, and~Deser \cite {deser70}
on a ``field'' derivation of Einstein's equations.
However, this ``repairing'' means
replacing the field-theoretical approach in Minkowski space by the geometrization principle of the geometrical~approach.

Note that, in all such derivations as the first step, they get the ``spin-2'' field equations (\mbox{i.e.,~FGT linear approximation including }the scalar part), and~they
are equivalent to Einstein's equations in the linear  approximation
(which also include the scalar part---the trace $h = \eta_{ik}h^{ik}$).
To perform the next step to obtain nonlinear equations, one needs to fix the EMT of the gravitational field (which is the basic concept in the consistent field approach). At~this step, one should use the physical concept of the gravitational field EMT, which is not uniquely defined by Lagrangian formalism and it must include additional physically necessary properties, such as localizability, positiveness for both static and variable field and zero trace (for massless gravitons). However, these crucial features of the gravitational field disappear in the ``top'' non-linear Einstein's equations (generating the problem of non-localizability of the energy-momentum pseudotensor in GRT).
Just this crucial step is still a controversial subject.
This is why many physicists feel a tenuity of such derivation and
try to get his personal
derivation of the geometry from field approach, although,
as we demonstrated above, it is impossible
on the conceptual~level.

\subsubsection{Conceptual Tensions between Quantum Mechanics and General~Relativity}

According to Wigner's theorem~\cite{wigner1939}, quantum physical particles are the irreducible unitary representations of the Poincare (inhomogeneous Lorentz) group and, hence, demand the existence of the total symmetry of the  Minkowski spacetime. Conceptual tensions between geometrical and quantum-field approaches to gravitation theory were noted by some physicists
(Wigner 1997~\cite{wigner1997};
Feynman 1995~\cite{feynman95}; 
Amelino-Camelia 2000~\cite{amelino00}; 
Chiao 2003~\cite{chiao03}; and,
Ehlers 2007~\cite{ehlers07}).
The most pressing problem in present-day theoretical physics is how to unify quantum theory with gravitation, i.e.,~``quantum gravity problem''.
The standard scheme of quantization applied to
general relativity gives a theory that is not renormalizable
(i.e.,~leads to infinities in physical quantities),
though, in principle, non-renormalizability is a temporary technical obstacle.
Quantization of space-time is now also under construction
\cite{rovelli04,approaches07}, including
the string/M theory,
canonical/loop quantum gravity, non-commutative geometry, and other
~\cite{pavsic02}.
However the difficulties on this way so large that after all
attempts there is still no quantum geometrical gravity theory (Amelino-Camelia 2000~\cite{amelino00}; Amelino-Camelia~et~al.,~2005~\cite{amelino05}).

Note that, if~in a physical theory, the energy-momentum tensor of the field is not defined, then~also the energy of the field quanta can not be defined properly.
General relativity is not quantizable in ordinary physical sense
because it has no energy-momentum tensor for the gravity field.
Additionally, it is~important that properly defined energy of the gravity field also excludes an appearance of singularity and horizon (Section~\ref{sec4.2}).

Additional inconsistencies of attempts to derive
Einstein's equations from the spin 2 field theory, was noted by
Straumann 2000~\cite{strauman00} (p. 16), who pointed out~that:
\begin{itemize}
\item[$\bullet$]{\emph{general relativity having black hole solutions
violates the simple topological
structure of the Minkowski space of the quantum field theory;}}
\item[$\bullet$]{\emph{general relativity has lost the energy-momentum tensor
of the gravity field
together with the conservation laws, while in the Standard Model
the EMT and its conservation is the direct consequences of the
global symmetry of the Minkowski space.}}
\end{itemize}

Padmanabhan 2004~\cite{padmanabhan04} gave a comprehensive review
of all such attempts and demonstrated
that all derivations of general relativity from a spin 2 field are based
on some additional assumptions that are equivalent to the geometrization of
the gravitational~interaction.

Indeed, as~we noted above,
general relativity and field gravity
rest on incompatible physical principles,
such as non-inertial frames and the Riemann geometry of curved space
on the one side, and~material tensor field in
inertial frames with Minkowski geometry of flat space
on the other side.
Geometrical approach eliminates the gravity force, as~already noted by
de Sitter~\cite{desitter16a}:
``Gravitation is thus, properly speaking, not a 'force' in the new theory''. This however leads to the problem of energy just because the work done by force changes the~energy.

Within the quantum-field approach
the gravity force is directly defined
in an ordinary sense as the fourth interaction  and
has quantum nature (Feynman~\cite{feynman71,feynman95}).
The question may be formulated, as following,
which is more general description of gravitation:
geometry of curved space (so a property of space-time itself) or relativistic quantum field (so a kind of matter) in space?

\subsubsection{Astrophysical Tests of the Nature of Gravitational~Interaction}

The relation between GRT and QFGT is still an open question. Intriguingly, due to different predictions for observations,
this question can be answered by means of astrophysical observations and lab experiments, so to test which theory has a wider region of applicability?

In physics, any mathematical theory has restricted region of applicability, i.e.,
exact mathematical equations and it's solutions actually have only approximative physical sense.
This is why in physics the last word belong to experiments, and~especially to the crucial experiments and observations, \mbox{when rival theories predict different} results for certain clearly stated experiment.
The geometrical and field approaches are not
equivalent experimentally,
though the classical relativistic
gravity effects in the weak field are identical in both theories.
Because of the common region of experimentally tested effects, \mbox{it is possible that}
geometrical approach can be an approximation of the true quantum field gravity
or vise~verse.

Geometrical approach of the classical general relativity predicts such specific objects as singularities, black holes, and~expanding space of Friedmann
cosmological~model.

The consistent field approach
predicts that the gravity force has an ordinary quantum nature. Actually, the gravity force is the sum of the attraction (spin-2) and
repulsion (spin-0) (as will be shown in Sections~\ref{sec-3} and \ref{sec4}). This prediction of the QFGT theory opens up possibilities for novel type of experiments in gravity physics.
Spin-2 plus spin-0 contribution to the gravity force,
scalar gravitational waves, the~translational motion of rotating bodies,
the atmosphere and the magnetic field
of the relativistic compact objects in ``black hole candidates''
are specific effects
of the field gravity that may distinguish QFGT from general relativity.
In cosmology within the frame of QFGT, there is a possibility of infinite flat static Minkowski space filled by evolving  baryonic and non-baryonic  matter and having linear Hubble law of cosmological redshift as the global gravitational redshift effect (see Section~\ref{sec5}).

It is a remarkable result of our considerations that the choice between two conceptually different gravity theories may be founded on the results of experiments/observations in a physical laboratory.
For example, the~problem of gravity quantization is directly linked to the choice of the nature of gravitational interaction. Indeed, if~gravity is
geometrical in nature (a property of curved space),
then one should develop methods of space-time quantization
\cite{rovelli04,approaches07}.
However, if gravity is a force that is mediated by gravitons (quanta of the tensor relativistic field), then one should find methods based on the concept of the energy of the gravity field and develops new principles for overcoming the non-renormalizability problem.
It will be shown below that future astrophysical observations of the compact relativistic objects, space experiments in the Solar System, and cosmologically  relevant observations of the Local and High Redshift Universe may distinguish between these two cardinally different (though having similar predictions within  common region of applications) approaches to the theory of gravitational~interaction.

\section{Einstein's Geometrical Gravitation~Theory}
\label{sec2}

The final mathematical
formulation of the main equations of general relativity
was done by Einstein~\cite{einstein15,einstein16a} and derived by Hilbert~\cite{hilbert15} from geometrical extension of
the stationary  action principle.
It is a mathematically exact non-linear theory without
any inner limitations to its physical applications and this is why in GRT  singularities and black holes exist.
Below, we shortly consider the basic steps in construction GRT and its main predictions for experiments/observations,  which
we shall compare with corresponding equations and predictions of the QFGT.
We use designations as in the textbook by  Landau \& Lifshitz~\cite{landau71}.
The fundamental physical constants $c, G, h$  are explicitly used because they are important parts of the gravity~physics.

\subsection{Basic Principles of~GRT }

\subsubsection{The Principle of~Geometrization}
General relativity is based on the
{\it principle of geometrization}, which states that all gravitational phenomena can be described by the metric  of the Riemannian space (Ehlers 2007~\cite{ehlers07}).
This means that Einstein's gravity theory has no ``prior geometry'', such as the flat Minkowski space in other fundamental interaction theories.
Gravity is not a material field in space, but~it is a property of the curved space itself. The~role of the gravitational ``potential''
is played by the metric tensor $g^{ik}$ which determines the four-interval of the corresponding Riemannian space:
\begin{equation}
  ds^2 = g_{ik} dx^i dx^k
\label{interval}
\end{equation}
A test particle moves along
a geodesic line  of the Riemannian~space.

Note that {\it geodesic motion} is a form of the {\it equivalence principle}, which actually has many ``non-equivalent'' formulations like universality of free fall or  philosophical equivalence of the inertial reference frames to the reference frames accelerated by homogeneous gravity field. Equivalence principle played
an important role when general relativity was born, while, now, the basic
principle is the principle of geometrization, having clear physical and mathematical formulation. The~most clear and concise presentation of GR is the textbook by Landau \& Lifshitz~\cite{landau71}.
The most comprehensive description of the geometrical view on gravity
is the textbooks Misner, Thorne \& Wheeler~\cite{misner73} and
Straumann 2013~\cite{strauman13}.

\subsubsection{The Principle of Least~Action}
Einstein's field equations are obtained from the principle of least (stationary) action
by the variation of the metric tensor $g_{ik}$ in the action $S$ of the system matter + gravitational field.
It is very important to note that instead of the three parts (field-interaction-matter) of the total action in  ordinary field theory, in~the GRT the total action contains only two parts (there is \textbf{no interaction Lagrangian}), because gravitation is not a matter in GRT
(while other fields contain the interaction part $S_{(\mathrm{int})}$, see~Equations~(\ref{act-em}) and (\ref{act-fgt})). So the GRT action is:
\begin{equation}
\label{sgr}
  S = S_{(\mathrm{g})}  + S_{(\mathrm{m})} =
  \frac{1}{c}
  \int \left(\Lambda_{(\mathrm{g})} + \Lambda_{(\mathrm{m})} \right)
  \sqrt{-g} d\Omega \,,
\end{equation}
where $S_{\mathrm{m}}$ and $S_{\mathrm{g}}$ are the actions for the matter and
gravitational field,
$\Lambda_{(\mathrm{m})}$ is the Lagrangian for the matter, and~the Lagrangian for the field is
%
\begin{equation}
\label{lgr}
   \Lambda_{(\mathrm{g})} = - \frac{c^4}{16 \pi G} \Re \,.
\end{equation}
Here, $\Re$ is the scalar curvature of the Riemannian~space.

\subsection{Basic Equations of General~Relativity}

\subsubsection{Einstein's Field~Equations}

Variation $\delta g_{ik}$, with~restriction $g_{ik}g^{ik}\equiv 4$
gives from $\delta (S_{(\mathrm{m})} + S_{(\mathrm{g})})=0$
the following field~equations:
\begin{eqnarray}
  \Re^{ik} - \frac{1}{2}\,g^{ik}\,\Re =
  \frac{8\,\pi\,G}{c^4}\,\,T^{ik}_{(m)}\,,
\label{efeq1}
\end{eqnarray}
where $\:\Re^{ik}$ is the Ricci tensor. $\:T^{ik}_{(\mathrm{m})}$
is the energy-momentum tensor (EMT) of the matter, which includes
all kinds of material substances, such as particles, fields,
radiation, and dark energy, including the vacuum
$T^{ik}_{(\mathrm{vac})} = g^{ik}\Lambda$
(where $\Lambda$ is the Einstein's cosmological constant).

Note that $\:T^{ik}_{(m)}$
does not contain the energy-momentum tensor of the gravity field itself,
because \textbf{gravitation is not a material field} in general relativity
(as was discussed in Introduction).

\subsubsection{The Equation of Motion of Test~Particles }
A mathematical consequence of the field Equation~(\ref{efeq1}) is
that, due to Bianchi identity,
the covariant derivative of the left side equals zero,
so for the right side we also have
\begin{equation}
  T^{ik}_{(\mathrm{m})\; ;~i} = 0\,.
\label{bianki-1}
\end{equation}
This \textbf{continuity equation} also gives the equations
of motion for a considered matter.
\footnote{
It is \textbf{not a conservation
of the energy-momentum }of the self-gravitating system because  $\:T^{ik}_{(m)}$
does not contain the energy-momentum of the gravity field
(Landau \& Lifshitz~\cite{landau71} p. 304; Ehlers 2007~\cite{ehlers07}).
}
It implies the geodesic equation of motion for a test particle:
\begin{equation}
 \frac{du^i}{ds}=-\Gamma^i_{kl}u^k u^l\,.
 \label{geodesic}
\end{equation}
$u^i=dx^i/ds$ is the four-velocity of the particle
and  $\Gamma^i_{kl}$ is the Christoffel~symbol.

\subsection{The Weak Field~Approximation}

All of the relativistic gravity effects that have been actually
tested by observations,
relate to the weak field,
where the Newtonian potential $|\varphi|<< c^2$.
This is why the weak field approximation has an important
role in gravity~physics.

\subsubsection{The Metric~Tensor  }
In the case of a weak gravity field, the metric tensor
usually is expressed in the form
\begin{eqnarray}
\label{weak-gr}
  g_{ik} = \eta_{ik} + h_{ik} \nonumber\\
  g^{ik} = \eta^{ik} - h^{ik} \nonumber\\
  g^i_k \:= \delta^i_k \;\;\;\;\;\;\;\;\;\;
\end{eqnarray}
As we discussed above, such a presentation of the metric tensor Equation~(\ref{weak-gr}) means that a tensor of Riemannian space $g^{ik}$ is presented by the sum
of two non-tensor quantities, because~$\eta ^{ik}$
and  $|h_{ik}|<<1$ are not tensors of the curved space.
e.g.,~in the third identity of Equation~(\ref{weak-gr}) the components
must be $h^i_k\equiv 0$ (though usually the convention is used
that $h^i_k = \eta^{li}h_{lk}$ and $h=\eta^{ik}h_{ik}$).
The different signs of the quantities $\hat{\textbf{h}}$ for covariant and contravariant
components of the metric tensor $\hat{\textbf{g}}$ are caused by the exact identity valid for the
\emph{trace of the metric tensor} of any Riemannian four-space:
\begin{equation}
\label{gtrace}
  g_{ik}\cdot g^{ik} = 4 \;\;\;\;\;
\end{equation}
As we shall see below, this is an essential difference
with the consistent field approach, where the sum of
two tensors is a tensor of Minkowski~space.

\subsubsection{The Field~Equations  }
In the linear GRT approximation, it is assumed that the metric tensor is
 $g_{ik} = \eta_{ik} + h_{ik}$ (Equation~(\ref{weak-gr})) where
$|h_{ik}|<<1$, so the Einstein's
Equation~(\ref{efeq1}) becomes \cite{maggiore08,strauman13}:
\begin{equation}
\label{feq1-grt}
  - h^{ik,l}_{\;\;\;l} +
  h^{il,k}_{\;\;\;l} +h^{kl,i}_{\;\;\;l} - h^{,ik}
  -\eta^{ik} h^{lm}_{\;\;\; ,lm} + \eta^{ik}h^{,l}_{\,l}  = \frac{16
  \pi G}{c^{4}} T^{ik}_{(m)}\,.
\end{equation}
The gauge freedom of the Equation~(\ref{feq1-grt})
allows for one to put four additional conditions on the
potentials, in~particular a Lorentz invariant gauge---the Hilbert-Lorentz gauge 
\footnote{Also called as the de Donder gauge}
 \cite{strauman13,maggiore08}:
\begin{equation}
\label{gauge-grhl}
 h^{ik}_{\;\;\; ,k} = \frac{1}{2} h^{\;,i}\,.
\end{equation}
With the gauge (\ref{gauge-grhl}), the field equations get
the form of the wave equation:
\begin{equation}
  \left(\triangle - \frac{1}{c^2}
\frac{\partial^2}{\partial t^2} \right)h^{ik} =
  \frac{16\pi G}{c^4}
     \left[ T^{ik}_{(m)}
 -\frac{1}{2} \eta^{ik}T_{(m)}\right] \,.
\label{efeqw}
\end{equation}
For the important case of a static spherically symmetric weak gravitational
field, the solution of these equations gives the metric tensor,
expressed in isotropic coordinates:
\begin{eqnarray}
\label{weak-sss-gr}
g_{ik} = \eta_{ik} + \frac{2\varphi_N}{c^2}diag(1,1,1,1) \nonumber\\
g^{ik} = \eta^{ik} - \frac{2\varphi_N}{c^2}diag(1,1,1,1) \nonumber\\
g^i_k = \delta^i_k \, ,\;\;\;\;\;g_{ik}g^{ik} = 4   \;\;\;\; \;\;\;\;\;\;\;
\end{eqnarray}
where $\varphi_{\mathrm{N}} =-GM/r $ is the Newtonian potential. Note that $h_{ik}$ is not a tensor quantity (see~Equation~(\ref{weak-gr-fg})).

\subsubsection{The Equation of Motion in the Weak~Field }
The post-Newtonian approximation of the weak field takes all terms of order $v^2/c^2$ and $\varphi_{\mathrm{N}}/c^2$ into account.
PN-geodesic equations are frequently used in relativistic
celestial mechanics. The~three-acceleration
of a test particle in the static spherically symmetric weak gravity
field \mbox{(e.g.,~a planet around the Sun)} is given by equation
(Brumberg 1991~\cite{brumberg91}; Kopeikin~et~al.,~2011~\cite{kopeikin11}):
\begin{eqnarray}
 (\frac{d\vec{v}}{dt})_{_\mathrm{GR}} = -
 \{1+(1+\alpha)\frac{v^2}{c^2} +
 (4-2\alpha)\frac{\varphi_{_\mathrm{N}}}{c^2}
 -3\alpha(\frac{\vec{r}}{r}\cdot\frac{\vec{v}}{c})^2
 \}\vec{\nabla}\varphi_{_\mathrm{N}} \nonumber\\
 +(4-
2\alpha)\frac{\vec{v}}{c}(\frac{\vec{v}}{c}
\cdot\vec{\nabla}\varphi_{_\mathrm{N}}
)\,,
 \;\;\;\;\;\;\;\;\;\;\;\; \;\;\;\;\;\;\;\;\;\;\;\;\;
\label{GR-accel}
\end{eqnarray}
where $\vec{v}=d\vec{r}/dt$, $\varphi_{_\mathrm{N}}=-GM/r$, and~$\vec{\nabla}\varphi_{_\mathrm{N}}=GM\vec{r}/r^3$.
An important quantity here
is the parameter
$\alpha$. It is determined by the choice of the coordinate system:
$\alpha = 2$ for the Painleve coordinates,
$\alpha = 1$ for the Schwarzschild coordinates, and~$\alpha = 0$ for harmonic and isotropic coordinates.
Hence  orbit of a particle will depend on the chosen coordinates.
To avoid this non-physical result it is suggested that observable physical quantities should not depend on the coordinate parameter $\alpha$ by taking into account
an ad~hoc procedure of measuring quantities involved in the orbital motion phenomenon. 
\footnote{The problem
of the physical meaning of coordinates in general relativity has been debated for a long time and up to now there are many opinions on the physical interpretation of the coordinate transformations in GRT.
E.g., Misner, Thorne \& Wheeler 1973~\cite{misner73}, p. 1097) wrote that Schwarzschild coordinates are ``wrong'' because ``physicists, astronomers and other celestial mechanics have adopted the fairly standard convention of using 'isotropic coordinates' rather than 'Schwarzschild coordinates' when analyzing the solar system''.}

It should be emphasized that directly from equation of motion (\ref{GR-accel}) follows the dependence of gravitational acceleration from the value and direction of the test particle velocity. Accordingly, this result contradicts that form of the equivalence principle, where one asserts the independence of the free fall on the velocity of a test~particles.

\subsection{Major Predictions for Experiments/Observations}
The number of predictions  of general relativity for both weak and strong fields were derived
from Einstein's field equations and the equations of motion.
The triumph of GR in physics and astronomy is caused by the experimental and observational
confirmation of Einstein's equations with  high accuracy. The application of GR to cosmology will be considered in Section~\ref{sec5}.

\subsubsection{The Classical Relativistic Gravity Effects in the Weak~Field}

The classical weak gravity effects have been tested with an accuracy of about $0.1 \div 1 $\%
(Will~\cite{will14}; Kopeikin~et~al.~\cite{kopeikin11}). Among~these experimentally verified effects~are:
\begin{itemize}
\item[$\bullet$]{\emph{universality of free fall for non-rotating bodies,}}
\item[$\bullet$]{\emph{the deflection of light by massive bodies,}}
\item[$\bullet$]{\emph{gravitational frequency-shift,}}
\item[$\bullet$]{\emph{the time delay of light signals,}}
\item[$\bullet$]{\emph{the perihelion shift of a planet,}}
\item[$\bullet$]{\emph{the Lense--Thirring effect,}}
\item[$\bullet$]{\emph{the geodetic precession of a gyroscope, and}}
\item[$\bullet$]{\emph{the emission and detection of the quadrupole gravitational waves.}}
\end{itemize}
In the next sections, we shall show that all these effects can be explained also within the field gravity approach with the same formulas for the observed quantities; hence, they can not distinct between GRT and QFGT.
However, in FGT, there are additional weak gravity effects that can be used as crucial tests for GR and FG theories: e.g.,
free fall of rotating bodies, attraction, and repulsion components of the
gravitational force, and~additional scalar gravitational radiation.
Recently detected GW signals by Advanced LIGO antennas are also
disccussed in Section~\ref{sec4.2}.

\subsubsection{Strong Gravity Effects in GRT: Schwarzschild~Metric }
General relativity predictions for the strong gravity
considered in many books, which contain many exact solutions
of the full non-linear Einstein's equations
(e.g.,~Landau \& Lifshitz~\cite{landau71},
Misner, Thorne \& Wheeler~\cite{misner73}, and
Straumann 2013~\cite{strauman13}).
Some observable constrains on strong gravity effects was considered by
Taylor et al. 1992~\cite{taylor92}

One of the basic exact solution of Einstein's Equation~(\ref{efeq1})
for any centrally symmetric mass distribution is called the
Schwarzschild metric. It has the following form for the four-interval
in the Schwarzschild system of coordinates $(t,r,\theta,\phi)$:
\begin{equation}
\label{schwarz}
 ds^2 =(1-\frac{r_{\mathrm{Sch}}}{r})c^2dt^2 - \frac{dr^2}{1-
\frac{r_{\mathrm{Sch}}}{r}} -
 r^2({\rm sin}^2\theta\,d\phi^2 + d\theta^2)\,.
\end{equation}
In other coordinate systems, this interval has a different form for the singular
term. The~metric in Equation~(\ref{schwarz}) only depends on the total mass $M$
of the gravitating body. The~quantity $r_{\mathrm{Sch}}$ is called the
Schwarzschild radius for the mass $M$, where the event horizon exist:
\begin{equation}
\label{rg}
  r_{Sch} = \frac{2GM}{c^2} = 3~\mbox{\rm km} \frac{M}{M _{\odot}}\,.
\end{equation}
This metric shows that, at
 $r=r_{\mathrm{Sch}}\;$, the 00-component is equal to zero
and the 11-component is
infinite. They say that the gravity ``force'' becomes so strong that
nothing, not even light, can escape a body whose whole mass $M$ is inside
$r_{\mathrm{Sch}}$ (a definition of the black hole).
For an extremely rotating Kerr BH, the event horizon radius can be two times
less.

An external observer within a static system of coordinates
will see matter collapsing eternally on the
black hole (infinite time formation of a BH). However,~if one chooses a non-static free-falling system of coordinates,
one finds that a co-falling matter
will cross the gravitational radius in a finite (\mbox{and~rather short)}
proper time, so the matter inevitably falls into the center of the
field ($r=0$), the~true singularity of the metric.
This demonstrates the crucial role of coordinate transformations
in general relativity, because~it leads to a paradox of the
``death before birth'' due to  finite time of
BH evaporation (via~Hawking radiation), while, for a distant observer,
the BH has not yet formed (Chowudhury and Krauss 2014
~\cite{chowdhury14}).

\subsubsection{Tolman-Oppenheimer-Volkoff~Equation}
Another important exact result within
general relativity is the  equation of
hydrostatic equilibrium:
\begin{equation}
\label{tov} \frac{dp}{dr} = - \frac{G(\rho + p/c^2)(M+4\pi
pr^3/c^2)} {r^2(1 - r_{\mathrm{Sch}}/r)}\,.
\end{equation}
According to this Tolman--Oppenheimer--Volkoff equation, the factor $1/(1-r_{\mathrm{Sch}}/r)$ leads to
an infinite pressure gradient for $r \rightarrow r_{\mathrm{Sch}}$. This has a deep consequence: there is an upper limit for the mass of static compact relativistic stars, around  2--3 $M_{\odot}$.
According to the standard GR, compact objects with larger masses may only exist as black~holes.

\subsection{Modifications of GRT to Aviod Field Energy~Problem}
\label{sec2.5}
In spite of the great success of the geometrical approach for the description of existing experiments/observations in gravitation physics, there are some conceptual problems of GRT, which already were discussed in the Introduction.
Among them the most important~are:
\begin{itemize}
\item[$\bullet$]{\emph{the physical sense of the energy-momentum of the space curvature,}}
\item[$\bullet$]{\emph{the physical sense of the black hole horizon and singularity, and}}
\item[$\bullet$]{\emph{the physical sense of the space creation in the expanding Universe.}}
\end{itemize}

\subsubsection{Geometrical Approach without Black Holes?}

In recent literature there is intriguing discussion regarding the physical impossibility
of black holes, horizons, and singularities. 
 Logunov 2002~\cite{logunov02};
Logunov \& Mestvirishvili \cite{logunov89,logunov01};
Kisilev~et~al.~\cite{kisilev10}; Mitra~\cite{mitra02}; and, Gershtein~et~al.~\cite{gershtein11}
emphasized the important role of additional physical conditions that should be used
for a physically reasonable solution of Einstein's equations.
For example, the Hilbert's causality principle
leads to the elimination of horizons and singularities~\cite{gershtein11}.

A re-analysis of the physical meaning of the coordinate transformation
in general relativity led Mitra\cite{mitra02,mitra06}
to the conclusion that a black hole should have zero mass.
Considering both the four-velocity and the
physical three-velocity of a co-moving observer he concluded
that instead of genuine black holes there is
a solution of Einstein's equations describing an
``eternally collapsing object'' (ECO) with a size close to $r_{Sch}$
and all of the time radiating energy, so that an event horizon
never~originates.

Robertson \& Leiter~\cite{robertson05} introduced
the strong principle of equivalence requiring that
``special relativity must hold locally for all time-like
observers in all of space-time''. They found solutions of Einstein's
equations that satisfy the requirement for time-like world line completeness
and introduced ``magnetospheric eternally collapsing objects''.
Such MECOs possess an intrinsic magnetic moment and they do not have
any event horizon and curvature~singularity.

If a substance has an unusual equation of state $p = p(\rho)$,
like that of the physical vacuum and dark energy,
it is possible to obtain non-singular
static GR solutions
for arbitrary large masses, which~are~stable and~have no singularity,
no event horizon, and no information paradox
(Dymnikova~\cite{dymnikova02}; Mazur \& Mottola~\cite{mazur04}; and, Chapline~\cite{chapline05}).

These works show that additional conditions
on the equation of state or
the coordinate transformations or the metric tensor
of Riemannian space can change the physical contents of the geometrical gravity~theory.

\subsubsection{The Energy-Momentum of the Space Curvature?}
As we already mentioned, together with Einstein's equations, the conceptual problem of the energy of the gravitational field was born.
The ``pseudo-tensor'' (actually non-tensor) character of the EMT of the gravity ``field''  in GRT has been discussed in many papers,
where different ways to avoid this obstacle were~suggested.

A mathematical
consequence of the Einstein's Equation~(\ref{efeq1}) is that the covariant
divergence  of the matter energy--momentum tensor equals zero:
\begin{equation}
  T^{ik}_{(\mathrm{m})\;\; ;~i} = 0\,.
\label{bianki2}
\end{equation}
One is tempted to see in this expression a usual conservation law, but~let us cite the famous, but~often ignored statement by
Landau \& Lifshitz (\cite{landau71}, sect.101 p. 304):
``however, this equation does not generally express any
conservation law whatever. This is related to the fact that in a
gravitational field the four-momentum of the matter alone must not
be conserved, but~rather the four-momentum of matter plus
gravitational field; the latter is not included in the expression
for  $\:T^{ik}_{(\mathrm{m})}$''. 
\footnote{
Mathematically this is because the integral
$\int T^{ik}_{(\mathrm{m})} \sqrt{-g} dS_k $ is conserved only
if the condition
$\partial ( \sqrt{-g} T^{ik}_{(\mathrm{m})} )/\partial x^k = 0\,$
is fulfilled,
while Equation~(\ref{bianki2}) gives relation
$ T^{ik}_{(\mathrm{m})\;\; ;~k} = (1/\sqrt{-g})
(\partial ( \sqrt{-g} T^{ik}_{(\mathrm{m})} )/\partial x^k )
- (1/2)(\partial g_{kl} /\partial x_i)T^{kl}_{(\mathrm{m})} = 0$.
}

To define a conserved total four-momentum for a gravitational
field plus the matter within it, Landau \& Lifshitz~\cite{landau71}
suggested the expression
\begin{equation}
 \frac{\partial}{\partial x^k}(\sqrt{-g})(T^{ik}_{(\mathrm{m})} +
t^{ik}_{(\mathrm{g})}) =
 0\,.
 \label{conserv-gr}
\end{equation}
Here, $t^{ik}_{(\mathrm{g})}$ is called the energy-momentum pseudo-tensor
(EMPT).
It is important that $t^{ik}_{(\mathrm{g})}$
does not constitute a tensor, i.e.,~it is not a generally covariant
quantity.
There are many variants of
the expressions suggested for the pseudo-tensor, among~them
Einstein's (non-symmetric) and Landau \& Lifshitz's (symmetric)
pseudo-tensors. Unfortunately, existing expressions for EMPT do not satisfy
to the all necessary field-theoretical conditions for EMT of a
massless boson field
(symmetry, positive localizable energy density, zero trace).

Moreover, this way of introducing the energy for the
geometrized  gravity field
within GRT is conceptually inconsistent, as~discussed, in detail, by
 Logunov 2002~\cite{logunov02};
Logunov \& Folomeshkin (1977) \cite{logunov77} and
Logunov \& Mestvirishvili (1989) \cite{logunov89}.
Additionally, Yilmaz (1992) \cite{yilmaz92} has shown that
for any pseudo-tensor
due to the Freud identity one has
$\partial_i (\sqrt{-g}\, t^i_k)=0$, which leads to
a difficulty with the definition of the
gravitational~acceleration.

\subsubsection{Non-Localizability of the Gravitation Field Energy in~GRT  }

The localizability of the field energy is a necessary feature
of  the fundamental physical interaction theory.
It means that the energy--momentum tensor of a field is a definite
function of the Minkowski space, i.e.,~at the classical level,
can be measured (detected) at each point of the~space.

For example, in
electrodynamics, there is localizable positive energy density
of the electromagnetic field $-$  for the relativistic vector field
$A^i(\vec{r},t)$ the energy density is localizable and positive
for both static and variable field
$T^{00}(\vec{r}, t) = (E^2+H^2)/\,8\pi >0$
(see Section~\ref{sec1.2}).

However in GRT, due to pseudo-tensor character of the energy-momentum of the
gravity field~\cite{landau71} (\S101, p. 307): ``...By choosing a coordinate
system which is inertial in a given volume element, we can make all the
$t^{ik}$ vanish at any point in space-time (since then all the
$\Gamma^i_{kl}$ vanish). On~the other hand, we can obtain values of the
$t^{ik}$ different from zero in flat space, i.e.,~in the absence of a
gravitational field, if~we simply use curvilinear coordinates instead
of cartesian. Thus, in any case, it~has no meaning to speak of a definite localization of the energy of the gravitational field in~space''.

Misner, Thorne \& Wheeler wrote about the energy of gravitational
field~\cite{misner73} (p. 467):
``It is not localizable. The~equivalence principle forbids.''
They also noted the following properties of the pseudo-tensor:
``There is no unique formula for it, ... , 'local gravitational
energy-momentum' has no weight. It does not curve space. It does not
serve as a source term ... It does not produce any relative
geodesic deviation of two nearby world lines ...
It is not observable".
The problem is also clearly seen in the case of the gravitational wave detection (as will be discussed in Section~\ref{sec4.2}).
Accordingly, the actual cause of the absence of the gravity field energy
(i.e.,~the pseudo-tensor character of the EMT of the gravitational
field in general relativity)
is the principle of geometrization.
Note that there is no such problem in electrodynamics and Feynman's field gravitation theory.
The LIGO detector's mirror has localized the GW energy well inside the GW wavelength, which is consistent with the classical limit of the~QFGT.

\paragraph{Attempts to Overcome the Energy Problem by
Using Simultaneously Minkowski and Riemannian~Spaces}
In the literature, there are
attempts to construct a gravity theory which based on
both flat and curved spaces
by accepting
some Lorentz-covariant properties of  Minkowski space
in ``effective'' Riemannian space
(this is comprehensively reviewed by Pitts \& Schieve~\cite{pitts01}).
As an example of such works, we mention three ``field-geometrical'' theories
developed by  Logunov, Yilmaz, Grishchuk and their~collaborators.

Logunov 2002~\cite{logunov02};
Logunov \& Mestvirishvili~\cite{logunov89,logunov01}
developed a field-geometrical
gravity theory, called the relativistic theory of gravitation (RTG),
where they accept ``geometrization principle'' for matter, while conserve Minkowski flat space for gravitation field.
They introduce the metric tensor $g^{ik}$ of the effective
Riemann space, and~also accept a ``causality principle'' as an additional
restriction on $g^{ik}$. Due to these assumptions, there is no black hole solution
in RTG. The~scalar part of gravitational tensor potentials exists only
in a static field and can not be radiated. The~cosmological solution
is the Friedmann expanding space with the critical matter density.
A recent development of the RTG also includes non-zero rest mass of the gravity~field.

Yilmaz~\cite{yilmaz92} constructed a field-geometrical theory
where the right-hand side of the field equation contains the EMT
of the gravity field in the effective Riemann space with the background Minkowski space.
The metric of the effective Riemann space has an
exponential form and it excludes the event horizon and singularity.
The existence of the EMT of the gravity field allows for one to consider N-body
solutions in this~theory.

Baback \& Grishchuk~\cite{babak00} claimed that they constructed a
field approach which is completely identical to general relativity :
``GR may be formulated
as a strict non-linear field theory in flat space-time. This is
a different formulation of the theory, not a different theory''.
They introduce the metric tensor $g^{ik}(x^l)$ of a curved space-time via
the field variables $h^{ik}(x^l)$ in the form
$g^{ik}=(\eta^{ik} + h^{ik})\sqrt{\gamma /g}$ with the
condition $g^{ik}g_{il}=g^k_l= \delta ^k_l$. Hence the tensor
of Riemannian space is presented
as a sum of two non-tensors, because~the Minkowski metric $\eta^{ik}$
is not a tensor of curved space. They developed a Lagrangian
theory, where they introduced an  energy-momentum tensor of the gravitational
field (close to LL-pseudotensor) and then  got black holes, quadrupole radiation, and expanding space of Friedmann's~cosmology.

However, the internal inconsistency of this approach follows from
incompatibility of the initial principles of the geometrical and field theories,
which we have discussed above in detail. Additionally, the expanding space of GR violates energy conservation, which is forbidden for the field-theoretical approach in Minkowski~space.

\paragraph{Absence of the Required EMT Physical Properties in the Metric Gravity Theories   }
Besides the true tensor character of the EMT, there are additional properties
of the energy-momentum tensor known from the quantum relativistic field theories
of other physical interactions.
For example, the EMT of the boson fields must have the
following features:
\begin{itemize}
\item[$\bullet$]{\emph{symmetry, $T^{ik}=T^{ki}$;}}
\item[$\bullet$]{\emph{positive localizable energy density, $T^{00}>0$; and,}}
\item[$\bullet$]{\emph{zero trace for massless fields, $T=0$.}}
\end{itemize}
e.g.,~in the case of a static electric field, its energy density is given by the expression:
\begin{equation}
\label{ed-t00}
\varepsilon_{el} (\vec{r}) =  T^{00}_{el}(\vec{r}) =
+\frac{1}{ 8\pi }\left( \vec{\nabla}
\varphi_{el}\right)^2 \,\,\,\frac{erg}{cm^3} .
\end{equation}
According to Equation~(\ref{ed-t00}), the definite positive energy of the field
exists in each point $(\vec{r})$ and it can be transformed (localized) in
other forms of~energy.

The attempts
to introduce  EMT of the gravity field within
geometrical and effective ``field'' approaches, though~could
obey the symmetry condition, but~they do not possess the other two necessary features of the EMT,
i.e.,~a positive localizable energy density and zero trace. These
properties of EMT
should  be fulfilled within the consistent field approach
for both static and free fields, as~in the case
of the electromagnetic field (Section~\ref{sec1.2}).

The energy problem can be demonstrated with the simplest case of a
spherically symmetric weak static gravity field.
Indeed, for~this case, like in a terrestrial laboratory,
one can easily calculate the predicted value of the energy density
of the gravitational field for different
energy-momentum pseudo-tensors (EMPTs).
For instance, in~harmonic coordinates, the Landau--Lifshiz
symmetric pseudotensor gives a negative energy density of
the static spherically symmetric gravity field
\begin{equation}
\label{LL-t00}
\varepsilon_{g} (\vec{r}) =
t^{00}_{\rm LL}(\vec{r}) =  -\frac{7}{ 8\pi G }\left( \vec{\nabla}
\varphi_{_{N}}\right)^2 \,\,\,\, \frac{erg}{cm^3} .
\end{equation}

The ``final'' energy-momentum tensor of the gravity
field, which was derived by
Grishchuk, Petrov \& Popova~\cite{grischuk84},
has a negative
energy density of the weak static field :
$t^{00}_{\mathrm{GPP}} =  -\frac{11}{ 8\pi G }\left( \vec{\nabla}
\varphi_{_{\mathrm{N}}}\right)^2 \,$,
while Einstein's pseudo-tensor gives
$t^{00}_{\mathrm{E}} =  +\frac{1}{ 8\pi G }\left( \vec{\nabla}
\varphi_{_{\mathrm{N}}}\right)^2 \,$.

Hence, according to the LL-pseudo-tensor and the GPP-tensor,
the energy density of the static gravitational field is negative,
which conflicts with the quantum
field theories of other fundamental interactions.
Additionally, the traces of all these EMPTs do not vanish
for static~fields.

\subsubsection{The Physical Sense of the Space/Vacuum  Creation in the Expanding~Universe}
In cosmology, GRT predicts that the homogeneous matter distribution expands together with space. The~linear Hubble law of the space-expansion velocity
$V_{exp} = H\times R$ is the strict consequence of the matter homogeneity.
The physics of the space/vacuum expansion (increasing scale factor $S(t)$ for distances between galaxies with increasing time) contains several~paradoxes.

Harrison~\cite{harrison00,harrison93,harrison95} demonstrated that the cooling of homogeneous hot gas (including photon gas of CMBR) in the standard cosmological model (SCM) actually means the violation of energy conservation in the expanding space. In~modern version of SCM ,the term ``space expansion'' actually means continuous creation of vacuum, something that leads to conceptual problems. Recent discussion by Francis~et~al.~\cite{francis07} on the physical sense of the increasing distance to a receding galaxy without motion of the galaxy is just a particular consequence of the arising paradoxes~\cite{baryshev15}.
In~Friedmann's cosmology, the absence  of the static Minkowski space
leads to the paradox of continuous creation/annihilation of matter
within any finite comoving volume of the expanding~space.

In Section~\ref{sec5}, we present analysis of the conceptual problems of the SCM: the violation of energy conservation for local comoving volumes, the~exact Newtonian form of the Friedmann equation (no~direct relativistic effects of expanding substances, e.g.,~the absence of an upper limit on the receding velocity of galaxies which can be greater than the speed of light), and~the presence of the linear Hubble law deeply inside very inhomogeneous large scale galaxy distribution of the Local~Universe.

\subsubsection{ Conclusions  }
The above discussion demonstrates that all  ``effective field-geometrical'' theories
that introduce a metric of an ``effective'' Riemannian space in the form
$g_{ik}\approx \eta_{ik} + h_{ik}$  must obey an exact equalities
$g^i_k\equiv\delta^i_k$ and $g^{ik} g_{ik} \equiv 4$ and so
exactly eliminate the internal scalar part---the trace of the true tensor potential.
Hence, such metric gravity theories lose some essential properties
of the field approach
and receive some nonphysical properties of the geometrical approach,
e.g.,~non-tensor character of the gravity field energy,
the negative energy of the static field, the~event horizon and singularity,~etc.

Accordingly, one can conclude that all attempts to derive ``geometry'' from
``gravitons'' explicitly or implicitly contain
propositions that reduce the field approach to geometrical one
\cite{ehlers07,strauman00,padmanabhan04}.
Hence,~the~question is: how can one construct a
consistent field gravitation theory (quantum gravidynamics)
based on relativistic
quantum principles and which, only as an approximation to reality,
contains geometrical interpretation, like geometrical optics in
electrodynamics.


\section{Feynman's Quantum-Field Approach to Gravitation~Theory}
\label{sec-3}
In Section~\ref{sec-1-3}, we have emphasized that
the field gravitation theory has its roots in papers by Poincar\'{e},
Fierz, Pauli,
Birkhoff, Moshinsky, Thirring, Kalman, Feynman, Maggiore, and other eminent physicists.
The field approach offers a natural solution to the energy problem, because~Minkowski space implies the invariance under the Poincar\'{e} group transformation and, hence, the usual definition of the energy-momentum tensor of the gravitational
field and conservation laws, as~it follows from the Noether's~theorem.

We stress that the construction of the  quantum-field gravity theory has not yet
completed and important questions are still open. 
Especially the problem of renormalization in the quantum field theory deals with infinities of such basic quantities as the electromagnetic mass of the electron in QED. In~the standard model the renormalization is a mathematical procedure, which allows for excluding divergent terms in measurable quantities (e.g.,~Bogolubov \& Shirkov 1982~\cite{ bogolubov93},
Ryder  1984~\cite{ryder1984}, and
Sadovskii  2019~\cite{sadovskii2019}).
Quantum electrodynamics  is a renormalizable theory, but~general relativity is a nonrenormalizable theory. In~the frame of QFGT, there are new types of physical arguments for divergences cancellation. 
For example, one must take the conservation of the gravitational field energy and the finiteness of the gravity force into account, which can help to overcome the problem of nonrenormalizability of the gravitational interaction.
The main strategy of the consistent field approach is not to
write down the final non-linear exact equations, but~to control each step of the iteration and understand the physical sense of the energy-momentum of the gravitational
field in the description of the gravitational~interaction.

The field-theoretical approach for the analysis of the gravitational interaction
was considered in many works
(e.g.,~Fierz \& Pauli 1939~\cite{fierz39};
Birkhoff 1944~\cite{birkhoff44};
Moshinsky 1950~\cite{moshinsky50};
Thirring 1961~\cite{thirring61}; Kalman 1961~\cite{kalman61};
Feynman 1971, 1995~\cite{feynman71,feynman95};
Bowler 1976~\cite{bowler76}; and, Maggiore 2008~\cite{maggiore08} Chapter~2).
In the frame of the field gravitation theory, the crucial role of the
intrinsic scalar part
(the trace $\psi(\vec{r}, t) = \eta_{ik}\psi^{ik}$)
of the reducible symmetric tensor
potentials $\psi^{ik}(\vec{r}, t)$
was discovered and studied by Sokolov and Baryshev
\cite{sokolov80,sokolov19,sokolov15,baryshev90,baryshev96,baryshev-soc97,baryshev03,baryshev08a,baryshev08b}.
Up to now, within~the quantum-field gravitation theory (QFGT)
the weak field approximation at the post-Newtonian level
has been studied in detail, although~some results for strong
field regime also exist.
The modern development of QFGT is enough to show the feasibility of the field approach and to give predictions,
which can distinguish between QFGT and GRT. Hence, in~contrast to many claims, the~quantum-field gravity theory is experimentally
different from the geometrical general~relativity.

\subsection{Initial~Principles}

\subsubsection{The Unity of the Fundamental~Interactions  }

As Feynman~\cite{feynman95} emphasized, the gravitational interaction can be
described as a non-metric quantum relativistic symmetric second rank tensor field in Minkowski space, which is based on the Lagrangian formalism of the field theory.
He discussed a quantum field approach to the gravity just as the next fundamental physical interaction and claimed that ``the geometrical
interpretation is not really necessary or essential to physics'' (\cite{feynman95}, p. 113).

The QFGT is constructed on the common basis with other fundamental physical interactions, which includes the Poincare symmetry group of the Minkowski space, the~Noether's conservation of fields energy-momentum tensor, and Wigner's classification of quantum particles. 
There are also several additional features that are specific for gravitational interaction. As~we noted in Section~\ref{sec1.3}, these basic principles include:
{\small
\begin{itemize}
\item[$\bullet$]{\emph{the inertial reference frames and Minkowski space with metric $\eta^{ik}$;}}
\item[$\bullet$]{\emph{the \textbf{reducible}  symmetric second rank tensor potential $\psi^{ik}(x^m)$;}} 
\item[$\bullet$]{\emph{two irreducible parts which correspond to spin-2 and spin-0 (trace) fields;}} 
\item[$\bullet$]{\emph{the Lagrangian formalism and Stationary Action principle;}}
\item[$\bullet$]{\emph{the principle of consistent iterations;}}
\item[$\bullet$]{\emph{the universality of gravitational interaction;}}
\item[$\bullet$]{\emph{the conservation law of the energy-momentum;}}
\item[$\bullet$]{\emph{the gauge invariance of the linear field equations;}}
\item[$\bullet$]{\emph{the positive localizable energy density and zero trace of the gravity field EMT;}}
\item[$\bullet$]{\emph{the quanta of the field energy as the mediators of the gravity force; and,}}
\item[$\bullet$]{\emph{the uncertainty principle and other quantum postulates.}}
\end{itemize}
}
These elements are the basis of the consistent field approach
to gravitation and form
a natural starting point for understanding the physics of gravity
phenomenon similarly to other fundamental~forces.

\subsubsection{The Principle of Consistent~Iterations  }
The gravity field has a positive energy density and this energy, in~turn, becomes a new source
of an additional gravity field and so on. This non-linearity is taken into account by the iteration procedure. It is usual in physics to first consider a linear approximation and then add non-linearity by means of~iterations.

The field gravity theory is constructed step by step using an iteration procedure, so that at each step all physical properties of the EMT of the gravity field are under control. Each step of iteration is described by
linear gauge-invariant field equations with fixed sources in the right-hand side,
as it is assumed in the derivation of field equations
from the stationary action principle.
An~important outcome of this procedure is that the
superposition principle also can be reconciled with the
non-linearity of the gravity~field.

\subsubsection{The Principle of Stationary~Action  }
The mathematical tool is the Lagrangian
formalism of the relativistic field theory. To~derive the
equations of motion for the gravity field and for the matter,
one uses the principle of stationary action, which states
that for the true dynamical  behaviour of the field and matter
the variation of the action~$\delta S =0$.

The action integral for the whole system of a gravitational
field plus particles (matter) consists of the three parts
(instead of two parts in GRT Equation~(\ref{sgr})):
\begin{equation}
  \label{act-fgt}
  S = S_{(\mathrm{g})} + S_{(\mathrm{int})} + S_{(\mathrm{m})} = \frac{1}{c}
  \int \left(\Lambda_{(\mathrm{g})} + \Lambda_{(\mathrm{int})} +
 \Lambda_{(\mathrm{m})} \right) d\Omega\,.
\end{equation}
The notations (g), (int), (m)
refer to the actions for the gravity field, the~interaction, and~the matter (particles or other sources), $d\Omega = dV\,cdt$ .
The physical dimension of each
part of the action is
\mbox{[S] = [energy density]$\times $[volume]$\times$[time]},
which means that the concept of energy density of the field should
be defined in the theory at conceptual level.
In general relativity the action integral Equation~(\ref{sgr})
has only two parts $S_{\mathrm{g}}$ and $S_{\mathrm{m}}$. There is no
interaction part in GR, because~of the  principle of~geometrization.

\subsubsection{Lagrangian for the Gravitational~Field  }
Within the Feynman's field approach, the~gravity field is
presented by the reducible symmetric second rank tensor potentials
$\psi^{ik}(\vec{r}, t)$ in Minkowski space with metric $\eta^{ik}$,
and $\psi(\vec{r}, t) = \psi^{ik}\eta_{ik}$ is its trace.
The Lagrangian for the  gravitational field,
related to considered fixed source in total action (\ref{act-fgt}),
we~take in the form
\cite{fierz39,sokolov80,maggiore08}:
\begin{equation}
\label{g-lag}
  \Lambda_{(g)} =  - \frac{1}{16 \pi G}
\left[\, \left( 2 \psi_{nm}^{\,,n}
  \psi^{lm}_{\,,l} - \psi_{lm,n} \psi^{lm,n}\right) -
\left(
  2 \psi_{ln}^{\,,l}\psi^{\,,n} - \psi_{\,,l} \psi^{\,,l}\right)
\right] \,.
\end{equation}
This differs from Thirring's~\cite{thirring61} choice by a divergent term,
which does not change the field equations,
but ir has the advantage that the canonical energy momentum
tensor is symmetric.
Here, $\psi^{ik}_{\,\,, l}=\partial \psi^{ik}/\partial x^l$ is
the ordinary partial derivative of the symmetric second rank tensor
potential.

An important property of the Lagrangian (\ref{g-lag}) is that a gauge transformation of the potentials (\ref{gauge}) leads to the addition of only divergent terms, which does not change the field equations.
This gauge freedom allows for one to choose gauge conditions in the Hilbert--Lorentz form
$\psi^{ik}_{\,\,,k} = \frac{1}{2}\psi^{\,,i}\,$,
then take into account the irreducible representation of the initial potential
\begin{equation}
    \label{fi-f}
     \psi^{ik} = \psi^{ik}_{\{2\}} + \psi^{ik}_{\{0\}} =
  \phi^{ik} +   \frac{1}{4}\eta^{ik}\psi\, 
\end{equation}
we can present the Lagrangian (\ref{g-lag}), as
\begin{equation}
\label{g-lag-2-0}
  \Lambda_{(g)} =  - \frac{1}{16 \pi G}
\left[\,  - \phi_{lm,n} \phi^{lm,n}
+ \frac{1}{4}  \psi_{,l} \psi^{,l} \right] \,.
\end{equation}
where the irreducible fields, bound to the source, satisfy the gauge condition
$\phi^{ik}_{\,\,,k} = \frac{1}{4}\psi^{\,,i}\,$.

The different signs for the spin-2 and spin-0 fields in the Lagrangian (\ref{g-lag-2-0})
demonstrate that the irreducible trace-free tensor potential corresponds to attraction, while the irreducible 4-scalar field (\mbox{the trace $\psi$}) corresponds to the repulsion force; however, the coupling constant is the same---the~Newtonian gravitational constant $G$.
Note that it does not mean the negative energy of the scalar field, but~it means that the repulsive force by the spin-0 field compensates the attractive force by the spin-2 traceless  tensor field. Both attrative and repulsive components have positive energy density for free fields, but~they are generated by the common source
$T^{ik}_{(m)}$ and this explains the different signs in action~$S$.

\subsubsection{Lagrangian for~Matter  }
The Lagrangian for matter depends on the physical problem
in question (\mbox{particles, fields, fluid, or
gas}). Gravity is also a kind of matter and at each iteration step it is considered as a source fixed by the preceding~step.

For relativistic point (structureless) particles, the Lagrangian is
\begin{equation}
\label{6}
  \Lambda_{(\mathrm{p})} =  - \eta_{ik}T_{(\mathrm{p})}^{ik} \,,
\end{equation}
where $T_{(\mathrm{p})}^{ik}$ is the EMT of the particles
\begin{equation}
T_{(\mathrm{p})}^{ik} = \sum_{a} m_a c^2
                 \delta({\bf r}-{\bf r}_a)
     \{1 - \frac{v_a^2}{c^2}\}^{1/2}
                         u_a^i u_a^k  .
\label{11}
\end{equation}
Here $m,~~v,~~u^i$ are the rest mass, 3-velocity, and~4-velocity of a particle.
For a relativistic macroscopic body, the EMT is
\begin{equation}
\label{emt-m0}
  T_{(\mathrm{m})}^{ik} =
  \left( \varepsilon + p \right) u^{i} u^{k} - p\eta^{ik}\,,
\end{equation}
where $\varepsilon$ and $p$ are
the energy density and pressure of a comoving volume~element.

\subsubsection{The Principle of Universality and  Lagrangian for~Interaction  }
In the field approach, the {\it principle of universality}
states that the gravitational field $\:\psi^{ik}$
interacts with all kinds of matter via their energy-momentum tensor
$T^{ik}$, so the Lagrangian for the interaction has the form \cite{moshinsky50,sokolov80,maggiore08}:
\begin{eqnarray}
\label{ugi}
  \Lambda_{(\mathrm{int})} =  -
  \frac{1}{c^{2}} \psi_{ik} T^{ik}
\end{eqnarray}
The principle of universality, Equation~(\ref{ugi}),
was introduced by Moshinsky 1950~\cite{moshinsky50}. It replaces
the equivalence principle that is used in the geometrical
approach. From~the principle of universality of gravitational interaction
(UGI) and the stationary action principle one can derive
those consequences of the equivalence principle, which
do not create paradoxes. As will be shown below, according to UGI, the free fall acceleration of a body does not depend on its total mass, but~it does depend on the direction and value of its~velocity.

An interesting aspect of the UGI is the composition structure of the interaction Lagrangian for the irreducible representation of the symmetric tensor potential
\begin{equation}
\label{psi2-psi0}
 \psi^{ik} =
\psi^{ik}_{\{2\}} +
\psi^{ik}_{\{0\}} =
(\psi^{ik} - \frac{1}{4}\eta^{ik}\psi) +
\frac{1}{4}\eta^{ik}\psi =
  \phi^{ik} +
  \frac{1}{4}\eta^{ik}\psi\, 
\end{equation}
and symmetric EMT of matter
\begin{equation}
\label{Tik-T2-T0}
T^{ik} = T^{ik}_{\{2\}} +
  T^{ik}_{\{0\}} = (T^{ik} - \frac{1}{4}\eta^{ik}T) +
  \frac{1}{4}\eta^{ik}T\,,
\end{equation}
where
$\eta_{ik}\psi^{ik} = \psi\, $,
$\eta_{ik}\psi^{ik}_{\{2\}} = 0\, $, and~$\eta_{ik}T^{ik} = T\,$,
$\eta_{ik}T^{ik}_{\{2\}} = 0\, $.
So the interaction Lagrangian will have the form
\begin{equation}
\label{ugi2-0}
  \Lambda_{(\mathrm{int})} =  -
  \frac{1}{c^{2}}\, \psi_{ik} (T^{ik}_{\{2\}} +
   T^{ik}_{\{0\}}) = -\frac{1}{c^{2}}\,
 ( \phi_{ik} T^{ik}_{\{2\}} +
  \frac{1}{4}\,\psi\,T\,)\,.
\end{equation}
According to Equation~(\ref{ugi2-0}), the gravitational interaction is essentially  different for massive (non-zero EMT trace)  and massless (zero EMT trace) particles/fields.

Note that the equivalence principle of GRT cannot be a basis of the field gravity, because~it eliminates the gravity force and accepts the equivalence between the inertial motion and the accelerated motion.
For example, the~equivalence principle creates a puzzle in a gedanken experiment with the electric charge resting in the gravity field on a laboratory table (which was debated in the literature for a long time).
In the frame of GRT, due to the equivalence
of the laboratory frame (with gravity) and the accelerated free falling frame with ``$a=g$'', the~charge at rest on the table is equivalent with an accelerated free falling charge. However,~a charge with a constant
acceleration ``$a$'' should radiate energy
according to the relation $P=(2/3)(e^2/c^3)a^2$ ergs/s
while resting charge in the lab has $a=0$, so radiated energy $P$ should be zero. 
\footnote{A discussion of the puzzle of the electron resting in the lab gravitational field presented in~\cite{logunov96}.
Note that the problem exists also for the free falling electron on a circular orbit around the gravitating mass ($a^2=const$). Actually this effect
is related to more general Unruh effect~\cite{ginzburg87}. }

In the field gravity theory, the charge at rest on the table does not radiate
and the free falling charge does radiate, just because the inertial frame is not equivalent to the accelerated frame.
The~concept of an inertial frame is fundamental for all physical interactions and it is preserved in the field gravity~theory.

Instead of the equivalence principle,
QFGT is based on the principle of universality of gravitational interaction, according to which gravity ``see'' only the energy momentum tensor
of any matter. This~point is also different from all ``effective geometry'' theories, where the universality of gravity is understood as geodesic motion in Riemannian~space.

\subsection{Basic Equations of the  Quantum-Field Gravity~Theory}

\subsubsection{Field~Equations }
Using the variation principle to obtain the field
equations from the action (\ref{act-fgt}), one must assume that the sources $T^{ik}$ of the field are fixed (or the motion of the matter given) and only vary the potentials $\psi^{ik}$ (serving as the coordinates of the system). On~the other hand, to~find the equations
of motion of the matter in the field, one should assume the field to be given and vary the trajectory of the particle (matter). Accordingly, keeping the total EMT of matter in (\ref{ugi}) fixed and varying
$\delta\psi^{ik}$ in (\ref{act-fgt}), we get the following field equations
\cite{fierz39,thirring61,feynman71,sokolov80,maggiore08}:
\begin{equation}
    \label{feq1-fgt}
  - \psi^{ik,l}_{\;\;\;l} +
  \psi^{il,k}_{\;\;\;l} +\psi^{kl,i}_{\;\;\;l} - \psi^{,ik}
  -\eta^{ik} \psi^{lm}_{\;\;\; ,lm} + \eta^{ik}\psi^{,l}_{\,l}  = \frac{8
  \pi G}{c^{2}} T^{ik}\,.
\end{equation}
 The trace of the field Equation~(\ref{feq1-fgt}) gives the scalar equation for generating the scalar part of the symmetric second rank tensor---its
trace $\psi(\vec{r}, t)$, in~the form
\begin{equation}
    \label{feq-trace}
-2\psi^{,l}_{\,l} + 2\psi^{lm}_{\;\;\; ,lm}
= -\frac{8 \pi G}{c^{2}} T\,.
\end{equation}
Note that the sign of the source term presented in Equation~(\ref{feq-trace})
is negative, while the source's sign in Equation~(\ref{feq1-fgt}) is positive
(the d'Alembertian operator has the same sign).
This means that the scalar part of the symmetric tensor field gives repulsion instead of attraction for the traceless tensor~field.

The field Equation~(\ref{feq1-fgt}) is similar  to the
linear approximation
of Einstein's field equations, and that is why there are many similarities
between GRT and QFGT in the weak field regime.
However, as~we discussed in Section~\ref{sec1.3} (see Equation~(\ref{weak-gr-fg})), there is an important difference between QFGT and GRT, which is that
$\psi^{ik}(\vec{r},t)$ and $\eta^{ik}$
(both and their sum too) are true tensors of the Minkowski space and the trace
$\psi(\vec{r}, t)$ is a true scalar of the Minkowski space.
However, in GRT, the quantities $h^{ik}$ and $\eta^{ik}$ are not tensors of
a general Riemannian space. In~consistent geometrical approach
these quantities should obey the following relations:
  $h^{ik}=-h_{ik}$, $h^i_k = 0$ and $h = 0$ due to the
exact relation  $g^{ik}g_{ik} = 4$
for the trace of any metric tensor.
Strictly speaking, it means that in GRT the scalar part of the symmetric tensor potential is~lost.

In QFGT, the second rang symmetric tensor obeys other relations
for their components:
 $\psi^{ik}= \psi_{ik}$, and~$\psi^{i}_{k} = \psi^{k}_{i}$,
and $\psi = \eta_{ik} \psi^{ik}$,
where the four-scalar trace $\psi = \psi(\vec{r},t) $ is a function of coordinates (not a constant).
Accordingly, the scalar part of the symmetric tensor potential plays
fundamental role in the gravity~physics.

\subsubsection{Remarkable Features of the Field~Equations  }

First,
the divergence of the left side of the field Equation~(\ref{feq1-fgt})
is zero, which implies the conservation~law
\begin{equation}
\label{div}
  T^{ik}_{\;\;\;\; ,k} = 0\,,
\end{equation} 
which corresponds to considered approximation for the related  step in the iteration procedure. It is important to note that
 the conservation law (\ref{div}) does not restricts the trace of the energy-momentum tensor, which means that there is the scalar part of the symmetric EMT as a real source of the scalar~gravitons.

In the zero approximation, the equation of  energy-momentum conservation (\ref{div})
does not include the EMT of the gravity field and the interaction. The~first iteration takes into account
the total source EMT, including the gravity field of the zero approximation.
In the first (post-Newtonian) approximation the total EMT of the self-gravitating system is equal to the sum of the EMT for the matter, interaction and
gravity field  (Kalman~\cite{kalman61}; Thirring~\cite{thirring61}; and, Baryshev~\cite{baryshev88}):
\begin{equation}\label{emt-total}
 T^{ik}= T_{(\mathrm{m})}^{ik} + T_{(\mathrm{int})}^{ik} +
T_{(\mathrm{g})}^{ik}\,.
\end{equation}
The divergence of the EMT (\ref{emt-total})  expresses
the conservation laws of the first approximation, \mbox{which gives the equations} of motion at the post-Newtonian~level.

Second,  Equation~(\ref{feq1-fgt}) (and its consequence (\ref{feq-trace}))
is gauge invariant, i.e.,~it is not changed under the following
transformation of the potentials:
\begin{equation}
    \label{gauge}
  \psi^{ik} \Rightarrow \psi^{ik} + \lambda^{i,k}+
  \lambda^{k,i}\,,
\end{equation}
and the corresponding transformation of the trace
\begin{equation}
\label{gauge-trace}
  \psi \Rightarrow \psi + 2\lambda^{m}_{\,\,\,,m}\,.
\end{equation}
The four arbitrary functions $\lambda^{i}$ are consistent with the
four restrictions on the source EMT due to energy-momentum conservation.
Note that conservation law (\ref{div}) is a strict consequence of the field
Equation~(\ref{feq1-fgt}),
which does not depend on the choice of a gauge transformation. Additionally, the conservation law (\ref{div}) does not restrict the
Equation~(\ref{feq-trace}) for the scalar part of the source, while it restricts the Equation~(\ref{feq1-fgt}) for the tensor part of the~source.

An important conceptual difference between the coordinates
transformation in GRT and the gauge transformation
of the gravitational potentials in QFGT
is that the (\ref{gauge}) (and its consequence~(\ref{gauge-trace}))
are performed in a fixed inertial reference frame. 
\footnote{
The gage transformation (\ref{gauge}) of the gravitational potentials
 can also be written as
$\psi^{ik} \Rightarrow \psi^{ik} + \lambda^{i,k}+
\lambda^{k,i} + 2\gamma^{\,,ik}$ which however does not change the number
of arbitrary functions because the arbitrary function $\gamma$
can be included in 4 arbitrary new functions
$\lambda'_i = \lambda_i + \gamma_{\,,i} $.}
The gauge freedom (\ref{gauge}) and (\ref{gauge-trace})
allows for one to put four additional conditions on the
potentials, in~particular a Lorentz invariant gauge---the Hilbert-Lorentz gauge 
\footnote{Also called as the de Donder gauge.}
\cite{thirring61}:
\begin{equation}
\label{gauge-hl}
 \psi^{ik}_{\;\;\; ,k} = \frac{1}{2} \psi^{\;,i}\,.
\end{equation}
With the gauge (\ref{gauge-hl}), the field equations get
the form of wave equations:
\begin{equation}
\label{feq2-fgt}
   \left(\triangle - \frac{1}{c^2}\frac{\partial^2}{\partial t^2} \right)
\psi^{ik}
  =  \frac{8 \pi G}{c^{2}} \left[
  T^{ik} - \frac{1}{2} \eta^{ik}T\right]\,,
\end{equation}
and, for the trace component, this gives the field equation for the scalar
part $\psi = \eta^{ik}\psi_{ik}$ of the gravitational potentials:
\be
\label{scalar-eq}
    \left(\triangle - \frac{1}{c^2}\frac{\partial^2}{\partial t^2} \right)
    \psi (\vec{r},t)
    = -  \frac{8 \pi G}{c^{2}} T (\vec{r},t)\,.
\ee
Note that the opposite signs in the right-hand sides of Equations~(\ref{feq2-fgt}) and (\ref{scalar-eq}).
This corresponds to the important fact that the pure \textbf{tensor part of the field corresponds to attraction},
while the \textbf{scalar part gives repulsion}.
This result is caused by the fact that in the Lagrangian (\ref{g-lag}) the tensor and scalar parts have opposite signs,
which does not mean negative energy of the scalar field,
but it reflects the opposite signs of the pure tensor and pure scalar forces.

\subsubsection{Scalar and Traceless Tensor Are Dynamical Fields in~QFGT}
The analysis of multi-component structure of different fields is one of
the most important parts of the quantum field theory
\cite{approaches07,maggiore05}. Especially, the Poincare symmetry plays a very important role in QFT. The~representations of the Poincare group allow to consider scalar, vector, and tensor fields in Minkowski space and provide the Wigner's classification of the elementary particles
\cite{wigner1939}.

In the frame of the Feynman's QFGT, the basic field describing gravity phenomena is the symmetric tensor $\psi^{ik}(\vec{r}, t)$
having 10 independent components (degrees of freedom---dofs). According to representation theory
( Fronsdal~\cite{fronsdal58}; Barnes~\cite{barnes65}; and,
Maggiore~\cite{maggiore05}),
the symmetric 2nd rank tensor $\psi^{ik}$ is a reducible representation,
which can be decomposed under the inhomogeneous  Lorentz (Poincare) group into a direct sum
of irreducible representations: traceless 4-tensor, 4-vector, and 4-scalar
fieds. In~terms of spins, these fields correspond to the direct sum
of  subspaces: one spin-2, one~spin-1, and~two spin-0 representations. 
\footnote{The decomposition and the appropriate projection operators are exhibited explicitly e.g.,~in Barnes~\cite{barnes65}.}
Accordingly, it corresponds to the decomposition of the reducible symmetric tensor
into the $5+4=9$ component (dofs) of the traceless part plus $1$ component (dof), which is a diagonal-trace part. The~single diagonal component subspace is the spin-0 representation of the Poincare 4-scalar---the trace $\psi$ of the initial symmetric tensor $\psi^{ik}$.

The relation between the number of independent components $n$ and value of the spin $s$ is $n=2s+1$, so the symmetric tensor $\psi^{ik}$ contains
$\,\,n=5+3+1+1=10$ independent components.
In the spin-symbolic form, we have:
\be \label{10comp}
  \{\psi^{ik}\} = \{2\} \oplus [\{1\}  \oplus
  \{0^{'}\}] \oplus  \{0\}\,.
\ee
The \textbf{four gauge conditions (\ref{gauge-hl})}
exclude four independent components (dofs) of the symmetric potential which corresponds to deleting the irreducible four-vector field ($3+1=4$
components. \mbox{Hence,~the initial reducible symmetric }tensor potential will only contain two irreducible parts corresponding to spin-2 tensor field (5 dofs) and spin-0 scalar field (1 dof):
\be \label{6comp}
  \{\psi^{ik}\} = \{2\} \oplus   \{0\}\,.
\ee

The same decomposition can be done for the source of the total symmetric gravitational potential $\psi^{ik}$---i.e.,~the symmetric energy-momentum
tensor of matter $T^{ik}$, which initially has
10 independent~components:
\be \label{10compT}
  \{T^{ik}\} = \{2\} \oplus [\{1\}  \oplus
  \{0^{'}\}] \oplus  \{0\}\,.
\ee
and after \textbf{four restrictions from
the conservation laws (\ref{div})} of the energy-momentum tensor,
which~delete four source components corresponding to particles
with spin-1 and spin-0', we get
\be \label{6compT}
  \{T^{ik}\} = \{2\} \oplus   \{0\}\,.
\ee

Following to the Schwinger's source approach~\cite{schwinger73}, the real particles corresponds to the source components after taking
into account conservation laws
(four additional conditions to delete the four source components).
Hence,
the field Equation~(\ref{feq1-fgt}) will describe
only two real sources of
the gravitational potentials $\psi^{ik}$ as the mixture
of two fields with spin-2 and spin-0
(there is no restriction on the trace component).
Therefore, the matter EMT as the source of the gravitational field
generates two corresponding parts of the potentials:
\vspace{6pt}
\be
\{T^{ik}\} = \{2\} \oplus \{0\}
\Longrightarrow
\{\psi^{ik}\} = \{2\} \oplus \{0\}
\label{16}
\ee

Now, we can present the EMT of the source
and the symmetric tensor potentials
as the sum of
pure tensor spin-2 and spin-0 (4-scalar)  parts:
\be
\label{2+0-t}
T^{ik} = T^{ik}_{\{2\}} + T^{ik}_{\{0\}} =
 (T^{ik} - \frac{1}{4}\eta^{ik}T)  +
\frac{1}{4}\eta^{ik}T\,
\ee
\be \label{2+0-psi}
  \psi^{ik} = \psi^{ik}_{\{2\}} +
  \psi^{ik}_{\{0\}} = (\psi^{ik} - \frac{1}{4}\eta^{ik}\psi) +
  \frac{1}{4}\eta^{ik}\psi =
  \phi^{ik} +
  \frac{1}{4}\eta^{ik}\psi\,
  = \phi^{ik} + \theta^{ik}\, ,
\ee
where
$\phi^{ik}=\psi^{ik}_{\{2\}}$,
$\,\,\eta_{ik}\psi^{ik}_{\{2\}}=0\,\,$, 
$\,\,\theta^{ik} = \psi^{ik}_{\{0\}}$,
$\,\,\eta_{ik} \psi^{ik}_{\{0\}}=\psi\,\,$
and
$\,\,\eta_{ik} T^{ik}_{\{2\}}=0$,  $\,\,\eta_{ik} T^{ik}_{\{0\}}=T$.

Both Equations~(\ref{feq1-fgt}) and (\ref{feq-trace}) are gauge
invariant; hence, for the Hilbert--Lorentz gauge (\ref{gauge-hl}),
they can be written in the~form

\be
\left(\triangle - \frac{1}{c^2}\frac{\partial^2}{\partial t^2} \right)
\psi^{ik}_{\{2\}} =
                     \frac{8\pi G}{c^2}\,
                                  T^{ik}_{\{2\}}\;
         \, \,\, \, \mbox{\rm or}\;\,\,\,
\left(\triangle - \frac{1}{c^2}\frac{\partial^2}{\partial t^2} \right)
\phi^{ik} =
                     \frac{8\pi G}{c^2}
     \left[ T^{ik}
                     -\frac{1}{4}
 \eta^{ik} \, T\right]
\label{17a}
\ee
and
\be
\left(\triangle - \frac{1}{c^2}\frac{\partial^2}{\partial t^2} \right)
\psi^{ik}_{\{0\}} =
                   - \frac{8\pi G}{c^2}\,
                                  T^{ik}_{\{0\}}\;
          \,\,\,\,  \mbox{\rm or}\;\,\,\,\,
\left(\triangle - \frac{1}{c^2}\frac{\partial^2}{\partial t^2} \right)
\psi \frac{1}{4} \eta^{ik} =
                    - \frac{8\pi G}{c^2}\,
           T
                       \frac{1}{4}
\eta^{ik}
\label{17b}
\ee
where the gauge conditions were used in the form
\be
\psi^{ik}_{\,\,,k} =\frac{1}{2}\psi^{\,,i} \,\,\,
\Longleftrightarrow  \,\,\,
\phi^{ik}_{\,\,,k} = \frac{1}{4}\psi^{\,,i}
\label{psi-phi-gauge}
\ee

This means that the field gravity theory is actually
a scalar-tensor theory, where the intrinsic scalar part of the field
is simply the trace of the tensor potentials
$ \psi  =\eta_{ik} \psi^{ik}$ generated by
the trace of the energy-momentum tensor of the matter
$T=\eta_{ik}T^{ik}$.
According to the wave
Equations~(\ref{17a}) and (\ref{17b})
for spin-2 and spin-0 fields, both kinds of gravitons are massless particles, moving with velocity of light (Sokolov \& Baryshev~\cite{sokolov80}).

Zakharov 1965~\cite{zakharov65} noted that the interacting gravitational
field $\psi^{ik}$ in Equation~(\ref{feq1-fgt}) describes spin-2 and spin-0 gravitons. However, the absence of the scalar part of the metric in geometric gravity theory
($g^{ik}g_{ik} =4$ and absence of the scalar waves in GRT) led him to rejection of the longitudinal scalar part of the gravitational propagator.
However, in~the frame of  QFGT,
one should take into account both quadrupole tensor and monopole scalar gravitational radiation. The~conservation law (\ref{div})
does not restrict the scalar part of the source,
and this is why the spin-0 field is real and not a constraint field.
\footnote{In the case of electrodynamics the conservation of 4-current
lead to exclusion of 1 dof of the source of the 4-vector potentials (\ref{4-field-eq-em}), so there is no source of the scalar photons.
The conservation law of EMT in QFGT restrict only four components and leaves 6 independent dofs $-$ pure  tensor and trace-scalar dynamical fields.
In the metric gravity theories there is an additional condition that the trace of the metric tensor equals to constant, so the 4-scalar wave is absent.}

The source of the scalar wave is the variable trace of the EMT source, e.g.,~for particles $T = mc^2 (1 - v^2/c^2)^{1/2}$ and variation of the particles kinetic energy will generate the scalar gravitational radiation, e.g.,~via spherical pulsations of a gravitating system.
The radiated scalar gravitational wave is
monopole and it has a longitudinal character in the sense that a test particle in the wave moves along the direction of the wave propagation (GWs are considered in Section~\ref{sec4.2}).

\subsubsection{The Energy-Momentum Tensor of the Gravity~Field  }

The standard Lagrangian formalism and the Lagrangian of the gravity field (\ref{g-lag}) give
the following expression for the canonical energy-momentum tensor:
\be \label{emt-g}
  T_{(g)}^{ik} = \frac{1}{8 \pi G}
 \left\{
      (\psi^{lm,i}\psi_{lm}^{\phantom{lm},k} -
 \frac{1}{2}
             \eta^{ik} \psi_{lm,n} \psi^{lm,n}) -
 \frac{1}{2}
     (\psi^{,i}\psi^{,k} - \frac{1}{2}
                 \eta^{ik}\psi_{,l}\psi^{,l})\right\}
\ee
Several important remarks should be made regarding this~expression.

First, the~canonical EMT of the gravitational field is a symstric
tensor of the Minkowski space  and so it is conceptually
well defined. However,
the Lagrangian formalism cannot give a unique
expression for an EMT of any field (e.g.,~Bogolyubov \& Shirkov~\cite{bogolubov93};
Landau \& Lifshitz~\cite{landau71}), because a term with zero divergence
can always be added to the Lagrangian, which does not change the field
equations, but will change the expression for the EMT.
For the final determination of the EMT of the field, some
additional physical requirements can be used to obey such EMT properties
as the positive
energy density, the~symmetry, zero value for the trace in the
case of a massless~field.

Second, the~expression (\ref{emt-g}) can be written in the form,
where the tensor $\psi^{ik}$ is presented as the sum of irreducible
pure spin-2 ($\phi^{ik}$) and 4-scalar spin-0 ($\psi$) parts
(\ref{2+0-psi}):
\be \label{emt-g-2-0}
  T_{(g)}^{ik} = \frac{1}{8 \pi G}
 \left\{
      (\phi^{lm,i}\phi_{lm}^{\phantom{lm},k} -
 \frac{1}{2}
             \eta^{ik} \phi_{lm,n} \phi^{lm,n}) -
 \frac{1}{4}
     (\psi^{,i}\psi^{,k} - \frac{1}{2}
                 \eta^{ik}\psi_{,l}\psi^{,l})\right\}
\ee
The two terms  in the canonical gravity EMT
(\ref{emt-g-2-0}))
have opposite signs and relate to the sources of the two parts of the symmetric tensor potential---pure tensor attractive field and scalar
repulsive field, which are determined by the total Lagrangian (\ref{g-lag}).
Because the spin-2 \textbf{attraction force} has opposite sign relative to
the spin-0 \textbf{repulsion force}, the~canonical symmetric gravity EMT
(\ref{emt-g-2-0}) of the field
$\psi^{ik} = \psi_{2}^{ik} + \psi_{0}^{ik}$
may be presented as the \textbf{difference between to positive EMTs} of the spin-2 and spin-0 fields:
\be
T_{(g)}^{ik} =  T_{(g)\{2\}}^{ik} - T_{(g)\{0\}}^{ik} \,.
\label{T2-T0}
\ee
The negative sign of the scalar part
(the 2nd term in (\ref{T2-T0}))
does not mean that the spin-0 field has negative energy.
It reflects the repulsive character of the force produced by the scalar field, which compensates the attraction force and, hence, decreases the effective total energy of the composite field. 
For the symmetric tensor field
$\psi^{ik}$ "tied" 
to the source,
the spin-2 and spin-0 fields interconnect through the source
(via gauge conditions (\ref{psi-phi-gauge})).
Note that the Lagrangian's freedom can be also used to obtain zero trace
of both $T_{(g)\{2\}}^{ik}$ and $T_{(g)\{0\}}^{ik}$.

Third, for~the free field the energy is positive for both
the pure tensor (spin-2) and scalar (spin-0) components.
Indeed, the~total Lagrangian (\ref{g-lag}) of the total gravity field
can be divided into \textbf{two independent parts} (in linear approximation) that correspond to two
independent particles with spin 2 ($\phi^{ik}$)
and spin 0 ($\psi$). Thus, we have two
following free field Lagrangians
\be
\Lambda_{\{2\}} =
             \frac{1}{16\pi G}
\phi_{lm,n}\phi^{lm,n} ,
                               \quad
            \mbox{\rm and}
                               \quad
\Lambda_{\{0\}} =
             \frac{1}{64\pi G}
\psi_{,n}\psi^{,n}       .
\label{30b}
\ee
Both of the signs are  positive  due  to  the positive  energy  density
condition for interger spin free particles.  Corresponding
EMTs for the tensor and scalar free fields are
\be
T_{(g)\{2\}}^{ik} =
             \frac{1}{8\pi G}
\phi_{lm}^{\phantom{lm},i}\phi^{lm,k} ,
                    \quad
            \mbox{\rm and}
                    \quad
T_{(g)\{0\}}^{ik} =
             \frac{1}{32\pi G}
\psi^{,i}\psi^{,k} \,\,.
\label{31b}
\ee
These EMTs are symmetric, have a positive energy density, and a zero trace
for the case of plane monochromatic~waves.

\subsubsection{The Retarded~Potentials}
In the frame of FGT, the solution of the field Equation~(\ref{feq2-fgt})
for the case of the weak field and slow motion can be presented in the
form of retarded potentials:
\begin{equation}\label{psi-retard-0+2}
  \psi^{ik} (\vec{r},t)=-\frac{2G}{c^2}
  \int{\frac{\hat{T}^{ik}(\vec{r}\,'\,,\,t-R/c)}{R}dV' }
  + \psi^{ik}_0\;,
\end{equation}
where $\vec{R}=\vec{r}-\vec{r}\,'$  - is the radius vector from the
volume element $dV' = dx'dy'dz'$ to the point $\vec{r}$, the~source
has the form $\hat{T}^{ik}=T^{ik}-(1/2)T\eta^{ik}$ in corresponding
approximation, $\psi^{ik}_0$ is the free field~solution.

There is important exact solution of the scalar field Equation~(\ref{scalar-eq}),  which (in the case of zero trace of the field and
interaction EMTs) has the form :
\begin{equation}\label{psi-retard-0}
  \psi (\vec{r},t)=\frac{2G}{c^2}
  \int{\frac{T_{(m)}(\vec{r}\,'\,,\,t-R/c)}{R}dV'} +\psi_0\;,
\end{equation}
where $T_{(m)}$ is the trace of the matter EMT, and~$\psi_0$ is
the solution of Equation~(\ref{scalar-eq}) without the right-hand side.
In particular, for~the case of a moving test particle along trajectory
$\vec{r}_0=\vec{r}_0(t)$, having the trace of EMT in the form
\begin{equation}\label{trace-point-part}
  T=mc^2\sqrt{1-v^2/c^2}\delta (\vec{r}-\vec{r}_0)\;,
\end{equation}
one gets from Equation~(\ref{psi-retard-0})
\begin{equation}\label{psi-grav-LV}
  \psi (\vec{r},t)=
  \frac{2Gm\sqrt{1-v^2/c^2}}{R-(\vec{v}\cdot\vec{R})/c}=
  \frac{2R_mc^2}{R}\mathcal{D}\;,
\end{equation}
where $R_m=Gm/c^2,\; R=|\vec{r}-
\vec{r}_0|\;,\;v=v(t')\;,\;R=R(t')\;$ - in the particle point
at the moment $t' = t-R(t')/c\;$, and~$\; \mathcal{D}=\sqrt{1-v^2/c^2}/(1-v\cos \theta/c)$ -
the relativistic~Doppler-factor.

\subsection{ Equations of Motion for Test~Particles}
\label{sec-3-3}
Let us consider the motion of a relativistic test particle with rest-mass $m_0$, four-velocity $u^i$, three-velocity $\vec{v}$ in the gravitational field described by the symmetric tensor
potential $\psi^{ik}$ in the flat Minkowski space-time, where the Cartesian coordinates always exist and the metric tensor is  $\eta^{ik}=diag(1,-1,-1,-1)$.

\subsubsection{Derivation from Stationary Action~Principle}
In order to derive the equation of motion in QFGT, we use the stationary action principle in the form
of the sum of the free particles and the interaction parts:
\begin{equation}
\label{delta1-S}
\delta S =\delta(\frac{1}{c}\int (\Lambda_{(p)} + \Lambda_{(int)})
d\Omega ) = 0
\end{equation}
where $d\Omega$ is the element of four-volume and
the variation of the action is made with respect to the
particle trajectories $\delta x^i$
for \textbf{fixed gravitational potential} $\psi^{ik}$.
The free particle Lagrangian is
\begin{equation}
\label{Lambda-p}
\Lambda_{(p)} = -\eta_{ik} T_{(p)}^{ik}
\end{equation}
and the interaction Lagrangian in accordance with principle of
universality of the gravitational interaction Equation~(\ref{ugi}) is
\begin{equation}
\label{Lambda-int-p}
\Lambda_{(int)} = -\frac{1}{c^2}\psi_{ik} T_{(p)}^{ik}
\end{equation}
It is important to note that the Lagrangian of the particles in gravitational field in (\ref{delta1-S}) can be written as
\begin{equation}
\label{Lambda-int-p-g}
\Lambda_{(p+int)} =
-(\eta_{ik} + \frac{1}{c^2}\psi_{ik})\,T_{(p)}^{ik}
= \hat{g}_{ik}\,\,T_{(p)}^{ik}
\end{equation}
where quantity $\hat{g}_{ik} = \eta_{ik} + \frac{1}{c^2}\psi_{ik}$ is usually
called as a metric of the effective Riemannian space.
However, the $\hat{g}_{ik}$ cannot be a metric tensor because its trace
does not equal to four (\mbox{$\hat{g}_{ik} \hat{g}^{ik} = 4 + 2\psi(\vec{r},t)$}
see (\ref{weak-gr-fg})). Accordingly, consistent QFGT cannot be a metric theory of gravitation. Demanding that the
\mbox{$\hat{g}_{ik} = \eta_{ik} + \frac{1}{c^2}\psi_{ik}$} is the metric tensor, equivalent to the ``geometrization~principle''. 

Taking the energy-momentum tensor of the point particle
in the form
\begin{equation}
\label{T-p-ik}
T_{(p)}^{ik}=m_0 c^2\delta(\vec{r}-\vec{r_p})\{1-\frac{v^2}{c^2}\}^{1/2}
u^i u^k
\end{equation}
and using Landau \& Lifshitz \cite{landau71} method
of four=coordinates variation one can derive the equation of motion.
Inserting Equations~(\ref{Lambda-int-p})  and (\ref{T-p-ik}) 
into Equation~(\ref{delta1-S})
and taking into account that $ds^2=dx_ldx^l$, we get
\begin{equation}
\label{var-S}
\int (m_0c\delta(\sqrt{dx_ldx^l})+
\frac{m_0}{c}\delta(\psi_{ik}\frac{dx^idx^k}{\sqrt{dx_ldx^l}} ))
= 0
\end{equation}
Performing the variation and integrating by parts, and~taking into
account that the variation is made for the fixed values
of the integration limits,  we find
\begin{eqnarray}
\label{var1-S}
\int ( m_0 cdu_i \delta dx^i +
\frac{2m_0}{c}d(u^k\psi_{ik})\delta x^i - \nonumber\\
\frac{m_0}{c}d(\psi_{lk}u^l u^k u_i)\delta x^i -
\frac{m_0}{c}u^k \delta \psi_{ik} dx^i ) = 0
\end{eqnarray}
Also consider that
$$
du^i = \frac{du^i}{ds}ds ; \qquad  dx^i=u^ids ; \qquad
\delta \psi_{ki} = \psi_{ki,l}\delta x^l ;    \qquad
$$
$$
d(u^lu^ku_i\psi_{lk}) = u^lu_iu^k\psi_{lk,n}dx^n +
u^lu^k\psi_{lk}du_i + 2\psi_{lk}u^lu_idu^k ;
$$
finally we obtain the following equation of motion for test
particles in the field gravity theory (Baryshev~\cite{baryshev86}):

\be
A^{i}_{k}\frac{d(m_0 cu^k)}{ds} =
 -m_0 cB^{i}_{kl}u^k u^l\,,
\label{18}
\ee
where $m_0 cu^k=p^k$ is the four-momentum of the particle, and~\be
   A_{k}^{i} =
            \left(1 -\frac{1}{c^2} \psi_{ln}
     u^l u^n \right) \eta^{i}_{k} -
 \frac{2}{c^2}\psi_{kn}u^n u^i + \frac{2}{c^2}\psi^{i}_{k}\,,
\label{19}
\ee
\be
  B^{i}_{kl} = \frac{2}{c^2} \psi^{i}_{k,l} -
      \frac{1}{c^2} \psi^{\phantom{kl},i}_{kl} -
  \frac{1}{c^2}\psi_{kl,n}u^n u^i \,,
\label{20}
\ee

Equation~(\ref{18}) is identical to the equation of motion that is derived by Kalman~\cite{kalman61} in another way,
by considering the relativistic Lagrange function
$L$ defined as $S=\int L \frac{ds}{c}$ and
relativistic Euler equation:
\begin{equation}
\frac{d}{ds}{(\frac{\partial L}{\partial u^k}u^k -L)u_i +
\frac{\partial L}{\partial u^i}} =
-\frac{\partial L}{\partial x^i}
\label{Kalman}
\end{equation}
Inserting in Equation~(\ref{Kalman}), the expression for the relativistic Lagrange function
\begin{equation}
\label{Lagrange-function}
L=-m_0c^2 - m_0\psi_{ik}\frac{dx^i}{ds}\frac{dx^k}{ds}
\end{equation}
one gets Kalman's equations of motion (Kalman~\cite{kalman61}),
which may also be presented in the form of Equation~(\ref{18}).

\subsubsection{Static Spherically Symmetric Weak~Field}

For a spherically symmetric static weak field
of a body with mass density $\rho_0$ and total mass $M$,
the zero (Newtonian) approximation of the total EMT
equals that of the matter
\be
\label{emt-m-pn-0}
T_{(m)}^{ik} =
{\rm diag}(\rho_{0} c^2,\,0,\,0,\,0)
\ee
and the field equations (Equation (\ref{feq2-fgt})) is
\begin{equation}
\label{feq2-fgt-sss}
  \triangle 
\psi^{ik}
  =  \frac{8 \pi G}{c^{2}} \left[
  T_{(m)}^{ik} - \frac{1}{2} \eta^{ik}T_{(m)}\right]\,,
\end{equation}
so, its solution, for~fixed boundary conditions, is the Birkhoff's~\cite{birkhoff44} potential
\be
\psi^{ik} = \varphi_{_{\rm N}} \,{\rm diag} (1,\,1,\,1,\,1),
\label{21}
\ee
where $\varphi_{_{\rm N}}= -GM/r$
is the Newtonian potential outside the SSS gravitating body.
We again note that $\psi^{ik}$ is a true tensor quantity in
Minkowski~space.

According to Equation~(\ref{2+0-psi}), the~Birkhoff's gravitational potential (\ref{21})
 can  be expressed as
the sum of the pure tensor and scalar parts
$\psi^{ik} = \psi^{ik}_{\{2\}} +  \psi^{ik}_{\{0\}}$,
 so that
\be
\psi^{ik} =
 \frac{3}{2}\varphi_{_{\rm N}} \,
 {\rm diag}
(1, \,\frac{1}{3},\, \frac{1}{3},\, \frac{1}{3})\,\, - \,\,\,
\frac{1}{2}  \varphi_{_{\rm N}} \,
 {\rm diag}
(1,\, -1,\, -1,\, -1) \,.
\label{22}
\ee
This corresponds to attraction by spin-2 and repulsion
by spin-0 parts of the Birkhoff potential.
Indeed, inserting potential (\ref{22}) to the equation of motion (\ref{18}) in the considered approximation we obtain expression for 3-force in the form:
\begin{equation}
\vec{F}_N=\vec{F}_{(2)} + \vec{F}_{(0)} =
-\frac{3}{2}m\vec{\nabla}\varphi_N + \frac{1}{2}m\vec{\nabla}\varphi_N
= -m\vec{\nabla}\varphi_N
\label{psiNew1}
\end{equation}
Hence, the Newton force of gravity is the sum of attraction due to the spin-2 tensor field and repulsion due to the spin-0 scalar field.
Thus, QFGT is a scalar-tensor~theory.

It is important to note that, according to the scalar-tensor Brans--Dicke theory (e.g.,~Capozziello \& Faraoni 2011~\cite{cap-far2011}), the~Lagrangin includes the additional scalar field $\phi$, which has a new coupling constant $\omega$. The~modern observational restrictions on this parameter is $\omega < 0.001$, so the Brans--Dicke scalar can only give a small correction to GRT observable~effects. 

In contrast, according to  QFGT, the~scalar field $\psi$
is the essential intrinsic part of the reducible symmetric tensor potential $\psi^{ik}$, i.e.,~its trace $\psi = \eta_{ik}\psi^{ik}$,
and it has the common universal coupling constant---gravitational constant $G$.

\subsubsection{The Role of the Scalar Part of the~Field  }
The scalar $\psi $ is an intrinsic part of the gravitational tensor potential $\psi^{ik}$ and does not relate to the extra scalar fields
introduced in the Jordan--Brans--Dicke theory.
Accordingly, the observational constraints existing for this extra scalar field
do not restrict the scalar part $\psi $ of the tensor field $\psi^{ik}$.
Moreover, without~the scalar $\psi $ it is impossible to explain the Newtonian gravity and the classical relativistic gravity~effects.

Inserting the scalar part of the gravitational potential (in the form
$\psi^{lm}_{\{0\}}=(1/4)\psi\eta^{lm}$)
to the equation of motion (\ref{18}), we obtain the equation of motion of a test particle in the scalar gravity
field $\psi=\psi_{lm}\eta^{lm}$, as
\begin{equation}\label{eq-motion-scalar}
  \left(1+\frac{1}{4}\frac{\psi}{c^2}\right)\frac{dp^i}{ds}=
  \frac{m}{4c}\left( \psi^{,i} - \psi_{,l}u^lu^i \right)\;.
\end{equation}
In the case of a weak field ($\psi/c^2<<1$), this equation gives,
for spatial components ($i=\alpha$), the expression for the gravity three-force
\begin{equation}\label{eq-motion-scal-N}
  \frac{d\vec{p}}{dt}=-\frac{m}{4}\vec{\nabla}\psi\,.
\end{equation}
In the case of the weak static field (\ref{21}),
the trace of the tensor gravitational potential
is equal to $\psi=-2\varphi_{_{\rm N}}$, hence we get for the gravity 3-force
\vspace{6pt}
\begin{equation}\label{eq-N-scal}
   \frac{d\vec{p}}{dt}=+\frac{1}{2}m\vec{\nabla}\varphi_{_N}\,.
\end{equation}
This means that the scalar spin-0 part of the tensor field leads to a repulsive force and only together with the attractive force from the pure
tensor spin 2 part the result is the Newtonian force
(\ref{psiNew1}).

The most intriguing consequence of the quantum-field gravity theory is that the scalar part (spin-0) corresponds to a repulsive force, while the pure tensor part (spin-2) corresponds to attraction.
This explains the "wrong" sign for the scalar part
in the EMT of the gravity field (\ref{emt-g-2-0}), because~total Lagrangian (\ref{g-lag}) is a part of the total action (\ref{act-fgt})
for the system gravity plus sources and
describes simultaneously attractive and repulsive parts of the total field
tired to the source part of the~action.

\subsection{Poincar\'{e} Force and Poincar\'{e} Acceleration in PN~Approximation }

\subsubsection{The EMT Source in~PN-Approximation}
In the first (post-Newtonian) approximation, the total EMT of the system is equal to the sum of the EMT for the matter, interaction, and
gravity field  (Kalman~\cite{kalman61}; Thirring~\cite{thirring61}; and, Baryshev~\cite{baryshev88}):
\begin{equation}\label{emt-12}
 T_{(\Sigma)}^{ik}= T_{(\mathrm{p/m})}^{ik} + T_{(\mathrm{int})}^{ik} +
T_{(\mathrm{g})}^{ik}\,.
\end{equation}
%
Taking into account (Equation~(\ref{21}) for Birkhoff's potential and accepting  the following
expressions for the interaction EMT
\be
\label{emt-int-pn-0}
T_{(\mathrm{int})}^{ik} = \rho_0 \varphi _{\mathrm{N}}
{\rm diag}(1,\,1/3,\,1/3,\,1/3)
\ee
and the EMT of the gravity field
\be
\label{emt-g-pn-0}
T_{(\mathrm{g})}^{ik} =+ \frac{1}{8\pi G}
\left(\nabla \varphi_{_{\rm N}}\right)^2
{\rm diag}(1,\,1/3,\,1/3,\,1/3)
\ee
we find the  total  energy  density
for the system gas + gravity in the form
\be
T_{(\Sigma)}^{00} = T_{(\rm p/m)}^{00} + T_{(\rm int)}^{00} +
   T_{(\rm g)}^{00} = \left(\rho_{0} c^2 + e \right) +
\rho_{0}\varphi_{_{\rm N}} + \frac{1}{8\pi G}
\left(\nabla \varphi_{_{\rm N}}\right)^2 \,.
\label{23}
\ee
Here, ($\rho_{0}c^2 + e$)
gives  the rest  mass  and  kinetic (or thermal)  energy densities,
$
\rho_{0}\varphi_{_{\rm N}}
$
is the negative interaction energy density and~$
(\nabla \varphi_{_{\rm N}})^2/8\pi G
$
is  the positive and localizable energy
density of the gravitational field
(equals to $T_{(g)}^{00}$ from (\ref{emt-g})).

\subsubsection{Relativistic Physical Sense of the Potential~Energy}
The total  energy  of  the system in PN approximation will be
\be
E_{(\Sigma)} =
\int T_{(\Sigma)}^{00} dV = E_0 + E_{\rm k} + E_{\rm p}
\label{24}
\ee
where
$
E_{0} = \int(\rho_{0}c^2)\,dV
$
is the rest-mass energy,
$
E_{\rm k} = \int(e)\,dV
$
is
the kinetic energy and~$E_{\rm p}$
is the classical potential energy that equals the sum of the interaction and
gravitational field energies:
\be
E_{\rm p} = E_{(\rm int)} + E_{(\rm g)} =
 \int(\rho_0 \varphi_{_{\rm N}} +
 \frac{1}{8\pi G} (\nabla \varphi_{_{\rm N}})^2)\,
 dV = \frac{1}{2} \int \rho_0 \varphi_{_{\rm N}}\, dV
\label{25}
\ee

\subsubsection{The PN Correction due to the Energy of the Gravity~Field}
In the quantum-field approach, a gravitating body is surrounded by a coat of virtual gravitons, i.e.,~a material gravitational field $\psi^{ik}$ whose
mass-energy density is given by the 00-component
of the EMT of the gravity field in Equation~(\ref{emt-g-pn-0}):
\begin{equation}\label{emt-g-00-pn}
 T_{(g)}^{00}=
  \varepsilon_g =\frac{ \left(\vec{\nabla }
  \varphi_{_{\rm N}}\right)^2}{8\pi G}=
  \frac{\,G\,M^2}{8\pi r^4}\,\,\,\mbox{\rm erg/cm}^3  \;.
\end{equation}
In the PN approximation, this corresponds to a nonlinear correction for the gravitational~potential.

When considering the energy density of the
gravitational field (the last term in Equation~(\ref{23})) as the source
in the field equation of the second order, we obtain
a nonlinear addition to Birkhoff's potential outside the SSS gravitating body:
\be
\psi^{ik} = \,{\rm diag} 
(\phi\,, \,\varphi_{_{\rm N}},\, \varphi_{_{\rm N}},\, \varphi_{_{\rm N}}),
\label{21-2}
\ee
where $\psi^{00}$ component is given by
\be
\psi^{00} =  \phi =
\varphi_{_{\rm N}} + \frac{1}{2}
\frac{ (\varphi_{_{\rm N}})^2}{c^2}\,\,.
\label{26}
\ee
where $\varphi_{_{\rm N}}= -GM/r$.
Corrections to other components do not influence the motion of particles in this~approximation.

\subsubsection{Post-Newtonian Equations of~Motion}
In the PN approximation, we keep terms down to an order of $v^2/c^2 \sim \mid\varphi_N/c^2\mid \ll 1$ in Equation~(\ref{18}). For~the PN accuracy we need calculations of the $\psi^{00}$
component with the same order, while other components of the tensor gravitational potential $\psi^{ik}$ can be calculated in the linear~approximation.

From the ($i=\alpha$) component of Equation~(\ref{18}),
under these assumptions, we get
the expression for the PN 3-dimensional
gravity force (that we shall call
the Poincar\'{e} gravity force remembering his pioneer work
in 1905 on the relativistic gravity force in flat space-time):
\begin{eqnarray}
\vec{F}_{Poincare}= \frac{d\vec{p}}{dt} =
-m_0\{(1+\frac{3}{2}\frac{v^2}{c^2} +
3\frac{\phi}{c^2})\vec{\nabla}\phi
-3\frac{\vec{v}}{c}(\frac{\vec{v}}{c}
\cdot\vec{\nabla}\phi)\} \nonumber\\
-m_0\{3\frac{\vec{v}}{c}\frac{\partial \phi}{c\; \partial t} -
2\frac{\partial \vec{\Psi}}{c\; \partial t} +
2(\frac{\vec{v}}{c}\times rot\vec{\Psi})\}
\label{Poincare-PN-force}
\end{eqnarray}
where $\phi= \psi^{00}$, $\vec{\Psi}=\psi^{0\alpha} = -\psi_{0\alpha}$.
Taking into account the relativistic relation between three-force and~three-acceleration
\begin{equation}
\label{force-accel}
\frac{d\vec{v}}{dt} = 
\frac{\sqrt{1-(\vec{v}^2/c^2)}}{m_0}
\left( \vec{F}  -
\frac{\vec{v}}{c}(\frac{\vec{v}}{c}\cdot\vec{F})
\right)
\end{equation}
we obtain the corresponding Poincare three-acceleration of the test particle :
\begin{equation}
\frac{d\vec{v}}{dt} = -
(1+\frac{v^2}{c^2} +
4\frac{\varphi_N}{c^2})\vec{\nabla}\varphi_N +
4\frac{\vec{v}}{c}(\frac{\vec{v}}{c}\cdot\vec{\nabla}\varphi_N)
+3\frac{\vec{v}}{c}\frac{\partial \varphi_N}{c\; \partial t} -
2\frac{\partial \vec{\Psi}}{c\;\partial t} +
2(\frac{\vec{v}}{c}\times rot\vec{\Psi})
\label{Poincare-PN-accel}
\end{equation}
From the ($i=0$) component of Equation~(\ref{18}),
we get the expression for the work of the Poincar\'{e}
force:
\begin{eqnarray}
\frac{dE_k}{dt}= \vec{v}\cdot\vec{F}_{Poincare} = 
-m_0\vec{v}\cdot \{
(1-\frac{3}{2}\frac{v^2}{c^2} + 3\frac{\phi}{c^2})\vec{\nabla}\phi -
3\frac{\vec{v}}{c}\frac{\partial \phi}{c\;\partial t} +
2\frac{\partial \vec{\Psi}}{c\;\partial t}\}
\label{Poincare-work-PN}
\end{eqnarray}

An important particular case is the static spherically symmetric
weak gravitational field for which $\vec{\Psi}=0$, $\partial
\phi/\partial t =0$, $\psi^{ik}=diag(\phi,\varphi_N,\varphi_N,\varphi_N)$;
hence, the  PN 3-acceleration will have the simple form:
\begin{equation}
(\frac{d\vec{v}}{dt})_{FGT} = -
(1+\frac{v^2}{c^2} +
4\frac{\varphi_N}{c^2})\vec{\nabla}\varphi_N +
4\frac{\vec{v}}{c}(\frac{\vec{v}}{c}\cdot\vec{\nabla}\varphi_N)
\label{FGT-accel}
\end{equation}
From the equation of motion (\ref{FGT-accel}), it is clear that the acceleration of a test particle depends on the value and the direction of its velocity, and~gravitational~potential. 

Importantly, all of these equations are coordinate independent and they correctly describe correctly all of the observationally tested PN relativistic gravity effect. In~particular, there are certain predictions for the difference between accelerations in tangential and radial motion. Accordingly, for the circular orbit, where $\vec{v}\perp \vec{\nabla}\varphi_N$, the~PN 3-acceleration is
\begin{equation}
(\frac{d\vec{v}}{dt})_{FGT}^{\perp} = -
(1+\frac{v^2}{c^2} +
4\frac{\varphi_N}{c^2})\vec{\nabla}\varphi_N
\label{FGT-accel-perp}
\end{equation}
and for radial motion, where $\vec{v}\uparrow \downarrow \vec{\nabla}\varphi_N$, 
the PN 3-acceleration is given by
\begin{equation}
(\frac{d\vec{v}}{dt})_{FGT}^{\parallel} = -
(1-3\frac{v^2}{c^2} +
4\frac{\varphi_N}{c^2})\vec{\nabla}\varphi_N
\label{FGT-accel-par}
\end{equation}

In GRT, as~we noted above, the PN equation of motion Equation~(\ref{GR-accel})
is dependent on the choice of a coordinate system
(due to parameter $\alpha$). 
In QFGT, the equation of motion (\ref{FGT-accel}) is valid for all coordinate systems in inertial frames, related to the center of mass of the main gravitating body. This allows for one in FGT to calculate the observable effects from coordinate independent
Equations~(\ref{Poincare-PN-force})  and (\ref{FGT-accel}).

\subsubsection{Lagrange Function in Post-Newtonian~Approximation}
According to expression (\ref{act-fgt}), for the action in the case of
a gravitating test particle the Lagrange function has the form
\begin{equation}\label{f-lagranga}
  L=-mc^2\sqrt{1-\frac{v^2}{c^2}} -
  \frac{mc^2}{\sqrt{1-v^2/c^2}}\left(
  \frac{\phi}{c^2} - \frac{2\vec{\Psi}}{c^2}\cdot\frac{\vec{v}}{c}
  +
  \frac{\theta_{\alpha\beta}}{c^2}\frac{v^{\alpha}v^{\beta}}{c^2}\right)\,,
\end{equation}
where
$ \phi=\psi^{00}\;,
  \vec{\Psi}=(\psi^{0\alpha})\;,
  \theta^{\alpha\beta}=\psi^{\alpha\beta}\;\,. $
Taking into account small parameters ($v/c$), we get
\begin{eqnarray}\label{L-4}
  L_{(4)} = - mc^2 + \frac{mv^2}{2} + \frac{mv^4}{8c^2}
  - m\phi - \frac{1}{2}m\phi\frac{v^2}{c^2}
  + 2m\left(\vec{\Psi}\cdot\frac{\vec{v}}{c} \right)
  -m\left(\theta_{\alpha\beta}\cdot\frac{v^{\alpha}v^{\beta}}{c^2}\right)
  \;.
\end{eqnarray}
For a system of N gravitationally interacting particles, we get
\begin{eqnarray}\label{L-4-PN-sigma}
  L_{(PN)} =\sum_a \frac{m_av_a^2}{2} +\sum_a \frac{m_av_a^4}{8c^2}
  +\sum_a\sum_b\frac{Gm_am_b}{2r_{ab}} +
  \sum_a\sum_b \frac{3Gm_am_b v_a^2}{2c^2 r_{ab}}
  \nonumber\\
  -\sum_a\sum_b\sum_c\frac{G^2m_am_bm_c}{2c^2r_{ab}r_{ac}}-
    \sum_a\sum_b\frac{Gm_am_b}{4c^2r_{ab}}
  \left[7(\vec{v}_a\vec{v}_b) +
  (\vec{v}_a\vec{n}_{ab})(\vec{v}_b\vec{n}_{ab})\right]
  \;,
\end{eqnarray}
where $\vec{n}_{ab}$ is the unite vector in direction
($\vec{r}_a -\vec{r}_b$).

The corresponding Euler--Lagrange PN equation of 
motion, for~$a$-th particle in the gravitational field of other N point masses,  are the 
Einstein--Infeld--Hoffmann equation
~\cite{strauman13}:
$$
 (\frac{d\vec{v}_a}{dt})_{(PN)} =
 -\,\,G\sum_{b\neq a} m_b \frac{\vec{r}_{ab}}{r_{ab}^3} 
\left[ 1 - \frac{4G}{c^2}\sum_{c \neq a} \frac{m_c}{r_{ac}}
+\frac{G}{c^2} \sum_{c \neq b} m_c \left(-\frac{1}{r_{bc}} +
\frac{\vec{r}_{ab} \cdot \vec{r}_{bc} }{2r_{bc}^3}
\right) \right]
$$
$$
 -\,\,\frac{G}{c^2} \sum_{b\neq a} m_b \frac{\vec{r}_{ab}}{r_{ab}^3}
\left[
v_a^2 -4\vec{v}_a\cdot \vec{v}_b +2v_b^2
-\frac{3}{2}\left(
\frac{\vec{v}_{b} \cdot \vec{r}_{ab} }{r_{ab}}
\right)^2 \right]
$$
\begin{eqnarray}\label{4-PN-sigma}
-\,\,\frac{7\,G^2}{2\,c^2}
\sum_{b\neq a} \frac{m_b}{r_{ab}}
\sum_{c\neq b}\frac{m_c \vec{r}_{bc}}{r_{bc}^3}
\,\,+\,\,
\frac{G}{c^2} \sum_{b\neq a}  m_b \left(\frac{\vec{r}_{ab}}{r_{ab}^3} \right)
\cdot \left(4\vec{v}_a - 3\vec{v}_b) \right)
\left(\vec{v}_a - \vec{v}_b) \right)\,\,.
\end{eqnarray}

Note that this Lagrange function in the frame of QFGT does not depends on coordinate system. It coincides with the corresponding
expression in GRT, just because, in the frame of GRT, the harmonic coordinates
were used to derive Equation~(\ref{L-4-PN-sigma}) .
This coincidence also explains many similarities in predictions of QFGT and GRT, including two-body~problem.

\subsection{Quantum Nature of the  Gravity~Force}
\label{sec-3-5}
The quantum electrodynamics uses virtual-particle description of the electromagnetic forces, as we discussed in Section~\ref{sec-1-2-2}. 
For the case of static forces,  QED explains the inverse-square behavior of the Coulomb law and predict the repulsive force between charges with the same signs and attractive force for charges with opposite~signs.

Feynman considered the quantum-field approach to the gravitational interaction, where the gravity force is also explained by the exchange of the massless spin-2 gravitons~\cite{feynman63,feynman71,feynman95}. In~the limit of static weak field, he got  Newtonian gravity law, which is an attraction between positive masses, and he also demonstrated that there is radiation of the gravitons carrying positive energy $\hbar \omega$.

In classical papers, van Dam \& Veltman 1970
~\cite{vDV1970} and Zakharov 1970~\cite{zakharov1970}   studied a
massive spin-2 field $\psi_{mn}$ in the flat Minkowski space, which couples to matter as the $\psi_{mn}T^{mn}$. 
They showed that, at~distances much smaller than the Compton
wavelength of the massive graviton, one recovers Newton’s law by an appropriate
choice of the spin-2 coupling constant.
However, in~the small-mass limit, the~bending angle of light by a massive
body approaches 3/4 of the Einstein result of general relativity theory (GRT). This
is the van Dam-Veltman-Zakharov (vDVZ) discontinuity paradox, which~means that one must take the exact zero mass for the spin-2 graviton in the Fierz-Pauli type
(Fierz~\& Pauli 1939~\cite{fierz39},
Veltman 1976~\cite{veltman1976}) quantum field theories having GRT as the weak field approximation. In~modern literature, there are several non-linear solutions of this massive gravity paradox (review~in~\cite{derham2014}).

In a recent paper~\cite{bar-oschepkov2019}, it was demonstrated that, in~the frame of the Feynman's Poincare-covariant quantum-field gravitation theory, the~problem of linear vDVZ mass discontinuity can be naturally solved, if~one takes into account the composite structure of the symmetric tensor field $\psi_{mn} $ and its conserved source---the matter energy-momentum tensor 
$T_{mn} $. 

\subsubsection{Propagators for Spin-2 and Spin-0 Massive~Fields}

In the frame of QFGT, three basic principles are fulfilled: Poincare symmetry of the Minkowski spacetime, Noether's conserved EMT of the field, and~Wigner's irreducible representations for quantum fields. 
According to our consideration of the QFGT Lagrangians and the main equations (Section~\ref{sec-3}), the~source $T^{mn}$ of the gravity and tensor field $\psi^{mn}$ can be presented as the sum of two irreducible parts (corresponding to spin-2 and spin-0) by
Equations~(\ref{2+0-t}) and (\ref{2+0-psi}). Here, the conservation of the EMT (Equation~(\ref{div})) and Hilbert-Lorentz gauge condition
(Equation~(\ref{psi-phi-gauge})) are~fulfilled. 

Let us now compute the gravitational exchange amplitude between two sources
$T^{\,\prime}_{mn} \leftrightarrow T_{mn}$
in both the massive and massless fields cases. For~this we rewrite Lagrangian 
and the field equations
by using the Hilbert--Lorentz
gauge conditions (Equation~(\ref{psi-phi-gauge}))
and presenting of the field $\psi^{\mu\nu}$ and source $T^{\mu\nu}$, according to (Equations~(\ref{2+0-t})) and  (\ref{2+0-psi}).
Subsequently, the total composite Lagrangian (Equation~(\ref{g-lag-2-0})) for the massive fields will be the difference between
spin-2 and spin-0  free field Lagrangians:
\be \label{lagr-HL-2+0} 
\Lambda_{(g)} = \Lambda_{(2g)} - \Lambda_{(0r)} =
 \frac{1}{16 \pi G}
\left[\,  \phi_{mn , \,s} \phi^{mn , \,s}
+ \hat{m}^2_g \phi_{mn} \phi^{mn}
\right] \,
 - \frac{1}{64\pi G} \left[ \psi_{,\,n} \psi^{\,,\,n}
+  \hat{m}^2_g \psi^2
\right] \,,
\ee
where the sign minus between Larangians means that the scalar spin-0 part corresponds to the repulsive force, which compensate the attractive force of the spin-2 field, so decreases the total energy of composite field. Note that the negative sign of a Lagrangian of a free field does not mean the negative energy of the free~field.

The interaction Lagrangian 
($\psi_{mn}T^{mn}$)  has the form
$\Lambda_{(int)} = \Lambda_{(2int)} + \Lambda_{(0int)}$:
\be \label{ugi2-0-2}
\Lambda_{(\mathrm{int})} =  -\frac{1}{c^{2}}
\, (\,\psi_{\{2\}mn} + \psi_{\{0\}mn})
(T^{mn }_{\{2\}} +    T^{mn}_{\{0\}}) =
-\frac{1}{c^{2}}\, ( \phi_{mn} T^{mn}_{\{2\}} +
\theta_{mn}T^{mn}_{\{0\}}\,).
\ee
Corresponding field equation for the traceless spin-2
field $\phi^{\mu\nu}$ is:
\be \label{feq-2-fgt-2}
 (\mathcal{D} - \hat{m}^2_g) \,\psi^{\,mn}_{\{2\}}
=  \frac{8 \pi G}{c^{2}}   T^{mn}_{\{2\}}
\qquad
\mathrm{\Rightarrow}
\qquad
(\mathcal{D} - \hat{m}^2_g) \,\phi^{\,mn}
=  \frac{8 \pi G}{c^{2}}  ( T^{mn} \, - \frac{1}{4}\eta^{mn} T )
\ee
The field equation for the scalar
part $\psi^{\,mn}_{\{0\}} = \theta^{mn} =\frac{1}{4}\eta^{\,mn}\psi$
will be:
\be \label{feq-0-fgt}
 (\mathcal{D} - \hat{m}^2_g) \,\psi^{\,mn}_{\{0\}}
= - \frac{8 \pi G}{c^{2}}   T^{mn}_{\{0\}} \,
\quad
\mathrm{\Rightarrow}
\quad
(\mathcal{D} - \hat{m}^2_g) \,\psi
= - \frac{8 \pi G}{c^{2}}  \, T
\ee
where $\mathcal{D}$ is the d'Alembertian:
$\mathcal{D} = -\partial_{\mu}\partial^{\mu}= -(\cdot)^{\,\,,\,\mu}_{\,\,\,\,\,\mu}
= \left(\triangle - \frac{1}{c^2}\frac{\partial^2}{\partial t^2} \right)$.

The sum of the field Equations~(\ref{feq-2-fgt-2}) and (\ref{feq-0-fgt})
gives the composite field Equation~(\ref{feq2-fgt})
in the form:
\be \label{feq2-fgt-2}
(\mathcal{D} - \hat{m}^2_g) \,(\psi^{mn} - \frac{1}{2}\eta^{mn}\psi)
=  \frac{8 \pi G}{c^{2}}   T^{mn} \,,
\quad
\mathrm{or}
\quad
(\mathcal{D} - \hat{m}^2_g) \,\psi^{mn}
=  \frac{8 \pi G}{c^{2}}   (T^{mn}
- \frac{1}{2}\eta^{mn}T)    \,,
\ee
In the massless limit, the field Equation~(\ref{feq2-fgt-2}) were used by Feynman for the graviton's propagator calculation
(\cite{feynman95}, Equation~3.7.9). In~particular, for the Feynman's
``bar operator'', there are two~relations:
\be \label{Tmn-2+0}
\overline{T^{\mu\nu}} = T^{\mu\nu} - \frac{1}{2}\eta^{\mu\nu}T =
(T^{\mu\nu} - \frac{1}{4}\eta^{\mu\nu}T) -
\frac{1}{4}\eta^{\mu\nu}T =
T^{\mu\nu}_{\{2\}} - T^{\mu\nu}_{\{0\}}
\ee
and
\be \label{Tmn-2+0-b} 
\,\,\,\,\,\,\,\,\,\,\,\,
\overline{\psi^{\mu\nu}} = \psi^{\mu\nu} - \frac{1}{2}\eta^{\mu\nu}\psi =
(\psi^{\mu\nu} - \frac{1}{4}\eta^{\mu\nu}\psi) -
\frac{1}{4}\eta^{\mu\nu}\psi =
\psi^{\mu\nu}_{\{2\}} - \psi^{\mu\nu}_{\{0\}} =
\phi^{\mu\nu} - \theta^{\mu\nu}
\,.
\ee
Hence, our decomposition 
(Equations~(\ref{2+0-t})) and  (\ref{2+0-psi})
onto irreducible traceless tensor and trace scalar components gives a physical explanation of the Feynman's bar operator
appearance due to the composite gravity source in
Equation~(\ref{feq2-fgt}) as the combination of the
traceless attractive (gravitons) and trace repulsive (repulsons)
forces.

In the weak field limit, the interaction energy is determined by
the single graviton exchange amplitude between two sources,
$T^{mn}_1 (x)$ and $T^{mn}_2 (y)$, which
takes the form
\be \label{Amplit-gen}
\mathcal{M} \sim \int d^4 x \int d^4 y
T^{mn}_1 (x)
G_{mnab}(x, y)
T^{ab}_2 (y)
\ee
where $G_{mnab}(x, y)$ is the corresponding propagator.
For the massless symmetric tensor field $\psi^{\mu\nu}$
(without decomposition),
the amplitude of the spin-2 particle exchange is 
\cite{feynman95,zakharov65,blago99,maggiore08}:
\be \label{amp-tot-massless} \qquad
\mathcal{M}(k) =
\lambda^2 T'^{mn}G_{mnrs}
T^{rs} =
\lambda^2 T'_{mn}
(\frac{1}{k^2}) \overline{T^{mn}}
= \lambda^2 T'_{mn}
(\frac{1}{k^2})
(T^{mn} - \frac{1}{2}\eta^{mn}T )
\ee
with graviton propagator
\be \label{prop-psi-massless}
G_{mnrs} (k) =
\frac{1}{2} [\eta_{mr}\eta_{ns} + \eta_{ms}
\eta_{nr} -
 \eta_{mn}\eta_{rs}]
\left(\frac{1} {k^2}\right)\,.
\ee
However, the reducible massless symmetric tensor field $\psi^{mn}$
cannot be smoothly extended to the massive case, which is the vDVZ~problem.

In the irreducible decomposition picture, where the potentials and the
source is presented by the irreducible traceless tensor plus scalar
components, there is natural possibility for the transition to smooth
massive case  for both partial fields.
Indeed, for~massive traceless symmetric tensor field 
$\psi^{mn}_{\{2\}} = \phi^{mn} $ (exclude extra degrees of freedom)
and scalar field $\psi^{mn}_{\{0\}} = \theta^{mn}$, there are simple
Lagrangians (Equation~(\ref{lagr-HL-2+0}))
and equations of motion
(Equations~(\ref{feq-2-fgt-2}) and (\ref{feq-0-fgt})) that have a smooth limit
to the massless~case.

Amplitudes for graviton and repulson exchange between
vertexes having conserved EMTs  ($\,T'^{rs}, T^{mn}$)
(up to terms that vanish when contracted
with a conserved tensor):
\be \label{amp-tot-mass-2+0}
\mathcal{M}(k, m_g) =
\lambda^2 T'^{rs}
G_{mnrs}^{\,(m)}(k, m_g) T^{mn}
=\lambda^2
(T'_{\{2\}\,mn} + T'_{\{0\}\,mn})
(\frac{1}{k^2-m_g^2})
(T^{mn}_{\{2\}} - T^{mn}_{\{0\}})
\ee
where the total propagator is the difference between the traceless massive graviton and massive scalar repulson
propagators:
\be \label{prop. 2+0-mass}
G_{mnrs}^{\,(m)} =
G^{\{2m\}}_{mnrs} (k,m_g) -
G^{\{0m\}}_{mnrs} (k,m_g) =
 \Biggl(\frac{1}{2}
[+\eta_{mr}\eta_{ns} +
\eta_{ms}\eta_{nr}] - \frac{1}{4}
[+\eta_{mn}\eta_{rs}] \Biggr)
\Biggl(\frac{ 1}{k^2 -  m_g^2 } \Biggr)
\ee
The amplitudes for each field contain two parts $-$ instantaneous ($\propto 1/\kappa^2 $) and retarded
($\propto 1/(\omega^2 - \kappa^2 $), which describe static and radiated fields.
In the massless limit $m_g \rightarrow 0$, we get smooth transition to massless fields; hence, there is no mass discontinuity problem in~QFGT.

We should emphasize that gravitons and repulsons have united source
$T^{\,mn}$ through which they related according to Hilbert--Lorentz
gauge conditions and describe combination of the attractive and repulsive forces. While outside the source  for free 
fields, they are independent, having helicity-2 and helicity-0 modes
with positive localizable energy~density.

\subsubsection{Composite Structure of the Quantum Newtonian Gravity~Force}

The quantum interpretation of the classical Newtonian gravity force
and testable relativistic gravity effects is based on the amplitude calculation for the corresponding EMT of gravity sources 
(Equation~(\ref{amp-tot-mass-2+0})).
In the QFGT linear weak field approximation, the Newtonian gravity
is generated by a point particle EMT
(Equation~(\ref{11})).

For the  decomposed symmetric tensor field $\psi^{ik}$
the interaction energy due to exchange by graviton and repulson
between two particles $m_1$ and $m_2$
can be calculated using propagator Equation~(\ref{prop. 2+0-mass}) and
corresponding normalization of the amplitude Equation~(\ref{amp-tot-mass-2+0}).
For static massive traceless spin-2 field, the interaction energy is:
\be \label{Newton-pot-2m}
U^{\{2\}}(r) = - \frac{i 8 \pi G}{c^4} \int \frac{d k^3}{(2\pi)^3}
T^{mn}_{(2)\,1}
G^{\{2\}m}_{mnrs} (k,m_g)
T^{rs}_{(2)\,2}
 e^{ - i k r}
= - \frac{3}{2} \,
\frac{G m_1 m_2}{ r }\,\,
e^{-\frac{m_gc}{\bar{h}}\, r}
\ee
For static massive spin-0 repulson, we get:
\be \label{Newton-pot-0m}
U^{\{0\}}(r) =
- \frac{i 8 \pi G}{c^4} \int 
\frac{d k^3}{(2\pi)^3}
T^{mn}_{(0)\,1}
 G^{\{0\}m}_{mnrs} (k,m_g)
T^{rs}_{(0)\,2}
 e^{ - i k r}
= +\frac{1}{2}\,
\frac{ G m_1 m_2}{ r }\,\,
e^{ -\frac{m_gc}{\bar{h}}\, r}
\ee
The total interaction energy (attraction plus repulsion) will be given
by the sum $U^{\{2\}}+U^{\{0\}}$, which~gives, for the massless limit
($m_g \rightarrow$ 0), the Newtonian gravitational potential and total
tensor potential in Birkhoff's form:
\be \label{Birkhoff-pot}
U_N (r) = -\,\frac{Gm_1 m_2}{r}
\quad
\mathrm{and}
\quad
\psi^{mn} = \varphi_N \,\, diag(1,1,1,1)\,,
\quad
\mathrm{where}
\quad
\varphi_N = -\, \frac{Gm}{r} \, .
\ee
Equations of motion for particles in Birkhoff's gravitational potential
and calculations of classical relativistic gravity effects were considered
in Section~\ref{sec-3-3}.

For a photon 
(energy density $T^{00}_{(em)} = \varepsilon\,\, \rightarrow \,\,
m_1 = \varepsilon/c^2$)
interacting with the Sun ($T_{(p)}^{00} = m_2c^2$), we have
exchange by only spin-2 gravitons (trace of the electromagnetic
EMT equals zero). Accordingly, the interaction energy
for the massless limit
($m_g \rightarrow$ 0)
will be:
\be \label{photon-pot}
U^{\{2\}}(r)
= - 2\,\, 
\frac{ G m_1 m_2} { r }
\quad
\mathrm{and\,\,bending\,\,angle}
\quad
 \theta_{\mathrm{QFGT}} = \theta_{\mathrm{GRT}} =
 2\theta_{\mathrm{N}} = \frac{4GM}{c^2 b}\,.
\ee
which corresponds to the observed bending angle by the Sun.

The retarded part of the amplitude Equation~(\ref{amp-tot-mass-2+0})
corresponds to gravitational wave
radiation for both 4-traceless tensor and 4-trace scalar irreducible parts
of the source EMT. The~amplitude of the transversal helicity-2 GW is the same in GRT and QFGT. 
Both tensor (helicity-2)
and scalar (helicity-0) waves have localizable positive energy density.
The dynamical longitudinal mode (helicity-0 repulson) obeys
the wave Equation~(\ref{feq-0-fgt}), which, in the massless limit, is:
\be
\label{scalar-eq-GW}
\left(\triangle - \frac{1}{c^2}\frac{\partial^2}{\partial t^2} \right)
\psi (\vec{r},t)
= -  \frac{8 \pi G}{c^{2}} T (\vec{r},t)\,,
\ee
so, it allows to perform an observational test by measuring sky localization, signal amplitudes, and polarizations of GW while using modern LIGO-Virgo facilities
\cite{derham2017,baryshev95,baryshev-p. 1,fesik-b-s-p. 7}.

Accordingly, in~the quantum interpretation, the Newtonian gravity force is a result of the
exchange by two kinds of virtual particles---spin-2 gravitons (universal attraction) and spin-0 repulsons (universal repulsion).
Experimental/observational testing of these
predictions of the QFGT is achievable with advanced LIGO-Virgo
gravitational wave~detectors.

The decomposition of the general symmetric second rank tensor field $\psi^{ik}$ onto the irreducible spin-2 and spin-0 parts (under the Poincare group)  is the central point of the consistent quantum-field gravitation theory. 
In order to understand the physical sense of the different signs of the spin-2 and spin-0 fields in the composite Lagrangian (or Hamiltonian) for the free composite field $\psi^{ik}$ 
\mbox{(Equations~(\ref{g-lag}) and (\ref{g-lag-2-0}))}
one needs to consider the composite structure of the source of the gravity field. Indeed, the~source is the energy-momentum tensor $T_{(m)}^{ik}$
of the matter and it also has the composite structure (Equation~(\ref{10compT})) and generates spin-2 and spin-0 real dynamical fields simultaneously. For~example, in electrodynamics, the source is the 4-vector (4-current) only, which generates one real dynamical spin-1~field.

In the gravity physics, we have a new situation when one source generates two real dynamical spin-2 and spin-0 fields
(actually the symmetric second rank tensor contains three irreducible parts, but~conservation laws and gauge invariance delete the additional 4-vector field). Hence, in the total action
(Equation~(\ref{act-fgt})), the interaction Lagrangian 
$\psi_{ik}T_{(m)}^{ik}$ contains the composite tensor potential 
(Equation~(\ref{psi2-psi0})),  where
both of the fields enter with the same~sign.

In the frame of the QFGT, the central basic concept is the gravity force (gradient of the potential) acting between particles via quanta of the gravity field. The~spin-2 field corresponds to the attracting forc,e while the spin-0 field corresponds to the repulsive force. This gives physical explanation why the negative sign of the spin-0 field (in the free field Lagrangian) does not mean the negative energy of this field.
To mathematically construct the gauge invariant field equations, one has to choose the structure of the total free field Lagrangian in the form Equation~(\ref{g-lag}),  which has different signs for spin-2 and spin-0~fields. 

The gauge invariance, which is consistent with the EMT conservation,  uniquely defines the field equations and equations of test particles motion, which demonstrate that the tensor spin-2 field gives attractive force, while the scalar spin-0 field gives repulsive force.
Hence, the~composite force acting between masses will be less than the sum of the absolute values of the spin-2 and spin-0 forces. 
This is why the energy (Hamiltonian) of the composite field $\psi^{ik}$
will be less than the sum of energies  of the spin-2 plus spin-0 fields considered~separately.  

In conclusion, this is why the spin-2 and spin-0 fields in QFGT is the real dynamical fields (wave~Equations~(\ref{17a}) and (\ref{17b})) with positive localizable energy density 
(Equation~(\ref{31b})).
For the case of the free field, the spin-2 and spin-0 particles are independent and the sign of corresponding parts of the free Lagrangian can be taken independently, so giving the energy (Hamiltonian) positive.

The Boulware--Deser scalar ghost field with negative energy was introduced, because they did not take the composite structure of the gravity source into account, which generates repulsive force by the trace of the real matter energy-momentum tensor. The~Fierz--Pauli scalar ghost field violates the general relativity normalization onto the Newtonian limit. However, for~QFGT, the~intrinsic scalar field is absolutely necessary for an explanation of the Newtonian limit and other relativistic gravity experiments.

\section{Relativistic Gravity Experiments/Observations in~QFGT}\label{sec4}

\subsection{Classical Relativistic Gravity~Effects }
The field Equation~(\ref{feq2-fgt}) and equations of
motion (\ref{18}), which, in the frame of QFGT, is contained in the expression of the conservation
of the total energy-momentum 
$T^{ik}_{(\Sigma)\,\,,k} = 0$
in corresponding iteration, lead to various observable consequences of the field gravity. It is important that the classical weak-field relativistic gravity effects are the same in both FG and GR  theories; hence, they cannot distinguish between GRT and QFGT.  These common predictions are~following:
\begin{itemize}
\item[$\bullet$]{\emph{universality of free fall for non-rotating bodies,}}
\item[$\bullet$]{\emph{the deflection of light by massive bodies,}}
\item[$\bullet$]{\emph{gravitational frequency-shift,}}
\item[$\bullet$]{\emph{the time delay of light signals,}}
\item[$\bullet$]{\emph{the perihelion shift of a planet,}}
\item[$\bullet$]{\emph{the Lense--Thirring effect,}}
\item[$\bullet$]{\emph{the geodetic precession of a gyroscope, and}}
\item[$\bullet$]{\emph{the quadrupole gravitational radiation.}}
\end{itemize}

\subsubsection{Universality of Free~Fall}
The rest mass $m_0$ of a structureless test particle appears in both sides of the equation of motion~(\ref{18}); hence, it is canceled off.
This demonstrate that, within QFGT, the universality of the free fall is a direct
consequence of the principles of stationary
action and universality of gravitational interaction. \mbox{Hence, the universality of free} fall is not a new ``principle of equivalence'' but is a particular case of the old stationary action principle.
The motion of a test particle in the gravity field of a massive body does not depend on the rest mass $m_0$ of the test particle and,
in fact, it checks the universality of the rest mass of a~particle.

For the case of an macroscopic extended body, when one probes the free fall of a real  body, \mbox{which includes internal structure and}
contributions from all interactions, contribution from thermal energy and pressure, and~also rotation of a body, should be analyzed separately
(this will be done~below).

\subsubsection{Light in the Gravity~Field}
Within the field gravity theory, the deflection of light and the time delay of light signals are consequences of the interaction Lagrangian 
$L_{\rm int} = \psi_{ik}T_{(\rm elm)}^{ik}$,
taken in the form that corresponds to the universality of gravitational interaction (UGI).
This gives the ``effective'' refraction index
in the PN approximation~\cite{thirring61,moshinsky50}:
\begin{equation}\label{index-refrac}
  n(r)= \sqrt{\varepsilon\mu}  = 1+\frac{2GM}{c^2r}\;.
\end{equation}
Hence, the velocity of a light signal will have the value
\begin{equation}\label{vel-light}
  c_g(r)= \frac{c}{n}=c\left(1-\frac{2GM}{c^2r}\right)\;,
\end{equation}
so the direction of light ray is changed and the time delay appears, both with the same amount as actually~observed.

For a photon moving at the impact distance $b$
from a point mass $M$, in~the weak field approximation, the
asymptotic deflection angle is:
\begin{equation}
\label{theta-gr}
 \theta_{\mathrm{FG}} = \theta_{\mathrm{GR}} =
 2\theta_{\mathrm{N}} = \frac{4GM}{c^2 b}\,.
\end{equation}
where $\theta_{\mathrm{N}} =2GM/v^2b$ is the Newtonian deflection angle. One of the most spectacular success of the GRT was
the observed value of the light bending for the Sun, which, according to Equation~(\ref{theta-gr}), equals $\theta_{\mathrm{GR}}=1.75''$.
The same value of light bending QFGT unambiguously~predicts.

Interestingly, while using a lift analogy (equivalence principle), Einstein 1911~\cite{einstein11} first derived, for the deflection angle, a value that was a half of really observed value, i.e.,~Newtonian result
$\theta_{\mathrm{N}}$. Later, it was claimed that, in GRT, an additional contribution from the curvature of space should be taken into account in order to obtain the observed value of the light~bending. 

In the frame of QFGT, the bending of light is the direct consequence of the universality of gravitational interaction principle. In~quantum interpretation, the bending of light in two times larger than the Newtonian value because the mass-less particles have two times larger interaction~energy.

From the equation of motion (\ref{FGT-accel}), one gets that a particle passing with velocity $v$ the central
mass $M$ at the impact distance $b$, will experience a small deflection angle:
\be
\label{GR-angl}
\theta_{\mathrm{FG}}  = (1+ \frac{v^2 }{c^2})\, \theta_{\mathrm{N}}\,,
\ee
hence, for the particle velocity $v=c$ one gets the same result as Equation~(\ref{theta-gr}), and there is a smooth transition from massive to massless~particles. 

One may verify, from the equation of motion
in GRT (\ref{GR-accel}), that, in order to derive these formulae, one should use isotropic or harmonic coordinates (i.e.,~$\alpha = 0$),
while in QFGT the equation of motion~(\ref{FGT-accel}) and deflection angle does not depend on~coordinates.

\subsubsection{The Time Delay of Light~Signals  }
In the frame of FGT, the~time delay phenomenon, or~the Shapiro effect,
is caused by the change of velocity of light according to Equation~(\ref{vel-light}).
If an emitter at a distance $r_1$ sends
a light signal to a mirror at a distance $r_2$
from a gravitating mass, and~ $R$ is the distance between the emitter
and the mirror, then the additional travel time is
\begin{equation}
\label{dt-gr}
  (\Delta t)_{\mathrm{FG}} = \frac{4GM}{c^3}\;ln(\frac{r_1 + r_2+R}
  {r_1+r_2-R})\,.
\end{equation}
For the case of the Sun, the value of $4GM_{\odot}/c^3$ is about $20$ $\upmu$s.

Though the time delay has the same value for both QFGT and GRT, but~the physical interpretation of this effect in the field gravity theory
has a different explanation without geometrical space-time properties.
Again, in the frame of QFGT, this effect does not depend on the~coordinates.

\subsubsection {Atom in Gravity Field and Gravitational Frequency Shift}
\label{sec-4-1-3}
The gravitational redshift of spectral lines in the frame of QFGT is a consequence of the shift of atomic levels. It is universal, because~gravitation changes the total energy and all energy levels of an
atomic system. In~the PN approximation,
$E_{obs}=E_0(1+\varphi_{_N} /c^2)$ and
hence $h\nu_{ik}^{obs}=\Delta E_{ik}^0(1+\varphi_{_N} /c^2)$.

Moshinsky~\cite{moshinsky50} was the first to consider the interaction Lagrangian (Equation~(\ref{ugi})) for the case of the interaction between
the gravity field and the spinor and electromagnetic fields of a hydrogen atom.
A spectral line with frequency $\nu_{\mathrm{em}}$ that is radiated by an atom at the distance $r$ from the
surface of a massive body with radius $R$ and mass $M$ will be observed at infinity from the body
to have a frequency $\nu_{\mathrm{obs}}$. This
gravitational redshift in the weak field approximation ($R>>R_g$) is given by
\begin{equation}
\label{dnu-gr}
z_{grav} =  \left(\frac{\nu_{\mathrm{em}} - \nu_{\mathrm{obs}}}
{\nu_{\mathrm{obs}}}\right) =
  \frac{GM}{c^2 r}\,.
\end{equation}
For the Sun, the value of $GM_{\odot}/R_{\odot}c^2$ is
$1.9 \times 10^{-10}$.
A more general formula for the gravitational redshift (obtained by~\cite{moshinsky50} method) is:
\be
\label{z-grav-fg}
1+z_{grav} = \frac{1}{\sqrt{1+\frac{2\phi}{c^2}}}
\ee
where $\phi = \psi^{00}$,
which gives the correct PN result
$z_{grav} \approx |\varphi_{_N}| /c^2 $. 
As it will be shown in the next section, the~limiting value  $\,\,\phi =-c^2/2$ can only be achieved in the center of a gravitating body having infinite radius (cosmological solution). Hence, all finite bodies have a finite gravitational~redshift.

In GRT, the observed frequency shift is due to
the clock that runs slower when it is closer to the gravitating body.
GRT general relation is $dt= d\tau /\sqrt{g_{00}}$, so the Einstein's gravitational redshift is
\begin{equation}
\label{dnu-gr-gen}
1+z_{_{GR}} = \frac{1}
{\sqrt{ 1-
  \frac{2GM}{c^2 r}}} \,.
\end{equation}
Note that, in GRT, there is an acute discussion regarding a correct interpretation of the gravitational redshift.
According to Will~\cite{will93} and Okun, Selivanov \& Telegdi~\cite{okun99,okun00}
the energy and frequency of the photon does not change during its radial motion in the gravity field, i.e.,~the photon does not lose or gain energy.
Some times, the gravitational redshift effect is considered to be a foundation of the strong equivalence principle of GRT~\cite{will14}. However it is actually a ``Schiff's hypothesis'' about strong equivalence principles, which belongs to the geometrization~principle.

\subsubsection{The Pericenter Shift and Positive Energy Density of Gravity~Field  }
It is well-known in celestial mechanics~\cite{brumberg91,kopeikin11}
that additional terms to Newtonian
equation of motion in the form of  the 
Equation~(\ref{FGT-accel}) leads to the formula for
the rate of the pericenter shift $\dot{\omega}$
of the orbit of a test particle (planet),
having semi-major axis $a$, eccentricity $e$ and period $P$,
in the form:
\be
(\dot{\omega})_{_{\rm FG}} = \frac{6\pi GM}{c^2 a(1 - e^2)P}\, .
\label{28}
\ee
This effect is derived in the frame of the QFGT as an analysis of the
small terms of the
equation of motion (\ref{FGT-accel}) using ordinary mechanics without geometrical
concepts of GRT.  For~Mercury, this gives 43``/century, while, for the binary pulsar
PSR1913+16, the effect is much larger, approximately $4^o$/year.

The formula (\ref{28}) is the same as in GRT, but~the interpretation is
different. e.g.,~the nonlinear contribution
(the second term in Equation~(\ref{26})) due to $T^{00}_{(\rm g)}$
provides 16.7\%
of  the total value~(\ref{28}).
Therefore, in the field gravity theory
the pericenter shift directly contains  the positive
energy density of the gravity field, which makes this physical
quantity experimentally~measurable.

\subsubsection{The Lense-Thirring~Effect  }
The Lense--Thirring (LT) precession is a direct consequence of the ordinary
mechanics for the system having additional terms in corresponding Lagrangian
without the geometrical concept of ``dragging of inertial~frames''.

In the frame of QFGT, the LT effect is a direct consequence of the Lagrange Function (\ref{L-4}).
An~elliptical orbit of a non-rotating test particle, moving
in the field of a central massive rotating body,
will revolve as a whole about the direction of the rotation axis of
the central body with the rate~\cite{landau71}
\vspace{6pt}
\begin{equation}
\label{omega-lt}
  \Omega_{\mathrm{LT}} = \frac{2GJ}{c^2 a^3 (1-e^2)^{3/2}}
  (\vec{j} - 3\vec{l}(\vec{l}\cdot \vec{j}))\,,
\end{equation}
where $\vec{j}=\vec{J}/J$, $\vec{l}=\vec{L}/L$, and
$\vec{L}$ aew the orbital angular momentum of the particle, and~$\vec{J}$ is the angular momentum of the central~body.

The recent Gravity Probe B experiment~\cite{Gp. 14}  confirmed that
for an Earth-orbiting satellite
LT effect is about 0.1''/year, meaning that the orbit will make a whole
rotation in about 13 million years. In~the case of pulsars in binary
systems and accreting RCO, this precession is much~larger.

\subsubsection{The Relativistic Precession of a~Gyroscope  }
The rate of precession of a gyroscope orbiting
a rotating massive body is the sum of two independent parts, one due to the
gravitational potential of the central
body, effectively non-Newtonian (the Weyl-effect),
and the second due to its rotation (the Schiff-effect).
This effect can be calculated as the contribution from the Lagrange function
of the system of gravitating point masses in the second approximation
(Equation~(\ref{L-4-PN-sigma})).
It does not
contain any reference on the geometrical concepts (\mbox{see~\cite{landau71} \S106})
Because this Lagrange function is the same for GRT and QFGT, the result is also the same:
\begin{equation}
\label{omega-vsh}
  \Omega_{\mathrm{WS}} =
  \frac{3GM}{2c^2R_o^2}\vec{n} \times \vec{V_o} + \frac{GJ}{c^2 R_o^3}
  (3\vec{n}(\vec{n}\cdot \vec{j})  - \vec{j})\,.
\end{equation}
Here, $\vec{R_o}\;$ is the radius vector of the center of inertia
of the gyroscope, $\;\;\vec{n}=\vec{R_o}/R_o$,
$\vec{V_o}$ is the orbital velocity, $\vec{J}$ and $M$ are the angular
momentum and the mass of the central body, and~$\:\vec{j}=\vec{J}/J$.

For a gyroscope orbiting the Earth over the poles,
this precession amounts to approximately 7''/year.
Recent measurement of the precession effects using the drag-free satellite
Gravity Probe B~\cite{everitt11} gave the value of the gyroscope precession, which is the same in GRT and QFGT and cannot distinguish between~them.

\subsubsection{The Quadrupole Gravitational~Radiation  }
The gravitational radiation is a natural consequence of the Quantum-Field Gravity approach, because relativistic gravitational field obeys the wave Equation~(\ref{feq2-fgt}).
In the weak field approximation by means of the usual
retarded potentials solution (\ref{psi-retard-0+2}) of the wave field equations one can infer \cite{baryshev03}
that a system of moving bodies will radiate energy in the form of tensor (spin 2) gravitational~waves.

Let us consider the generation of gravitational wave by means of a system of gravitating bodies, which have slow motion ($v<<c$). In~the wave zone ($R>>\lambda$), where the distance
$R$ to the field point much larger than the size $b$ of the system ($R >> b$), the~retarded potentials (\ref{psi-retard-0+2})
can be presented in the form of $\;(\vec{n}\cdot\vec{r}')/c\;$ series
\begin{eqnarray}\label{psi-ik-retard-ser}
  \psi^{ik} (\vec{r},t)\approx -\frac{2G}{c^2 r}
  \int \hat{T}^{ik}(\vec{r}',\,t)d^3r'-
  \frac{2G}{c^3r}\frac{\partial}{\partial t}\int
  (\vec{n}\cdot\vec{r}')\hat{T}^{ik}(\vec{r}',\,t)d^3r'
  \nonumber\\
  -\frac{G}{c^4r}\frac{\partial^2}{\partial t^2}\int
  (\vec{n}\cdot\vec{r}')^2\hat{T}^{ik}(\vec{r}',\,t)d^3r'
  +\ldots \;.
\end{eqnarray}
The standard calculations, which takes into account the traceless
character of the pure tensor free wave $\phi^{ik}$,  gives
\begin{equation}\label{phi-gw}
  \phi_{23}=\phi_{32}=-\frac{G}{3c^2r}\ddot{D}_{23}\;,\quad
  \phi_{22}-\phi_{33}=-\frac{G}{3c^2r}\left(\ddot{D}_{22}-
  \ddot{D}_{33}\right)\;\;,
\end{equation}
where
$D_{\alpha\beta}=\int\varrho(3x^{\alpha}x^{\beta}-r^2
\delta^{\alpha\beta}dV$ is the reduced tensor of
quadrupole mass moment. Accordingly, the tensor gravitational radiation
(spin 2 field) is quadrupole (the third term in (\ref{psi-ik-retard-ser})).

According to Equation~(\ref{31b}), the positive and localizable energy density
of the tensor quadrupole gravitational wave is
\vspace{6pt}
\begin{equation}\label{gw-T-00-2}
  T_{(g)\{2\}}^{00}=\frac{G}{36\pi
  c^6r^2}\left[{\stackrel{\ldots}{D}}_{23}^2+
  \frac{1}{4}\left({\stackrel{\ldots}{D}}_{22}-
  {\stackrel{\ldots}{D}}_{33}\right)^2\right]\;\;
  {\rm{\frac{ergs}{cm^3}}} \;.
\end{equation}
The total radiation in all directions gives the quadrupole luminosity
$c T_{\{2\}}^{00}$:
\begin{equation}
\label{gwlum2}
  L^{\mathrm{FG}}_{\{2\}} = \frac{G}{45c^5}\,
{\stackrel{\ldots}{D}}_{\alpha\beta}^{\phantom{\alpha\beta}2}\,\,\;
{\rm{\frac{ergs}{sec}}} \, .
\end{equation}
The tensor gravitational waves in the frame of QFGT are transversal and correspond to a particle with spin-2.
The quadrupole luminosity (\ref{gwlum2}) is identical to the corresponding formula in GRT.
A binary system  will lose orbital energy via
quadrupole gravitational radiation with luminosity
an order
\mbox{$L_{\{2\}} \approx 2\times 10^{32}(M_1/M_{\odot})^2
(M_2/M_{\odot})^2 (M_1+M_2/2M_{\odot})(a/R_{\odot})^{-5}$
ergs/s}. $M_i$ is the mass of a component, $a$~is
the semi-major axis. 
\footnote{It is important to note that to
calculate the loss of energy (\ref{gwlum2})
one should use in GRT an expression for the energy-momentum ``pseudotensor''
of the gravitational field, ill-defined  in general
relativity.
This difficulty originated a long-time discussion about the reality of
gravitational waves in GRT (Trautman 1966~\cite{trautman66}, Cervantes-Cota~et~al.,~2016~\cite{cervantes2016}, Chen~et~al.,~2016~\cite{chen2016}). }

\subsection{New QFGT  Predictions Different from~GRT}\label{sec4.2}
\subsubsection{The Quantum Nature of the Gravity~Force}

Existing variants of quantum geometry predict the violation of the equivalence principle,
possible violation of the Lorentz invariance, and~time-varying fundamental physical constants
at such a level that their detection may be realistic in the near future (Amelino-Camelia~et~al.~\cite{amelino05};
Bertolami~et~al.~\cite{bertolami06}).

However, up~to now, increasingly strong limits have been derived on variations of fundamental constants
(Uzan~\cite{uzan2010}; Pit'eva \& Pit'ev~\cite{pitev12, pitev13}).
Additionally, the first observations of sharp images of a very distant supernova
did not confirm the predicted quantum structure of space-time
at Planck scales (Ragazzoni~et~al.~\cite{ragazzoni03}).
There is also no deflection from the Newtonian gravity law at distances down to $\upmu$m scales (Nesvizhevsky \& Protasov~\cite{nesvizhevsky04}, and~even to nm scales
Klimchitskaya, \mbox{Kuusk \& Mostepanenko}~\cite{klim-kuu-most2020}).

Evidence on the similarity of the gravity force to other physical
forces was obtained in recent experiments by
Nesvizhevsky~et~al.~\cite{nesvizhevsky02,nesvizhevsky05}. Using freely falling ultra-cold neutrons, they showed that the gravity force acts similarly to the usual
electric force producing quantum
energy levels for the micro-particles moving in the gravity field
(Westphal at al.~\cite{westphal06}).

As we derived above from the equation of motion (\ref{18}) for the case of  a test particle
in the symmetric tensor potential $\psi^{ik} $, the~Newtonian limit gives the usual Newtonian force as the sum of the attractive (spin-2 part) and the repulsive (spin-0 part) force
(see Equation~(\ref{psiNew1})):
\be
\label{F_N_2+0}
\vec{F}_N=(\vec{F}_{\{attr\}}+\vec{F}_{\{repuls\}})=
\frac{3}{2}\vec{F}_N  - \frac{1}{2} \vec{F}_N \,\,.
\ee
This new understanding of the Newtonian force and
potential opens new ways for experiments on the nature of the
gravitational interaction, e.g.,~to measure
the scalar ``antigravity'', even in weak-field laboratory conditions.
A change of balance between the scalar and tensor parts of the
gravitational potential could in principle explain the (debated)
gravity-shielding  experiments with high-critical-temperature ceramic
superconductors reported by
Podkletnov \& Nieminen~\cite{podkletnov92} and Podkletnov~\cite{podkletnov97}.
Modanese~\cite{modanese99} concluded that there is no convincing physical
understanding of the experiments. Recently, an analogous
effect of a small change in the weight of a rotating
superconducting disc was detected by Tajmar~et~al.~\cite{tajmar06}.

\subsubsection{Translational Motion of Rotating Test~Body}
The  gravity force that was acting on a rotating test body
was considered by Baryshev~\cite{baryshev02,baryshev02a} in the frame of QFGT.
From Equation~(\ref{Poincare-PN-force}), in~the case of a gyroscope motion in a static spherically symmetric gravitational field, it follows the expression for the elementary Poincar\'{e}
force $dF_P$ acting on each elementary mass $dm$ of the gyroscope:
\begin{equation}
d\vec{F}_P = -\{
(1+\frac{3}{2}\frac{v^2}{c^2} +
4\frac{\varphi_N}{c^2})\vec{\nabla}\varphi_N -
3\frac{\vec{v}}{c}(\frac{\vec{v}}{c}\cdot\vec{\nabla}\varphi_N)\}dm
\label{dm-force}
\end{equation}
For a rotating rigid body, the total gravity force is the sum of
elementary forces acting on elementary masses:
\begin{equation}
\vec{F}_P =\int d\vec{F}_P
\label{total-force}
\end{equation}
Taking into account that the velocity $\vec{v}$ of an element $dm$
may be presented in the form
\begin{equation}
\vec{v} = \vec{V} + [\vec{\omega}\vec{r}]
\label{v}
\end{equation}
where $\vec{V}$ is the translational velocity of the body,
$\vec{\omega}$ is the angular velocity, and $\vec{r}$ is the radius
vector of an element $dm$ relative to its  center of inertia,
so that $\int \vec{r}dm=0$.

Inserting Equation~(\ref{v}) into Equations~(\ref{dm-force}) and~(\ref{total-force}),
we get
\begin{eqnarray}
\vec{F}_P = -M\{
(1+\frac{3}{2}\frac{V^2}{c^2} +
4\frac{\varphi_N}{c^2}+
\frac{3}{2}\frac{I\omega ^2}{Mc^2}
)\vec{\nabla}\varphi_N       
\nonumber\\
-3\frac{\vec{V}}{c}(\frac{\vec{V}}{c}\cdot\vec{\nabla}\varphi_N) -
\frac{3}{Mc^2}\int [\vec{\omega}\vec{r}]
([\vec{\omega}\vec{r}]\cdot \vec{\nabla}\varphi_N) dm \}
\label{gyroscope-force}
\end{eqnarray}
where $M=\int dm = M_0$ is the total rest mass of the body and
$I$ is its moment of inertia.
Note that the assumption of rigid rotation
 of the test body Equation~(\ref{v}) is
used by Landau \& Lifshitz (\cite{landau71}, p. 331) for a post-Newtonian
derivation of rotational  relativistic effects in GR.
The non-rigidity of a body does not play an important role in our~case.

The relativistic relation between force $\vec{F}$ acting on a body
and momentum $\vec{p}$  of the body is:
\begin{equation}
\vec{F} =\frac{d\vec{p}}{dt} = m_I \frac{d(\vec{V}/(1-V^2/c^2))}{dt}
\label{relforce}
\end{equation}
where $m_I$ is the inertial mass of the body and
$\vec{V}$ is the translational velocity of the body.
From this relation, it follows that
the three-acceleration is given by:
\begin{equation}
\frac{d\vec{V}}{dt}=\frac{\sqrt{1-V^2/c^2}}{m_I}
(\vec{F} - \frac{\vec{V}}{c}(\frac{\vec{V}}{c}\cdot \vec{F}))
\label{accel-force}
\end{equation}
The inertial mass $m_I$ of the rigidly rotating test body
may be found from the relation
\begin{equation}
m_I =\frac{1}{c^2} \int T^{00} d^3x = M_0  + E_{rot}/c^2
\label{m-I}
\end{equation}
where $T^{00}$ is the 00-component of the energy momentum tensor
of the rotating body, $M_0$ is its rest mass, and~the last equality
is obtained using Equation~(\ref{T-p-ik}) for energy momentum tensor
of particles comprising the body. For rigidly rotating ball
$E_{rot} = (1/2) I \omega^2 $.
Note that, in the general case of a self-gravitating macroscopic body,
the energy density is
$T^{00} = T_{(m)}^{00} + T_{(int)}^{00} + T_{(g)}^{00}$, which also gives the correct contribution from classical potential energy
(see Thirring~\cite{thirring61}; Baryshev~\cite{baryshev88}).

Under the gravity force Equation~(\ref{gyroscope-force}) the rotating body
will get the three-acceleration according to the general relation of
Equation~(\ref{accel-force}) where the inertial mass is given by
Equation~(\ref{m-I}). Hence, the acceleration may be written in the form:
\begin{eqnarray}
\frac{d\vec{V}}{dt} = -
(1+\frac{V^2}{c^2} +
4\frac{\varphi_N}{c^2}+
\frac{I\omega ^2}{M_0c^2}
)\vec{\nabla}\varphi_N     \nonumber\\
+4\frac{\vec{V}}{c}(\frac{\vec{V}}{c}\cdot\vec{\nabla}\varphi_N) +
\frac{3}{M_0c^2}\int [\vec{\omega}\vec{r}]
([\vec{\omega}\vec{r}]\cdot \vec{\nabla}\varphi_N) dm
\label{gyroscope-accel}
\end{eqnarray}
The Equation~(\ref{gyroscope-accel})
of motion of a small rotating body having the angular velocity $\vec{\omega}$ and the rest mass $M_0$ around of the  central mass $M$
shows that the translational orbital velocity of the body
will have additional perturbations due to its rotation.
Note that one should also add conditions of energy and
angular momentum conservation of the rotating body.
The last term in Equation~(\ref{gyroscope-accel}) depends on
the direction and value of the angular velocity $\vec{\omega}$ of the body and
it has an order of magnitude $v_{rot}^2/c^2$.
In principle, this effect may be measured
in laboratory experiments and astronomical observations,
including Lunar Laser Ranging and pulsars in binary 
systems~\cite{baryshev02,baryshev02a}.

\subsubsection{Testing the Equivalence  and Effacing~Principles}
Important conceptual problem in discussion of the
equivalence principle (EP) is how to give proper
relativistic definitions for inertial and gravitational masses without referring to the non-relativistic Newtonian
equation of motion and without non-verifiable statements (Bertolami, Paramos \& Turyshev 2006~\cite{bertolami06}, 
Unzicker 2007~\cite{unzicker07}).

\emph{In the frame of general relativity} the definition of the weak equivalence principle(WEP) based on consideration of an inertial frame of Newtonian dynamics and the test body equation of motion in  Newtonian gravity field 
$\varphi_N$ :
\begin{equation}
\label{Newton}
(\frac{d\vec{v}}{dt})_N =- \frac{m_G}{m_I}\,\, \vec{\nabla} \varphi_N
= - (1+\eta)\,\,\vec{\nabla} \varphi_N
\end{equation}
where $m_I$ is the inertial mass and $m_G$ is the gravitational (passive) mass of the body and $d\vec{v} /dt=\vec{a} $ is
the Newtonian acceleration of the body under the action of the Newtonian gravity force
$\vec{F}_N = -m_G\vec{\nabla} \varphi_N $.
The ratio
\begin{equation}
\label{wep}
\frac{m_G}{m_I} = 1+\eta
\end{equation}
is not generally restricted by Newtonian mechanics, and~the statement that $\eta = 0$ is called the weak equivalence principle or universality of free fall (UFF). 

An additional assumption of GRT is that the Newtonian equation of motion (Equation~(\ref{Newton})) can also be applied to an extended body (up to tidal force).
It is called the ``effacing principle'', which claims that, for the test body parameter $\eta = 0$, and~ the acceleration under gravity force
does not depend on the velocity and internal structure of the~body.

The Lunar Laser Ranging experiment gives the value $\eta \sim 10^{-13}$
Williams et al. 2009~\cite{williams09}.
From modern tests, which use non-rotating test bodies having different compositions, the~achieved precision
in the inferred equality of the inertial  and gravitational masses is  
$\eta = m_{\rm G} /m_{\rm I} - 1  = 1.3 \times 10^{-14}$, according to the first MACROSCOPE mission data 
(Touboul~et~al.,
~2019~\cite{touboul2019}).
 
\emph{Within the quantum-field gravity theory}, the basic concept is the universality of gravitational interaction (UGI), which is determined by
the relativistic interaction Lagrangian (Equation~(\ref{ugi})), from~which a certain form of
equations of motion is derived and can be tested by experiment/observations.

In QFGT, according to the relativistic PN
equation of motion (\ref{FGT-accel}) for a test body in the spherically symmetric static field,
the three-acceleration  is
\begin{eqnarray}
\frac{d\vec{V}}{dt} = -(\frac{m_0}{m_0})
\left\{(1+\frac{V^2}{c^2} +
4\frac{\varphi_N}{c^2}+
)\vec{\nabla}\varphi_N
+4\frac{\vec{V}}{c}(\frac{\vec{V}}{c}\cdot\vec{\nabla}\varphi_N)
\right\}
\label{fgt-V-accel}
\end{eqnarray}
where $\vec{V}$ is the velocity of the test body, $\varphi_N$ is the
Newtonian gravitational potential.
In the right side of Equation~(\ref{fgt-V-accel}), the rest mass of the body
$m_0$ is canceled due to the same energy-momentum tensor in both
(interaction and free particle)  Lagrangian.
Hence, the gravitational acceleration of the test~body
\begin{itemize}
\item[$\bullet$]{\emph{does not depend on the rest mass $m_0$ of the test body, and}}
\item[$\bullet$]{\emph{does depend on its velocity $\vec{V}$ (both on value and direction) and on the value of the gravitational potential $\varphi_N$ at the location of the body.}}
\end{itemize}
Hence, in the Newtonian limit ($v^2/c^2 = \phi/c^2 = 0$), we get that the inertial and gravitational masses equal to rest mass of the particle
$m_{\rm I} = m_{\rm G} = m_{\rm 0}$. However, already in PN approximation, the relation between the force and acceleration is more complex.
This means that there are different ways in relativistic regime to define the inertial $m_{\rm I}$ and the gravitational $ m_{\rm G}$ masses, which open new possibilities for  neperformingw kinds of experiments/observations for
testing the principle of UGI (Equation~(\ref{ugi})) in~QFGT.

For example, in~the case of rotating test body,
according to GRT equivalence principle, the free fall acceleration of a body does not
depend on its internal structure 
(effacing principle).
Hence, differently rotating bodies will have the same
gravitational acceleration~\cite{will93}
(if one neglects the tidal effect).

However in QFGT, according to the equation of translational motion of rotating body (\ref{gyroscope-accel})
there is an orientation-dependent contribution in the free fall acceleration.
Hence, rotating the body can give new tests  of possible
violation of EP due to orientation and magnitude of the linear
velocities in rotating test body.
For example, rotating spherical body has
$E_{rot} = (1/2)I\omega^2$ and deflection from EP will be
at the level of $E_{rot}/M_0c^2$ which
for radius $R_0$ and angular velocity $\omega$ is
\begin{equation}
\frac{E_{rot}}{Mc^2} \approx 2\times 10^{-12}
(\frac{R_0}{10cm})^2 (\frac{\omega}{10^4 rad/sec})^2
\label{E-rot}
\end{equation}

The most straightforward application of Equation~(\ref{gyroscope-accel})
is to perform a ``Galileo-2000'' experiment (which is an improved 21st century version of the famous Stevinus-Grotius-Galileo experiment
with freely falling bodies in the Earth's gravity field) just taking into account the rotation of the bodies.
Instead~of the Newtonian equation of motion Equation~(\ref{Newton}),
in frame of QFGT we have Equation~(\ref{gyroscope-accel}) and the motion of a rotating body differs from that of non-rotating one~\cite{baryshev02}.

Indeed, let us consider three balls on the top of a tower (like the 110-m Drop Tower of the Bremen University).
The first ball is non-rotating and, according to
Equation~(\ref{gyroscope-accel}), its free fall acceleration is:
\begin{eqnarray}
\vec{g}_1=
(\frac{d\vec{V}}{dt})_1 = -
(1-3\frac{V^2}{c^2} +
4\frac{\varphi_N}{c^2}
)\vec{\nabla}\varphi_N
\label{g1}
\end{eqnarray}

Let the rotation axis of the second ball be parallel to
the gravity force, i.e.,~$\vec{\omega}\|\vec{\nabla}\varphi_N$;
hence, its free fall acceleration is:
\begin{eqnarray}
\vec{g}_2=\vec{g}_1 \times
(1+\frac{2}{5}\frac{R_0^2\omega^2}{c^2} )
\label{g2}
\end{eqnarray}
where one takes into account that, for a homogeneous ball
with radius $R_0$ and mass $M_0$,
the moment of inertia is $I=\frac{2}{5}M_0R_0^2$.

Let the rotation axis of the third ball be orthogonal to the gravity
force, i.e.,~$\vec{\omega}\perp\vec{\nabla}\varphi_N$;
hence, its free fall acceleration is:
\begin{eqnarray}
\vec{g}_3=\vec{g}_1 \times
(1-\frac{1}{5}\frac{R_0^2\omega^2}{c^2})
\label{g3}
\end{eqnarray}

Equations~(\ref{g1})--(\ref{g3}) imply that the
considered three balls will reach the ground at different
moments. True, the~difference is very small, for~example, if the
radius of the ball is $R=10$ cm and its angular velocity
$\omega =10^3$ rad/sec, then the expected difference in
the falling time from 110 m tower
will be $\Delta t \approx (1/2)(\Delta g/g)t
\approx 2.5\times 10^{-13}$ s.

For NASA's Gravity Probe-B experiment~\cite{everitt11}
with the drag-free gyroscope orbiting the Earth,
the expected in QFGT perturbation of the acceleration
of the translational orbital motion of the
gyroscope (having radius 2 cm and rotational speed about 80 Hz) is about $\delta g/g \propto 10^{-15}$, which is yet too small for
detecting the signal in this~experiment.

Another type of laboratory experiment, for~direct testing the velocity
dependence of the Poincar\'{e} gravity force
Equation~(\ref{gyroscope-force}), is to weigh the rotating
bodies. If~two bodies are at the balance and at a moment they start to rotate with different orientations of the rotation axes
then the balance will be violated and, hence, measured by a scale. The~expected difference in forces is again regarding the value given by Equation~(\ref{E-rot}).

In these laboratory experiments, there are no problems
with choosing a coordinate system at all. The~height of a tower and the moments of the contact of the rotating bodies with the ground, and~also the readings of a balance scales are directly measurable quantities. Hence, the equation of motion
Equation~(\ref{FGT-accel}) in the field gravity theory gives a uniquely defined value for these laboratory experiments. Note that, at the microscopic level, the spin orientation dependence
of the gravity force also should be tested.
Materials that are composed with regularly oriented
spins of particles or regularly directed internal motion of particles
can have different free fall accelerations,
which, in principle, may be tested by~experiments.

\subsubsection{Scalar and Tensor Gravitational~Radiation}
Gravitational field Equations~(\ref{17a}) and (\ref{17b}) describe the radiation of two types---pure tensor (\mbox{traceless, spin-2}) ``gravitons`` and scalar (trace of the tensor potential, spin-0) ``repulsons'' (or~``levitons'') 
\footnote{The name ``leviton'' was suggested by V.V. Sokolov for spin-0 scalar
gravitons which corresponds to the repulsive force}.

In the frame of QFGT, the~scalar wave generation can be calculated from the
Equation~(\ref{psi-retard-0}) for retarded potentials, which gives in the case of the wave zone approximation the following expression
~\cite{baryshev82,baryshev90,baryshev03}:
\begin{equation}\label{psi-0-retard-ser}
  \psi (\vec{ r}, \, t) \approx
                         \frac{2GM_{0}}{r}
                        - \frac{2GE_{k}}{c^2r} +
                           \frac{2GM_{0}}{cr}
      \left( \vec{n} \cdot \dot{ \vec{R}} \right)
                            + \frac{G}{c^2r}
  n_{\alpha}n_{\beta} \ddot{I}_{\alpha \beta}
   +\ldots \; ,
\end{equation}
where $ M_{0} =\sum m_{a} \;$, $\; E_{k} = 1/2 \sum m_{a}v_{a}^{2}
\;$, $\;\vec{R} = \sum m_{a} \vec{r}_{a} / \sum m_{a} \;$ , $\;
I_{\alpha\beta} = \sum m_{a} x_{a}^{\alpha} x_{a}^{\beta}\; $.
Taking the derivative of~(\ref{psi-0-retard-ser}) over time (at fixed point $r$)
and excluding non-contributing terms, we get following equation for the time
derivative of the scalar potential:
\begin{equation}\label{psi-dot-Ekin}
   \dot{\psi} ( \vec{r},\, t) =
    - \frac{2G\dot{E}_{k}}{c^2r}\;.
\end{equation}
It means that the scalar gravitational radiation is the second order monopole radiation, and~there is no first order monopole radiation (due to mass conservation). Additionally, dipole and quadrupole scalar radiation are absent.
Using the expression Equation~(\ref{31b}) for the energy density in the scalar wave, we~obtain
\begin{equation}\label{T-00-psi-dot-Ekin}
   T_{(g)\{0\}}^{00} =
    \frac{G\dot{E}_{k}^2}{8\pi c^6r^2}
    \;\;\; {\rm{\frac{ergs}{cm^3}}}\;.
\end{equation}
The energy flux is $cT_{\{0\}}^{00}$, so
the additional loss of energy (in $4\pi$ steradian) due to the scalar monopole radiation~\cite{baryshev95} is
\begin{equation}\label{Lum-0-fgt}
  L_{\{0\}} = \frac{G}{2c^5} \dot{E}_{k}^{2}
  \;\;\;{\rm{\frac{ergs}{sec}}} \;.
\end{equation}
so the scalar gravitational (actually ``anti-gravitational'') radiation
has the same order $1/c^5\,$ as the tensor quadrupole~radiation.

Binary pulsar systems offer a test of the validity of the gravitational radiation formulae. For~a binary system the loss of energy due to the pure tensor gravitational radiation is given by the quadrupole luminosity Equation~(\ref{gwlum2})  (which is the same in QFGT and GRT)
$ L_{(2)\mathrm{FG}} = (G/45c^5)\,
{\stackrel{\ldots}{D}}_{\alpha\beta}^{\,\,2}
\,\,$ (ergs/sec),
where $D_{\alpha\beta}$ is the quadrupole moment of the system.
The quadrupole tensor (spin-2) gravitational wave in the frame of QFGT is transversal and it has localizable positive~energy.

For a binary star system, the quadrupole luminosity is
\be\label{lum-2-double}
 <\dot{E}>_{\{2\}} = \frac{
                     32 G^{4} m_{1}^{2} m_{2}^{2}
    \left(m_1 + m_2 \right)
    \left(1 + \frac{73}{24}e^2 +
\frac{37}{96}e^4
                     \right)}
                    {5c^5 a^5
    \left(1 -e^2
                     \right)^{7/2}}
    \;\;{\rm{\frac{ergs}{sec}}}   \,,
\ee
here $m_1,m_2$ are masses of the two stars, $a$ is the semimajor axis and
$e$ is the eccentricity of the relative orbit.
For the binary system, the orbital additional energy loss via scalar waves (Equation~(\ref{Lum-0-fgt})) will be:
\be\label{lum-0-double}
  <\dot{E}>_{\{0\}} =   \frac{G^4 m_1^2 m_2^2
    \left(m_1 + m_2 \right) \left(e^2 +
        \frac{1}{4}e^4 \right) }
    {4c^5 a^5 \left(1 -e^2 \right)^{7/2}}
    \;\;\;{\rm{\frac{ergs}{sec}}}   \,.
\ee
Hence,the ratio of the scalar to tensor luminosity is given by relation:
\be \label{ratio-0-2-db}
  \frac{<\dot{E}>_{\{0\}}}
         {<\dot{E}>_{\{2\}}} =
         \frac{5}{128} \cdot
         \frac{ \left(e^2 +
                           \frac{1}{4}e^4
           \right) }
                  { \left(1 + \frac{73}{24}e^2 +
                             \frac{37}{96}e^4
           \right) }.
\ee
The value of this ratio lies in the interval 
$\,\,0\div 1.1$\%, depending on the value of the eccentricity $e$,
e.g.,~for the circular orbit it equals zero. However for a pulsating spherically symmetric body there is no quadrupole radiation, so the scalar radiation becomes~dominating.

\subsubsection{The Binary NS System with Pulsar PSR1913+16}
According to Weisberg~et~al.,~2003~\cite{weisberg03},
2010~\cite{weisberg10},  2016~\cite{weisberg16}
the observed rate of change of the PSR1913+16 binary system orbital period was measured more and more precisely.
The main result of the observations is the time derivative of the orbital period~\cite{weisberg16}:
\be \label{obs-P}
\dot{P}_b^{obs} = (-2.423 \pm 0.001)\,\cdot\, 10^{-12}
\ee
The GRT and QFGT predict for the corresponding change of binary period, due to positive energy loss in quadrupole gravitational waves the very precise value~\cite{weisberg16}:
\be \label{grt-P}
\dot{P}_b^{quad} = (-2.40263 \pm 0.00005)\,\cdot\, 10^{-12}
\ee
A comparison of the observed value Equation~(\ref{obs-P}) with 
theoretically predicted value Equation~(\ref{grt-P}) shows hat the observed excess  of energy loss (relative to
quadrupole  radiation) is
\be \label{delta-P-obs}
\Delta_{obs}^{\{2\}} = (+ 0.848 \pm 0.041)\%
\ee
Accordingly, the observational fact is that
(at the level of 1\% )
the energy radiated by the PSR1913+16 binary system \textbf{larger}
than the predicted value of
the energy radiated  by pure tensor gravitational~waves.

The orbit of the binary pulsar PSR1913+16 has an eccentricity
$e = 0.6171334(5)$; hence, the expected additional energy loss due to scalar gravitational radiation (Equation~(\ref{ratio-0-2-db})) is~\cite{baryshev95}:
\be \label{fgt-0-rad-PSR}
\Delta_{scalar}^{\{0\}} = + 0.735 \%
\ee
Hence, after~taking into account the additional scalar waves luminosity,  the~remain observed excess relative to the sum of spin-2 plus spin-0 gravitational radiation will be only
\be \label{fgt-2+0-rad-PSR}
\Delta_{obs}^{\{2\}+\{0\}} = (+ 0.113 \pm 0.041)\% .
\ee

It has been shown by Damour \& Taylor 1991~\cite{damour91} that
the observed rate of the orbital period change $\dot{P}_b^{obs}$
includes the kinematic ``Galactic effect'' of the relative acceleration of the pulsar and the Sun in the Galaxy. In~the model of the planar circular motion of the Sun and the pulsar, the~Galactic contribution is given by the relation~\cite{damour91}:
\be
\label{gal-eff}
 \left( \frac{\dot{P}_b}{P_b} \right)^{Gal} =
 -\frac{V_0^2}{cR_0}\cos l -
 \frac{V_1^2}{cR_1} \left[ \frac{R_0}{R_1}
 \left( \frac{d}{R_0} - \cos l \right) \right] +
 \mu^2 \frac{d}{c}
\ee
where
$V_0$, $V_1$ are the circular velocities at the Sun's $R_0$
and the pulsar's  $R_1$ positions in the Galaxy,
$l=49.\,97^{\,\circ}$ is the pulsar's galactic longitude,
$\mu$ is the proper motion of the pulsar, $d$ is the distance
to the pulsar, and $c$ is the velocity of~light.

However, there is large uncertainty in the Galactic effect due to the adopted
model of the Sun-pulsar relative motion, adopted distance to the
pulsar, and errors in the proper motion  of the pulsar.
The~distance $d$ to the pulsar PSR1913+16 and proper motion $\mu$ are critical parameters in the calculation of the Galactic effect.
Unfortunately, the~line of sight to the
pulsar passes through a complex region of our Galaxy, and~one must be very careful when deriving the value of the distance to the pulsar from the
dispersion~measure.

Intriguingly, in the PSR1913+16 analysis
the adopted values for these crucial parameters were essentially changed
from paper-to-paper. For~example, according to
Weisberg~et~al.,~2003~\cite{weisberg03} the~distance to the pulsar is
$d = 5.9 \pm 0.94$ kpc and the proper motion is $\mu = 2.6 \pm 0.3$ mas/yr.
%
While Weisberg~et~al.,~2016~\cite{weisberg16} adopted
$d = 9.0 \pm 3$ kpc and the proper motion $\mu = 1.48 \pm 0.04$~mas/yr.
It is possible to choose such Galaxy model parameters, which allow for
compensating the observed energy loss excess (\ref{delta-P-obs})
and claim the perfect coincidence between observations and pure
quadrupole~radiation.

Accordingly, the distance to the pulsar PSR1913+16 requires further
careful determination using different methods. 
\footnote{If the distance to the pulsar PSR 1913+16
has the critical value $d = d_{crit} =5.4$ kpc,
then the second term in Equation~(\ref{gal-eff}) equals zero.
For distances $d < d_{crit}$ this term even changes its sign.}
A direct determination of the pulsar distance,
together with the more accurate proper motion,
may be regarded as a test of fundamental physics, related to the nature of the gravitation. Additionally, distances to other binary pulsars will be crucial for gravity physics. According to~\cite{weisberg16}, now the rate of orbital period change
has been measured for other eight binary pulsars with accuracy about 5\%  so in the near future the scalar GW contribution will be tested
more~reliably.

\subsubsection{Detection of GW Signals by Advanced LIGO-Virgo~Antennas}
The recent detection of GW signals by Advanced LIGO-Virgo antennas~\cite{abbott16a,abbott16b,abbott2020} was a
breakthrough discovery, which had opened new possibility for study the fundamental physics of the gravitational interaction
\footnote{The list of LIGO-Virgo publications see
~\cite{abbott16c} }.

The LIGO Scientific Collaboration and the Virgo Collaboration team presents the interpretation of the detected GW events
as the binary black hole merger signals~\cite{abbott16d}.
They used models of the waveform covering the inspiral, merger and ringdown phases based on combining post-Newtonian theory, the~effective-one-body formalism, and numerical relativity simulations.
Consequently, they found that the observed GW signals corresponds to coalescence of the RCO binaries with masses from 7 to 100  $M_{\odot}$,
at distances about 1000~Mpc.

Up to now from observations of ~50 events, only one
reliable optical (and other electromagnetic bands)
identification was done for GW170817 event, which coincides with the short gamma-ray burst GRB170817A in the galaxy NGC4993~\cite{abbott2020}. Interesting suggestion for the explanation of small number of optical identifications was done by 
Broadhurst, Diego, and Smoot 2020~\cite{broadhurst2020}. They demonstrated that 
gravitational lensing plays important role in the detections of GW events. It allows for the detection of binary black hole candidates (BBHC) at cosmological distances with chirp masses that appear to be enhanced by 1 + z in the range 1 < z < 4, in~good agreement with the BBHC masses observed in our~Galaxy.

An important application of the GW observations is the testing of alternative gravitation theories~\cite{fesik-b-s-p. 7}.
As was discussed in Introduction,
in the frame of the GRT, there is an important conceptual obstacle that forbids the localization of the GW energy due to the pseudo-tensor character of the energy-momentum of the gravitational field. According to
(Trautman 1966~\cite{trautman66}),
\mbox{Landau \& Lifshitz} 1971~\cite{landau71}, and
Misner, Thorne, and Wheeler 1973~\cite{misner73} 
the energy of the gravitational field cannot be localized inside the gw-wavelength because of the equivalence principle.  However, the LIGO gw-detectors have localized the gw-energy by measuring the oscillating wave-form of the gw-signal well inside the~gw-wavelength.

In the QFGT the energy-momentum of the gravitational field is the true tensor, so the GW energy is localizable and carries positive energy, which is given by Equation~(\ref{31b}). Hence, the detection of the GW signals can be considered as a new confirmation of the Feynman’s field gravitation theory, which is based on the fundamental concept of localizable positive energy of the gravitational~field.

The radiation of the quadrupole GW is given by the same formulas in both QFGT and  GRT, so the observation of the oscillating form of the LIGO-Virgo signals has the same interpretation  as the coalescence of the binary Relativistic Compact Objects (or Black Hole Candidates). However,~in QFGT additional scalar radiation
exists and this can be tested in the near future by means of modern GW detectors. The~scalar gw-radiation arises from the spherical pulsations
of the collapsing bodies (in massive core collapse SN) and can also be detected and identified by related optical SN \mbox{explosion~\cite{baryshev90,baryshev03,fesik-b-s-p. 7}}.

According to Equation~(\ref{31b}), the flux of the gw-energy in the flat monochromatic scalar GW is given by
\be
\label{gw-0-flux}
S_{\{0\}} =
c T_{(g)\{0\}}^{00}=
\frac{c^3}{32\pi G}(\dot{h})^2 \,\,\,\,\,
\rm{\frac{erg}{sec\,cm^2}}\,\,,
\ee
where $h=A/c^2$ is the dimensionless gravitational potential $A$ of the wave.
Let us consider the ``standard''  gw-puls
that was introduced by Amaldi \& Pizzella 1979~\cite{amaldi79},
which is a sinusoidal wave $A(t,x)=A_0 \cos (\omega t - kx)$
with amplitude $A_0$,
frequency $\omega_0 = 2\pi \nu_0$ and duration $\tau_g$.
For the scalar GW, the~amplitude $h_0 = A_0/c^2$
of the signal on the Earth due to
the gw-puls that occurs at a distance $r$, with~total energy
$E_{gw}$ is
\begin{equation}
 h_0 =1.95 \times 10^{-21} \left( \frac{100 Mpc}{r}\right)
 \left(\frac{10^2 Hz}{\nu_0}\right)
 \left(\frac{E_{gw}}{1 M_{\odot}c^2}\right)^{1/2}
 \left(\frac{0.5s}{\tau_g}\right)^{1/2}\,.
\label{ho}
\end{equation}
The expected rate of gravitational wave signals from core collapse SN
explosions and binary RCO coalescence within inhomogeneous Local Universe
($r \leq 100$ Mpc) was considered by Bayshev~\&~Paturel 2001~\cite{baryshev-p. 1}. Sensitivity $h \simeq 10^{-22}$ is enough
for the detection of such gw-events from the Virgo galaxy cluster and the Great~Attractor.

Interaction of gravitational wave with gw-antenna has different physics
in QFGT and \mbox{GRT~\cite{baryshev95,baryshev-p. 1}}.
If we substitute the expression
for the scalar plane monochromatic gravitational wave,
which~propagates along the $x$-direction
($k = 2\pi / \lambda$, $\,\omega =2\pi \nu $),
\be
\label{scalar-wave-cos}
\psi^{ik}_{\{0\}} =
A(t, x) \eta^{ik} = A_0 \cos(\omega t - k x) \eta^{ik}
\ee
into Equation~(\ref{eq-motion-scalar}) and leave the main terms, we get the
following equations of motion of test particle in scalar wave
\begin{equation}
 \frac{dv_y}{dt}=\frac{dv_z}{dt}=0\,,
\label{vyz}
\end{equation}
\begin{equation}
 \frac{dv_x}{dt}=\frac{1}{4c}\frac{\partial \psi}{\partial t}\,.
\label{vx}
\end{equation}
and the equation for the work that is produced by the scalar wave ($i=0$ in Equation~(\ref{eq-motion-scalar})):
\vspace{6pt}
\begin{equation}
 \frac{dE_{kin}}{dt}=\frac{m_0}{4c}v_x\frac{\partial \psi}{\partial t}=
-\frac{m_0 c^2 }{32 c^4}\,h_0^2 \,\omega\, \sin(2\omega t + \alpha)
 \,,
\label{E-x-kin}
\end{equation}
so the kinetic energy of the test particle in the scalar wave is
\begin{equation}
 E_{kin}(t) = \frac{m_0c^2}{64}\,h_0^2 \,
 \left(1 + \cos(2\omega t + \alpha) \right) \,.
\label{E-kin-t}
\end{equation}
Hence, according to Equations~(\ref{vyz})--(\ref{E-x-kin}), the scalar wave accelerates the test particle and the gravitational force produces the work
that changes the kinetic energy of the~particle.

Additionally, the scalar wave is longitudinal in the sense, that  the test particle oscillates along the direction of the wave propagation (around an initial position) with the velocity amplitude $\Delta v$ and the distance amplitude $\Delta x$
\be
\label{delta-v-x}
\Delta v = \frac{cA}{c^2} \,\,\,\,\,   {\rm{and}}
\,\,\,\,\,  \Delta x = \frac{A}{kc^2} =
\frac{\lambda A}{2\pi c^2}.
\ee
For two test particles at a distance $l_0 \ll \lambda$ along x-axis,
we get the dimensionless amplitude of oscillation in the form
$\Delta l_0/l_0 = A/c^2 = h$.

It is important to note that scalar gravitational wave does not interact with the electromagnetic field, because the interaction Lagrangian equals zero
\be
\label{L-int-0-em}
\Lambda_{(int)} =
-\frac{1}{c^2}\psi^{ik}_{\{0\}}
 T^{(em)}_{ik} =
 -\frac{1}{4c^2}\psi \,
 T_{(em)} =  0 \,.
\ee
It means that the detection of the scalar wave can be achieved by means of laser interferometric antenna, because~the GW only affects the test masses and has no action on the photon beam.
Additionally, it is very important to take into account the position of the GW source on the sky for testing the longitudinal and transversal character of the GW
~\cite{fesik-b-s-p. 7}.
This is why the optical-x-ray-gamma identification of GW at the sky plays the crucial role for the physical interpretation of the GW radiation.
As it was demonstrated many times in the history of astronomy (e.g.,~radio sources, x-ray sources, gamma bursts), it is impossible to obtain the correct model of radiation process without the identification of source at least in several different wave-bands (multimessenger observations)---too many concurrent models are possible. In~the near future, the next runs of the several GW antennas will start to operate and the localization of the GW sources by means of optical identification of this violent astrophysical event will be possible with accuracy about 1 arcsec. Only after such multimessenger identifications can one can speak about correct understanding the GW events that are detected by~gw-antennas.

The observed transversal GW events due to black hole candidates coalescence can only describe part of physical processes of GW radiation. For~example, two orbiting relativistic compact objects (RCO) (possible in QFGT) can emit scalar gravitational waves during the late inspiral, merger and pulsations of the resulting RCO. The~oscillating decay form of the GW signal is possible during the late  merging state of coalescent binary RCO and for the core  pulsations during supernova collapse.
Even the debated detection of GW signal from SN1987A by resonance GEOGRAV bar-antenna can be explained by the scalar gravitational waves from spherical pulsations of collapsing SN core~\cite{baryshev97}.

\subsubsection{The Riddle of Core Collapse Supernova~Explosion}
The problem of supernova explosion is one of the most intriguing in modern relativistic astrophysics
\cite{paczinski01,imshennik10,burrows13,burrows2020}.
The expected amplitudes and forms of gravitational wave (GW) signals 
from supernovae explosions detected on the Earth by gravitational antennas essentially depend on the adopted scenario of core-collapsed explosion of massive stars and relativistic gravity theory. This is why the expected detections of GW signals from SN will give, for the first time, experimental limits on possible theoretical models of gravitational collapse, including the strong field regime and even the quantum nature of the gravity force.
For the estimates of the energy, frequency, and duration of supernova GW emission, one needs a realistic theory of SN explosion which can explain the observed ejection of massive envelope. Unfortunately, for~the  case of the massive core-collapse supernovae explosion, such a reliable theory does not exist now, though~first 
three-dimensional (3D) supercomputer's calculations has recently been performed~\cite{burrows2020}.
However, as was sadly noted by Paczynski 1999
~\cite{paczinski01}, if there were no observations of SNII, then it would be impossible to predict them from the first~principles.

Modern theories of the core collapse supernova are able to explain all stages of evolution of a massive star before and after the explosion. However, the~theory of the explosion itself, which~includes the relativistic stage of collapse, where a relativistic gravity theory should be applied for the calculation of gravitational radiation, is still controversial and unable to explain the mechanism by which the accretion shock is revitalized into a supernova explosion
(see discussion by
Burrows 2013~\cite{burrows13}, Imshennik 2010~\cite{imshennik10}, and recent~\cite{burrows2020}).

Burrows~\cite{burrows13} in his review ``Perspectives on Core-Collapse Supernova Theory'' emphasized that one of the most important, yet frustrating, astronomical question is “What is the mechanism of core-collapse supernova explosions?” Fifty-years history of CCSN theory, which uses advanced hydrodynamics and shock physics, convection theory, radiative transfer, nuclear physics, neutrino physics, particle physics, statistical physics, thermodynamics, and
gravitational physics have not definitively answered that question.
Intriguingly, up to now,
there is no theoretical understanding how to extract such energy from the relativistic collapse of
the iron core and produce observed kinetic energy of the expanding stellar envelope~\cite{burrows13,dolence15,imshennik10}.
Recent 3D FORNAX simulations reveal large variety of model behaviour, even for a fixed initial mass of the collapsing star, so the conclusion was that the explosion strongly depended on the combination of many complex theoretical suggestions~\cite{burrows2020}.

According to the review~\cite{burrows13} for all trustworthy models of the core-collapse SN (CCSN), the energy of the explosion is never higher than a few tenths of Bethe (1 Bethe $= 10^{51}$ ergs), which is not enough for overcoming the gravitational binding energy of the “canonical” neutron star mass $\sim$$1.5 M_{sun}$.
For many~years~theorists have been presented with a stalled accretion shock at a radius near $\sim$$100$--200 km and have been trying to revive it (a review of the literature see~\cite{burrows13,imshennik10}).
This bounce shock should be the CCSN explosion. However, both simple theory and detailed numerical simulations universally indicate that, due to neutrino burst and photodissociation of the in-falling nuclei, debilitate the shock wave into accretion within $\sim$5 milliseconds of bounce. What is more, if~the shock is not revived and continues to accrete, all cores will collapse to black holes, which contradict observations of NS in SN remnants. Modern neutrino-driven mechanism in 3D simulations of the proto-neutron star can produce, in some cases, the needed explosion. However,~there is still no universal basic principle for CCSN explosion, which can produce observed massive neutron stars in the frame of GRT, because~of infinite gravity near black hole~horizon.

Rapid rotation with magnetic fields (e.g.,~\cite{bisnovatyi14}) and 3D MGD simulations taking different instabilities into account need to be studied more carefully in future.
The true model should also explain such observational properties of the CCSN as two stage collapse and simultaneous burst of gravitational waves and neutrino burst~\cite{imshennik10} (as it was in the case of SN1987A~\cite{galeotti16}).
However, up to now, athough many different revival mechanisms were considered, there has been no successful model yet, because~the problem of CCSN explosion exists at a very fundamental level of the gravity theory.\mbox{ Note that in the frame of} QFGT there is a new possibility for CCSN explosion because of finite gravity~force.

A possibility to revive the bounce shock essentially depends on the gravity force acting within the pre-neutron star (PNS), where at least post-Newtonian relativistic gravity effects should be taken into account~\cite{wallace16}.

According to QFGT, the general physical concepts of force, energy-momentum, and energy-quanta are working as in other theories of fundamental physical interactions. Accordingly, the finite gravity force and positive energy density of the gravitational field exist inside and outside a collapsing massive body. An~important new element of the QFGT is the principal role of the scalar part of the symmetric tensor field (generated by the matter EMT trace). It presents the repulsive force, which was missed in Feynman's lectures on gravitation
~\cite{feynman71,feynman95}.

Within the field approach to gravitation besides the tensor (spin-2) waves, there is the scalar (spin-0) ones. Though~in the field gravity theory, there is no detailed calculations of the relativistic stages of the massive core collapse, but, in~principle, the~repulsive scalar part of gravitational potential could lead to revive the bounce shock. The~released energy of the scalar GW during spherically symmetric pulsations/collapse, may reach values up to one solar rest mass, with~characteristic frequency \mbox{100--1000 Hz} and durations up to several seconds
(Baryshev~\cite{baryshev97};
Baryshev \& Paturel~\cite{baryshev-p. 1}).

The CCSN explosion within QFGT has essentially different scenario than in GRT. Post-Newtonian equations of relativistic hydrodynamics in the frame of QFGT were derived in Baryshev~\cite{baryshev88}, according to which the gravity force essentially depends on the value and direction of the gas flow~velocity.

This gives the possibility for pulsation of the inner core of the pre-NS star and the formation of main explosion shock wave together with jet-like outflow along the rotation axis of collapsing~star.

The powerful relativistic jets are common phenomenon in the physics of relativistic compact objects (SN explosions, galactic BHC, gamma-ray bursts, active galactic nuclei, quasars). The~AGN relativistic jets transfer matter from the central energy machine, having size about 1 a.e.($= 1.5\times 10^{13}$~cm), to~the very large distances up to 100 kpc($= 3\times 10^{23}$ cm)
with kinetic power about $10^{45}$ ergs/sec 
\mbox{(see the example of M87 jet in~\cite{goddi2019}).}
Note that GRMGD simulations of the M87*  EHT observations give maximal possible jet power less than $10^{43}$erg/sec~\cite{akiyama2019b}.

In the frame of the QFGT, there is the possibility for additional ``gravitational'' origin and the collimation of  jets from rotating relativistic compact objects, due to the strong dependence of the gravity force acting on particles moving in transversal and radial directions
~\cite{baryshev86, baryshev90, baryshev02}.
According to PN equations for Poincare force and Poincare acceleration 
Equations~(\ref{Poincare-PN-force}) and (\ref{FGT-accel}),
for a test particle  there is the critical value of the radial velocity $v_{rad} \simeq c/\sqrt{3} \approx 0.577\,c$. For~$v > v_{crit}$ the radial gravitational force goes to zero and radiation pressure can accelerate particles along the RCO rotation~axis.

\subsubsection{Self-Gravitating Gas~Configurations}

The PN equations of the self-gravitating gas motion was analyzed  in~\cite{baryshev88,baryshev90, baryshev03} 
from the conservation laws of the total energy-momentum, which is fulfilled for  the system gas plus the gravitational field considered in the first iteration:
\begin{equation}\label{emt-tot-cons-2}
   \left( T_{(gas)}^{ik} + T_{(int)}^{ik} +
   T_{(g)}^{ik}\right)_{\;,\;i}  = 0\,.
\end{equation}
From Equation~(\ref{emt-tot-cons-2}) we obain the equation of gas motion:
\begin{eqnarray}
  \frac{d\vec{v}}{dt} =\frac{\partial\vec{v}}{\partial t}+
  (\vec{v}\cdot\vec{\nabla})\vec{v}=
 -(1+\frac{v^2}{c^2} +
 3\frac{\phi}{c^2})\vec{\nabla}\phi
 +4\frac{\vec{v}}{c}(\frac{\vec{v}}{c}
 \cdot\vec{\nabla}\phi)\quad \nonumber\\
 +3\frac{\vec{v}}{c}\frac{\partial \phi}{c\; \partial t} -
 2\frac{\partial \vec{\Psi}}{c\; \partial t} +
 2(\frac{\vec{v}}{c}\times rot\vec{\Psi}) -
 \frac{1}{\varrho_0}\frac{\vec{v}}{c}
 \frac{\partial p}{c\;\partial t}\nonumber \\
 -\frac{1}{\varrho_0}(
 1-\frac{e+p}{\varrho_0 c^2} -\frac{v^2}{c^2}+\frac{\phi}{c^2})
 \vec{\nabla}p\;,
\label{gas-PN-accel}
\end{eqnarray}
where $\vec{v},\,\rho_0,\,p,\,e$ are the velocity, rest mass density, internal energy of the gas element, and~gravitational potentials  $\phi= \psi^{00}$, $\vec{\Psi}=\psi^{0\alpha} = -\psi_{0\alpha}$.

For the static configuration, the gas velocity $v=0$ and post-Newtonian equation of hydrostatic equilibrium of a spherically symmetric body in QFGT will be:
\begin{equation}\label{hydrostatic-eq}
  \frac{dp}{dr}=-\frac{G(\varrho_0 + \delta\varrho)\;
  M_r^*}{r^2}\,,
\end{equation}
where
\begin{equation}\label{delta-pho}
  \delta\varrho=\frac{e+p}{c^2}+2\varrho_0\frac{\phi}{c^2}\;,
\end{equation}
\begin{equation}\label{Mr*}
  M_r^*=\int_0^r{4\pi r'^2(
  \varrho_0 +\frac{e+3p}{c^2}+2\frac{\varrho_0\phi}{c^2}+
  2\frac{(d\phi/dr)^2}{8\pi Gc^2})
  }dr'\;.
\end{equation}
The most important difference between equation Equation~(\ref{hydrostatic-eq})
of hydrostatic equilibrium in QFGT and
the Tolman--Oppenheimer--Volkoff equation
(Equation~(\ref{tov})) in GRT,  is that, within FGT, the relativistic gravity corrections lead to a decrease of the gravitating mass (and so gravitational force) relative to its  Newtonian value (due to the negative value of the gravitational potential ($\phi =\psi^{00} < 0$)). According to Equation~(\ref{hydrostatic-eq}), a hydrostatic equilibrium is possible for any large mass. Another important prediction of the QFGT is that the supermassive stars (suggested as  a possible source of energy in quasars) are stable to small adiabatic pulsations (\cite{baryshev92,oschepkov-raikov95}).

Core-collapse supernova explosions, gamma-bursts, neutrino, and gravitational
bursts have a common origin. Hence, a direct test of the strong gravity effects would be the detection of a gravity wave signal from the relativistic collapse. The~absence of black holes in the QFGT makes dramatic changes in the physics of supernova explosions. The~collapse of the iron core of massive pre-supernovae stars will have a pulsation character and leads to long duration gravitational signals, comparable with neutrino signals and gamma ray bursts, i.e.,~several seconds.
The relation of the gamma-ray-burst (GRB) phenomenon to relativistic core-collapse supernovae has become a generally accepted interpretation of the GRBs (Paczynski~\cite{paczinski01}, Sokolov~\cite{sokolov14}).
If the compact GRB model that was suggested by Sokolov~et~al.~\cite{sokolov06} obtains further confirmation, then there should be a correlation of the gamma-x-ray signal with neutrino and gravitational bursts. First, the detection of the GW170817/GRB170817A gravitational-gamma event from  the galaxy NGC4993~\cite{abbott2020} is consistent with such~scenario.

Note that the gravitational antenna GEOGRAV observed a signal from SN1987A together with the neutrino signal observed by the Mont Blanc Underground Neutrino Observatory (Amaldi~et~al.,~1987~\cite{amaldi87};
Aglietta~et~al.,~1987~\cite{aglietta87}, Imshennik 2010~\cite{imshennik10}, and
Galeotti \& Pizzella 2016~\cite{galeotti16}).
This has been interpreted by Baryshev 1997~\cite{baryshev97} as a possible detection of the scalar gravitational radiation (if the bar changes its length) from the spherical core-collapse of the supernova.
Another possibility to explain the gw-signal from SN1987A in metallic bar antenna (fixed to the ground) is to take the relative difference in motion of free electrons and proton lattice under the action of the scalar~GW into account.

An observational strategy to distinct between scalar and tensor gravitational waves using sidereal time analysis was considered in
~\cite{baryshev-p. 1,fesik-b-s-p. 7}.
There is an evidence for possible detections of gravitational signals by Nautilus and Explorer antennas (Astone~et~al.,~2002~\cite{astone02}).
Though they was not confirmed by later observations (after ``improving'' sensitivity, which actually excluded resonance), it is needed to develop detectors of GW signals that are based on principles compatible with~QFGT.

\subsubsection{Relativistic Compact Objects Instead of Black~Holes}
\label{sec-4-2-10}

In the case of strong gravity, QFGT and GRT 
predictions diverge dramatically, mainly because of the positive localizable energy density of the gravitational field and the crucial role of the scalar potential
component (trace of the symmetric tensor potential) generated by the trace of EMT of the gravitational field sources.
The scalar field is repulsion and only in combination with pure tensor
part (which is attraction) gives the classical Newtonian~gravitation.

In QFGT, there is no black holes, horizons and singularities, and~no such limit as the Oppenheimer-Volkoff mass. This means that compact massive objects in binary star systems and active galactic nuclei are good candidates for testing GRT and QFGT.
According to QFGT for a static weak field conditions, the positive energy density of the gravitational field around an object with mass M and radius R is given by $T^{00}$ component of the EMT gravity field:
\begin{equation}\label{emt-g-00-2}
  \varepsilon_g =\frac{ \left(\vec{\nabla }
  \varphi_{_{\rm N}}\right)^2}{8\pi G}=
  \frac{\,G\,M^2}{8\pi r^4}\,\,\,\mbox{\rm ergs/cm}^3  \;.
\end{equation}
Accordingly, around a neutron star, there is a ``coat'' of gravitational field with
mass density
\vspace{6pt}
\begin{equation}\label{rho-g}
  \varrho_g = 1.1\times10^{13}
  \left(\frac{M}{M_{\odot}}\right)^2
  \left(\frac{10\; km}{R}\right)^4 \mbox{\rm g/cm}^3  \;.
\end{equation}
It is positive, localizable and~does not depend on a choice of the coordinate system. On~the surface of a neutron star, the mass density of the gravity field is about the same as the mass density of the nuclear~matter.

A very general mass-energy argument shows that in QFGT there is the
limiting radius of any self-gravitating body and there is no
singularities.
This argument is a precise analogue to that of the classical radius of electron.
Indeed, the~total mass-energy of the gravitational field existing around a body is given by
\begin{equation}\label{M-g-00}
  E_{fg} =\int_{R_0}^{\infty} \frac{ \left(\vec{\nabla }
  \varphi_{_{\rm N}}\right)^2}{8\pi G}4\pi r^2dr=
  \frac{G\;M^2}{2 R_0}\;.
\end{equation}
This energy should be less than the rest mass-energy of the body, which includes the energy of the gravity field. From~this condition, it follows that:
\begin{equation}\label{R-m}
  E_{fg}<Mc^2\quad \Rightarrow \quad R_0 >  \frac{G\;M}{2 c^2}\;.
\end{equation}
If one takes into account the non-linearity of the gravity field and
the internal energy-part inside the object,
then the value of the limiting radius further increases, because~``the energy of the field energy'' should be added. As~the limiting gravitational radius $R_g$ for any massive body in the field gravity, we define the radius, where mass-energy of the gravitational field equals to half of its
mass-energy measured at infinity, so:
\begin{equation}\label{R-m-g}
  R_g = \frac{G\;M}{c^2} = \frac{1}{2}R_{Sch}\;,
\end{equation}

As we have discussed in Introduction,
very recent surprising observational fact is that the estimated radius of the inner edge ($R_{in}$) of the accretion disk around
black hole candidates has sizes around $(1.2 - 1.4) R_g = (0.6 - 0.7) R_{Sch}$
(Fabian 2015~\cite{fabian15}, Wilkins \& Gallo 2015~\cite{wilkins15},
King~et~al.,~2013~\cite{king13}).
This points to a suggestion that, instead of a Kerr BH rotating with velocity about 0.998c, we observe ordinary RCO having a radius close to its limiting QFGT value $R_g$ (Equation~(\ref{R-m-g})).

Additionally, VLBI observations, using  submm wavelength  Event Horizon Telescope (EHT), will have a unique angular resolution that will achieve
event-horizon-scale structure in the supermassive black hole candidate at the Galactic Centre (SgrA*) and M87. The~first results of EHT SgrA* observations  at 1.3mm surprisingly point to the absence of the light ring at radius $5.2R_{Sch}$ (Doeleman 2008~\cite{doeleman08}). Recent EHT observations of RCO  M87* \cite{akiyama2019a,akiyama2019b} discover the radiation from a ring, which is interpreted as a combination of the BH light ring plus radiation of the accretion disk. However, the angular resolution is not enough for distinction between them, so this observations is also consistent with the existence of an accretion disk around the limiting QFGT RCO
having finite gravity force,  which does not produce light~ring. 


Observations of the stellar mass BH candidates surprisingly discovered
a preferred value of RCO mass approximately $7 M_{sun}$ \cite{sokolov15}.
Intriguingly a quantum consideration of the macroscopic limiting high density quark-gluon bag gives self-gravitating configurations with preferred mass $6.7 M_{sun}$ and a radius 10 km~\cite{sokolov19}. Accordingly, quantum gravidynamics predicts two peaks in the mass distribution of the stellar-mass relativistic compact objects:
$1.4 M_{sun}$ for neutron stars and $6.7 M_{sun}$ for quark~stars.

An important consequence of the positive energy density $\varepsilon_g$ of the gravitational field is the existence of the limiting gravity force for objects having limiting size $\sim R_g$ \cite{baryshev90},
\cite{baryshev95},
\cite{baryshev03}.
The static spherically symmetric massive gravitating body has the coat of gravitational energy around it. Let us consider two cases: (1) positive energy density of the gravity field 
$T^{00}_g = \varepsilon_g = 
(\vec{\nabla}\varphi)^2/8\pi G > 0 $ 
(as predicted in QFGT), and~(2) negative energy density
$t_{g}^{00} = - (\vec{\nabla}\varphi)^2/8\pi G < 0$, as~predicted by GRT~pseudotensor.

For static spherically symmetric body at distances outside the body $r>R_0$ (where $\rho_0 =0$), the field Equation~(\ref{feq2-fgt}) gives the equation for the $\phi$-component
\begin{equation}\label{feq-Phi-nl-fgt}
  \frac{1}{r^2}\frac{d}{dr}\left( r^2
  \frac{d}{dr}\phi (r)\right)=
  \frac{1}{c^2}\left(\frac{d\phi(r)}{dr}\right)^2\;.
\end{equation}
The solution of this non-linear equation, with~condition  $\phi(r)\rightarrow - GM/r$ for $r\rightarrow \infty$, will be
\begin{equation}\label{Phi-nl-fgt}
  \phi(r)=-c^2\ln \left( 1+ \frac{GM}{c^2r}\right)\;,
\;\quad \;\;
  \phi (r)= -\frac{GM}{r} + \frac{1}{2}\frac{G^2M^2}{c^2r^2}
  -\frac{1}{3}\frac{G^3M^3}{c^4r^3} +\ldots\,,
\end{equation}
where the first term in the series is the Newtonian potential and the second term is the PN addition, which is really measured by the observations of the relativistic pericenter shift 
(\mbox{Equations~(\ref{26}), (\ref{FGT-accel}) and (\ref{28})).}
The potential gradient and absolute value of the gravity force predicted by QFGT is
\begin{equation}\label{F-m-g}
  F_g = m\frac{d\phi}{dr} = \frac{GmM}{r^2}\frac{1}{(1+GM/rc^2)}\;,
\end{equation}
hence, for $r \rightarrow R_g$ the gravity force decreases relative to its
Newtonian value. Accordingly, for $r=R_g$ 
\begin{equation}\label{F-m-g-lim}
  g_{max} \leq \frac{c^4}{GM} = \frac{c^2}{R_g}\;,\,\,\,\,\,\,
  and \,\,\,\,\,\,\,
  F_{g(max)} \leq \frac{mc^4}{GM} = \frac{mc^2}{R_g}
  <\frac{c^4}{G}\;,
\end{equation}
where the last inequality is written for the case m = M .
Importantly, in~the frame of FGT, the supermassive RCO have very small gravitational acceleration (and force for unite mass $m$) at their surface, so excluding~singularities.

In the case of a negative energy density of the gravitational field $\varepsilon_g < 0$,
the corresponding field equation is
\begin{equation}\label{feq-Phi-nl-grt}
  \frac{1}{r^2}\frac{d}{dr}\left( r^2
  \frac{d}{dr}\phi (r)\right)=-
  \frac{1}{c^2}\left(\frac{d\phi(r)}{dr}
  \right)^2\;,
\end{equation}
the solution of Equation~(\ref{feq-Phi-nl-grt}) gives gravitational potential in the form
\begin{equation}\label{Phi-nl-gr}
  \phi(r)=c^2\ln \left( 1- \frac{GM}{c^2r}\right)\;,\quad \;\;\;
  \phi (r)= -\frac{GM}{r} - \frac{1}{2}\frac{G^2M^2}{c^2r^2}
  -\frac{1}{3}\frac{G^3M^3}{c^4r^3} -\ldots\,,
\end{equation}
and the gravity force will be
\begin{equation}
 F_g = m\frac{d\phi}{dr} = \frac{GmM}{r^2}\frac{1}{(1-GM/rc^2)}\;, 
 \end{equation}
This gives the infinite gravity force at finite radius $r=R_g$, as~it happens in GRT (at Schwarzschild radius). Hence, the origin of BH is related to the negative energy density of the gravitational~field.

Let us consider the case of SSS ball having radius $R_0$ and constant rest-mass density $\varrho_0$ in order to illustrate the role of the interaction energy in the internal structure of relativistic compact objects.
The~gravitation field equation for the 
$\,\phi = \psi^{00}$ component will be
\begin{equation}\label{delta-phi-sss}
  \Delta \phi=4\pi G (
  \varrho_0 +2\frac{\varrho_0\phi}{c^2}+
  2\frac{(d\phi/dr)^2}{8\pi Gc^2})
  \;.
\end{equation}
To see the action of the interaction energy density and taking into account Equation~(\ref{25}), we consider linear field equation (neglecting the last term in Equation~(\ref{delta-phi-sss})) in the form
\begin{equation}\label{delta-phi-sss-lin}
  \Delta\phi(r) -  \frac{8\pi G\varrho_0}{c^2}
  \phi(r)   = 4\pi G \varrho_0\;\,\,.
\end{equation}
The solution of this equation inside the ball 
($r<R_0$) gives the gravitational potential and gravity force  in the form
~\cite{baryshev-k91}:
\begin{equation}
\label{phi-hom-ball}
\phi(x) = -\frac{c^2}{2} +
 \frac{c^2 \,sh(x)}{2x\, ch(x_0)} \,\,,
\,\,\,\qquad\,\,\,\,
 F_g(x) = m\frac{d\phi(x)}{dx} = \frac{c^2}{2}\,\frac{x\,ch(x) - sh(x)}
 {x^2\,ch(x_0)}\;, 
 \end{equation}
where $x=r/a$, $\,\,a=c/(8\pi G \varrho_0)^{1/2}$, and~$x_0 = R_0/a$. 

The important property of the QFGT potential 
Equation~(\ref{phi-hom-ball}) inside the body is its finite lower limit in the center of the ball $\phi(0) > -c^2/2 $, where the gravity force equals zero and also finite value of the gravity force at the surface of the ball (no singularities).

\section{Cosmology in GRT and~QFGT}\label{sec5}

Modern physics considers the observable Universe as a part of ``the cosmic laboratory'', \mbox{herew~all basic principles and} main physical fundamental laws must be tested with increasing accuracy. 
In particular, in~the spirit of modern theoretical physics, such cosmological basis as the constancy of fundamental constants, the~equivalence principle, the~Lorentz invariance, the~cosmological principle, the~general relativity and its alternatives,  the~space expansion paradigm must be tested in the cosmic laboratory~\cite{turner02,uzan2003, uzan2010,baryshev-t12, baryshev15,clifton2012,derham2014, derham2017,debono2016,ishak2018,lusso2019,perivol2019,riess2020,shirokov20a, shirokov20b}.

In fact, in~the beginning of the 21st century, a New Cosmology emerges and a new set of questions arises. In~particular, the~famous Turner’s list of new cosmological problems contains the
following puzzles: what is the physics of underlying inflation? How was the baryon asymmetry produced? What is the nature of the nonbaryonic dark matter particles? Why is the composition of our Universe so “absurd” relative to the lab physics? What is the nature of the dark energy? Answering these questions will reveal
deep connections between fundamental physics and cosmology: “There may even be some big surprises – time variation of the constants or a new theory of gravity that eliminates the need for dark matter and dark energy” Turner 2002~\cite{turner02}.

In number of recent papers, the problem of tension between the observed parameters in the Local and Global Universe was raised
~\cite{divalentino2020a,divalentino2020b,handley2019,riess2020,verde2019,lin2019}.
They conclude that either LCDM needs to be replaced by a drastically different model, or~else there are significant, but still undetected, systematics. The~new theoretical suggestions call for new observations and stimulate the investigation of alternative theoretical models and solutions in~cosmology.

\subsection{General Principles of~Cosmology}

\subsubsection{Practical~Cosmology}
Cosmology is a science on the infinite spatial matter distribution
and its evolution in time.
Cosmology as a physical science is based on observations, experiments,
and theoretical interpretations.
Sandage 1997~\cite{sandage97} used the term ``Practical Cosmology'' in order to denote the observational study of the largest achievable scales of the Universe and the search for the world model that best describes observations.
Our understanding the Universe is growing with gradually deepening
sample of its observable part that delivers possibilities for testing
alternative hypotheses in the bases of cosmological~models.

Allan Sandage, in~his famous lecture ``Astronomical problems for the next three decades''~\cite{sandage97}, formulated 23 primary astronomical problems for the period 1995--2025 years, whose solution seem possible. Intriguingly, the~first problem in the Class C  (The Universe, practical cosmology) was named: ``Is the expansion real?''. Actually, this problem has not solved yet because observed cosmological redshift could have the nature different from Lemaitre's ``space expansion''. \mbox{e.g.,~global gravitational} redshift by cosmologically distributed matter also has properties that are similar to the Doppler effect. The~problem is how to measure the  distance and its increasing with time  between galaxies without using the cosmological~redshifts. 

Sandage~\cite{sandage62} suggested the test of measuring the change of redshift with time $dz/dt$, which can distinguish between expansion (accelerated) and non-expansion, which is a task for ELT~\cite{pasquini05, liske2008}). In~the near future, additional samples of the Local Universe galaxies will also be available thanks to the projects
Cosmicflows~\cite{tully2019}  
and WALLABY~\cite{koribalski2020}.

Baryshev \& Teerikorpi 2012~\cite{baryshev-t12} presented
the  ``practical cosmology'' as a meta-science of testing alternative cosmological models. 
In fact, the~main task of practical cosmology is to develop different methods that aimed for testing the initial assumptions and basic observable predictions of alternative world models.
The application of the Sandage's practical cosmology approach to forthcoming observations of the high redshift gamma-ray bursters was done by Shirokov~et~al.,~2020a,b~\cite{shirokov20a, shirokov20b}.

\subsubsection{Empirical and Theoretical~Laws}
Cosmology deals with a number of empirical facts, among which one hopes to find fundamental laws.
This process is complicated
by great limitations and even under the paradigmatic grip of any current standard cosmology. One should distinguish between two kinds of cosmological
laws:
\begin{itemize}
  \item experimentally measured empirical laws, and
  \item logically inferred theoretical laws.
\end{itemize}

The major empirical steps in modern cosmology are connected with advances in
instrumentation during the 20th century. The~logically inferred theoretical laws
(theoretical interpretations) are made on the basis of an accepted
cosmological model, e.g.,~the standard or an alternative cosmological model.
Three fundamental cosmological empirical laws were then~unveiled:
\begin{itemize}
  \item the cosmological redshift-distance law $cz=Hr$,
  \item the thermal law of isotropic cosmic background radiation $B_{\nu}(T)$, and
  \item the power-law correlation of galaxy clustering.
  $\Gamma(r)\sim r^{-\gamma}$.
\end{itemize}

The empirical laws, which are based on repeatable observations, are independent of existing or future cosmological models.
The theoretical laws are valid only in the frame of a specific model. A good example is the empirical Hubble's
redshift-distance ($z \propto r$) law 
(Hubble 1929~\cite{hubble29}, Sandage~et~al.,~2010~\cite{sandage10}, Paturel~et~al.,~2017~\cite{paturel2017}) 
and the corresponding Lemaitre's theoretical linear space expansion velocity-distance ($V \propto r$) law within the Friedmann model (Lemaitre 1927
~\cite{lemaitre1927}, Peebles 1993
~\cite{peebles93}).  


\subsubsection{Global Inertial Rest Frame Relative to Isotropic~CMB}

The discovery of cosmic microwave background radiation is a breakthrough event in cosmology; it was marked by the Nobel Prize in Physics 2006 to 
John C. Mather and George F. Smoot “for their discovery of the blackbody form and anisotropy of the cosmic microwave background radiation”.

In cosmology, the~usual lab suggestion regarding the isolation of a local system
is not valid at all, because~there is no external empty space.
In the infinite mass distribution, all problems are internal and division
on local and global physics should be studied carefully.
New specific physical relativistic quantum effects can appear at
cosmological distances and time scales---it is called Feynman's principle of the ``responding   Universe''
(Kim~\cite{kim2020}).

For example, the~definition of a fundamental inertial reference frame in cosmology can be made on the basis of the Holtsmark theorem  of exact cancellation of all external gravity forces $\sum \vec{F}_i = 0$ in the infinite Poisson's
 mass distribution (see e.g.,~\cite{baryshev-t12}). Additionally, in the frame of the quantum-field gravitation theory, the global gravitational potential
 equals constant, so global cosmological force is zero (Section~\ref{sec5.3}).

 A practical realization of such
globally rest inertial frame can be based on the observation of the cosmic microwave background radiation isotropy.
Hence, in cosmology inside the infinite mass distribution
of the Universe there is well defined
\textbf{global inertial rest frame} 
(GIR). Very important that
relative to this GIR reference frame it is possible to measure both
the velocity and acceleration of any body in the Universe. If~a two bodies are at rest relative to CMBR (i.e.,~they see isotropic global background radiation), then they are at rest relative to each other (for non-expanding space-vacuum).
This fact delivers new relativistic and quantum physical situation in the 21st century cosmology and, in general, physics too.

%

\subsubsection{Gravitation Theory as the Basis of Cosmological~Models}

A particular cosmological model always starts with the application of adopted gravitation theory to the cosmologically distributed matter. This is why the 
physics of gravitational interaction
lies in the heart of cosmology and, in~fact, it determines the structure and all main predictions for observations.
\mbox{A comparison of initial principles and main predictions for some alternative}
cosmological models, based on alternative gravitation theories, is presented in Milgrom 2020~\cite{milgrom2020},
Debono \& Smoot 2016~\cite{debono2016},
Baryshev \& Teerikorpi 2012
~\cite{baryshev-t12}, Clifton~et~al.,~2012~\cite{clifton2012}, and
Pavsic 1975~\cite{pavsic75}.

For example, the~Friedmann's equation is the direct consequence of the GRT and it is the basis of the standard cosmological model, together with homogeneity of matter distribution 
(Peebles 1993~\cite{peebles93}), 
Straumann 2013~\cite{strauman13}.  
Accordingly, this is why the gravity theory and  observations of cosmologically distributed matter both are principally important for testing possible cosmological~models. 

Modern elementary particles theoretical and experimental physics also tests the foundations of the Standard Cosmological Model (SCM). In~particular, the~modified theories of gravity change the study of cosmic structure formation
(Slosar~et~al.,~2019~\cite{slosar2019},
Bartelmann~et~al.,~2019~\cite{bartelman2019},
Ishak 2018~\cite{ishak2018}, and
Clifton~et~al.,~2012~\cite{clifton2012}).

\subsection{ Friedmann's Homogeneous Model as the Basis of the~SCM}

\subsubsection{Initial Assumptions: General Relativity, Homogeneity, Expanding~Space}

The geometrical approach of general relativity leads to the Friedmann cosmological model, the~frame for modern cosmological research. The~expanding homogeneous universe explains the available data, although there are some problems and paradoxes, which are discussed in the next~section.

Nowadays, the expanding Big Bang cosmological model is generally accepted as the standard cosmological model (SCM) for the description of the structure and evolution of the physical Universe
(Peebles~\cite{peebles93}, Weinberg~\cite{weinberg08}, and
Baryshev \& Teerikorpi~\cite{baryshev-t12}).
SCM is based on the geometrical gravity theory (general relativity) and it uses
the description of all \emph{physical processes in expanding space}.
\mbox{The fundamental assumptions of the SCM~are:}
\begin{itemize}
\item[$\bullet$]{\emph{General relativity can be applied to the whole Universe
($g_{ik}$; $\:\Re_{iklm}$; $\:T^{ik}_{(\mathrm{m+de})}$).}}
\item[$\bullet$]{\emph{Homogeneity and isotropy of matter distribution in the expanding Universe ($\rho=\rho(t)$;  
$g^{ik} = g^{ik}(t)$).}}
\item[$\bullet$]{\emph{Laboratory physics works in the expanding space.}}
\item[$\bullet$]{\emph{Inflation in the early universe is needed for explanation of the flatness, isotropy and initial conditions of large scale structure formation.}}
\end{itemize}

The fundamental basic element of the SCM is the Einstein's Cosmological Principle, \mbox{which states that }the matter distribution is spatially homogeneous and isotropic at ``enough  large scales''. The~term ``enough large scales'' relates to the fact that the universe is obviously  inhomogeneous at scales of galaxies, clusters
and superclusters of galaxies~\cite{baryshev-t12}. 
\footnote{There is more general Mandelbrot's Cosmological Principle which states the fractality of matter distribution together with isotropy. Fractal cosmological models can be build on the basis of MCP also in the frame of GRT.}
The hypothesis of homogeneity and isotropy of the matter distribution in space means that starting from certain scale $r_{\mathrm{hom}}$, for~all scales
$r>r_{\mathrm{hom}}$, we can consider the total energy density
$\varepsilon=\rho c^2$  and the total pressure $p$ as a function of time only, i.e.,~$\varepsilon(r,t) = \varepsilon(t)$ and $p(r,t) = p(t)$. Here, the total energy density and the total pressure are the sum of the energy densities for matter and dark energy:
$\varepsilon = \varepsilon_{\mathrm{m}} + \varepsilon_{\mathrm{de}}$, and~$p = p_{\mathrm{m}} + p_{\mathrm{de}}$.

An ideal fluid equation of state
$p=\gamma \varrho c^2$
is usually considered for cosmological fluid, \mbox{where  matter and dark energy have} following partial equations of state:
$p_{\mathrm{m}} = \beta \varepsilon_{\mathrm{m}}$ with  $0\leq \beta \leq 1$, and~$p_{\mathrm{de}} = w \varepsilon_{\mathrm{de}}$ with  $-1 \leq w < 0$.
Recently, values $w<-1$ were also considered for description the ``fantom''~energy.

An important consequence of homogeneity and isotropy is that the line element
$ ds^2 = g_{ik} dx^i dx^k$
may be presented in the 
Friedmann--Lemaitre--Robertson--Walker (FLRW) form:
\begin{equation}
 ds^2 = c^2dt^2 - S^2(t)d\chi^2 - S^2(t) I_k^2 (\chi) (d\theta^2 +
 sin^2 \theta d\phi^2)\,,
\label{rw1}
\end{equation}
where $ \chi,\,\theta,\,\phi $ are the "spherical" comoving space coordinates,
$t$ is synchronous time coordinate, and~$I_k(\chi) = (sin(\chi),~\chi,~sinh(\chi))$,
corresponding to curvature constant values $k = (+1,~0,~-1)$, respectively.
$S(t)$ is the scale factor, which determines the time dependence of the~metric.

The expanding space paradigm states that the proper (internal) metric distance $r$ to a galaxy with fixed co-moving coordinate $\chi$ from the observer is given by relation $r(t) = S(t)\cdot \chi $
and increases with time $t$ as the scale factor $S(t)$.
Note that physical dimension of metric distance [r] = cm , hence,~if~physical dimension [S] = cm, then $\chi$ is the dimensionless comoving coordinate distance. In~direct mathematical sense $\chi$ is the spherical angle and S(t) is the radius of the sphere \mbox{(or pseudosphere)} embedded in the four-dimensional Euclidean space.
It means that the ``cm'' \mbox{(the measuring rod) }itself is defined as unchangeable unit of length in the embedding 4-d Euclidean~space.

It is important to point out that the hypothesis of homogeneity and isotropy of space implies that, for a given galaxy, the expansion (recession) velocity is proportional to distance (exact linear velocity-distance relation for all FLRW metrics Equation~(\ref{rw1})):
\begin{equation}
\label{expvel}
 V_{exp}(r) = \frac{dr}{dt} =  \frac{dS}{dt}
 \chi = \frac{dS}{dt}\cdot \frac{r}{S} = H(t) r = c \frac{r}{r_{H}}
\end{equation}
where $H=\dot{S}/S$ is the Hubble constant (also is a function of time) and
$\,r_{H} = c/H(t)$ is the Hubble distance at the time $t$. Note that,
for $\,r> r_{H} $, one gets expansion velocity more than velocity
of light $V_{exp}(r) > c$
(Harrison 1993, 2000~\cite{harrison93, harrison00}).

\subsubsection{Friedmann's Equations for  Dark Energy and~Matter }
In SCM, the dark energy is included in the Einstein's field equations in the form:
\begin{equation}
  \Re^{ik} - \frac{1}{2}\,g^{ik}\,\Re =
  \frac{8\,\pi\,G}{c^4}\,\,(T^{ik}_{(\mathrm{m})} + T^{ik}_{(\mathrm{de})})\,,
\label{efeq2}
\end{equation}
where $\:\Re^{ik}$ is the Ricci tensor, $\:T^{ik}_{(\mathrm{m})}$
is the energy-momentum tensor (EMT) of the matter, \mbox{which includes
all kinds of material substances, such as particles, fields,}
radiation, and~$T^{ik}_{(\mathrm{de})}$ is the EMT of dark energy,
in particular, the~cosmological vacuum is described by
$T^{ik}_{(\mathrm{vac})} = g^{ik}\Lambda$, where
$\Lambda$ is Einstein's cosmological constant.
Usually, $\:T^{ik}_{(\mathrm{m})}$ and $T^{ik}_{(\mathrm{de})}$ are
considered to be independent quantities, qlthough~there are models with
interacting matter and dark energy~\cite{gromov04}.

Note that $\:T^{ik}_{(\mathrm{m})}$ does not contain the energy-momentum tensor of the gravity field itself
(problem of pseudotensor),
because gravitation, in general relativity, is a property of space and~gravity
is not a material field.
A mathematical consequence of the field equations (Equation~(\ref{efeq2})) is
that the covariant divergence of the left side equals zero
(due to Bianchi identity), so, for the right side, we also have
\begin{equation}
  (T^{ik}_{(\mathrm{m})} + T^{ik}_{(\mathrm{de})} )_{\; ;~i} = 0\,.
\label{bianki}
\end{equation}
The continuity equation (Equation~(\ref{bianki})) also gives the
consistency relation with other equations.
As we discussed in Section~\ref{sec2}, this equation leads to the ``pseudotensor paradox'' of the geometrical gravity~theory.


In comoving coordinates, the total EMT has the form
$ T_{k}^{i} = diag(\varepsilon,-p,-p,-p)$
and, for the case of unbounded homogeneous matter distribution given by metric
Equation~(\ref{rw1}), the~Einstein's equations (Equation~(\ref{efeq2})) are directly
reduced to the Friedmann's equations (FLRW – model). From~the initial set of 16 equations, we only have two independent equations for the (0,0) and (1,1) components, to~which we must add the continuity equation (Equation~(\ref{bianki})), which has the form $$3\dot{S}/S=-\dot{\varepsilon}/(\varepsilon + p).$$
Using the definition of the Hubble constant $H=\dot{S}/S$, the~Friedmann's equations get the form:
\begin{equation}
\label{friedmann00}
 H^2 - \frac{8\pi
 G}{3}\varrho=-\frac{kc^2}{S^2}\,,\,\,\,\,\,\,\mathrm{or}\,\,\, \,\,\,
  \Omega -1 = \Omega_k\,,
 \end{equation}
and
\begin{equation} \label{friedmann11}
 \ddot{S}= - \frac{4 \pi G}{3} \left(
 \varrho+ \frac{3p}{c^{2}}\right)S\,,\,\,\,\,\,\,\mathrm{or} \,\,\,\,\,\,
q = \frac{1}{2}\Omega \left(1 + \frac{3p}{\varrho c^2} \right)\,,
\end{equation}
where $\Omega = \varrho / \varrho_{crit}$,
$\varrho_{crit}=3H^2/8\pi G$, $\Omega_k = kc^2/S^2H^2$ and
$ q = - \ddot{S}S/\dot{S}^2 $, and~$\Omega, p ,\varrho$
are the total quantities, i.e.,~the sum of corresponding components for matter and dark~energy.

Note that the Friedmann's equations Equations~(\ref{friedmann00}) and (\ref{friedmann11}), in terms of the metric distance $r(t)=S(t)\cdot\chi$, obtain the exact Newtonian form:
\begin{equation} \label{newton-friedman}
 \ddot{r}= - \frac{G M_{\mathrm{g}}(r) } {r^2} \,,\,\,\,\,\,\,\,\,\,\mathrm{and} \,\,\,\,\,\,\,\,\,
\frac{V^2_{\mathrm{exp}}}{2} - \frac{GM}{r} = \mathrm{const}\,,
\end{equation}
where  $M_{\mathrm{g}}(r) = - \frac{4 \pi G}{3} \left(
 \varrho+ \frac{3p}{c^{2}}\right) r^3 $ is the gravitating mass of a comoving ball with radius $r(t)$.
Solving~the Friedmann's equations, one finds the dependence on time the scale factor $S(t)$ or the metric distance $r(t)$, which is the mathematical presentation of the space~expansion.

The fundamental conclusions of the SCM include many explained astrophysical phenomena, \mbox{such as cosmological redshift} of distant objects, cosmic microwave background radiation, Big Bang nucleosynthesis of light elements,  large scale structure formation, chemical composition of matter, and other.
The main conclusions of the LCDM cosmological model~are:
\begin{itemize}
\item[$\bullet$]{\emph{Cosmological redshift $ (1+z) = \lambda_{0}/(\lambda_{1}) =S_{0}/S_{1}\,,$ is the Lemaitre effect
and the linear expansion velocity-distance relation $V_{exp}=H\times r\,\,$ is the consequence of the space expansion $r(t)=S(t)\times \chi \,\,$ of the  homogeneous Universe.}}
\item[$\bullet$]{\emph{Cosmic microwave background radiation is the result of the photon gas cooling in the expanding space and the CMBR temperature is $T(z) = T_0 (1+z)$.}}
\item[$\bullet$]{\emph{Small anisotropy $\Delta T/T(\theta)$ of the CMBR is determined by the initial spectrum of density fluctuations which are the source of the large scale structure of the Universe.}}
\item[$\bullet$]{\emph{The physics of the expanding Universe is described by the LCDM model which predicts the following matter budget at present epoch: 70\%
of unobservable in lab dark energy, 25\%
unknown nonbaryonic cold dark matter, 5\%
ordinary matter . Visible galaxies contribution is less than 0.5\%.}}
\end{itemize}

\subsubsection{Observational and Conceptual Puzzles of the~SCM}

Although the successes of the LCDM is generally known, there are also some deep observational and conceptual problems for the basis of the SCM.
We emphasize here several such observational puzzles that were recently discussed in the~literature:
\begin{itemize}
\item[$\bullet$]{\emph{\textbf{Tension between measured cosmological parameters}, which derived from the observational data in the Local Universe (Hubble constant, gravitational lensing, large scale structure) and Global Universe (uncertainties in the CMBR data interpretation)   \cite{riess2020,divalentino2020a,handley2019,sylos11, sylos-bao09}.}}
\item[$\bullet$]{\emph{\textbf{Absurd Universe. }The visible matter of the Universe, the~part which we  can actually observe, is a surprisingly small (about 0.5\%) piece of the predicted matter content and this looks like an ``Absurd Universe''~\cite{turner03}. What is more, about 95\% of the cosmological matter density, which determine the dynamics of the whole Universe has unknown physical nature. Turner~\cite{turner02} emphasized that modern SCM predicts with high precision the values for dark energy and nonbaryonic cold dark matter, but~``we have to make sense to all this''.}}
\item[$\bullet$]{\emph{\textbf{The cosmological constant problem.}  One of the most serious problem of the LCDM model is that the observed value of the cosmological constant $\Lambda$ is about 120 orders of magnitude smaller than the expectation from the physical vacuum (as discussed by Weinberg~\cite{weinberg89} and Clifton~et~al.~\cite{clifton2012}). In~fact, the critical density of the $\Omega =1$ universe is $\varrho_{\mathrm{crit}} = 0.853 \times 10^{-29}$ g/cm$^3$, while the Planck vacuum has $\varrho_{\mathrm{vac}} \approx 10^{+94}$~g/cm$^3$..}}
\item[$\bullet$]{\emph{\textbf{The cold dark matter crisis on galactic and subgalactic scales.} There are number of problems with predicting behavior of baryonic and nonbaryonic matter within galaxies. It was discussed by  Kroupa~\cite{kroupa12} that there are discrepancies between observed and predicted galaxy density profiles (the cusp problem), small~number of observed satellites galaxies  (missing satellites problem), and~observed tight correlation between dark matter and baryons in galaxies, which is not expected within LCDM galaxy formation theory.}}
\item[$\bullet$]{\emph{\textbf{The LCDM  crisis at super-large scales.} The most recent observational facts which contradict the LCDM picture of the large scale structure formation, come from:  the 2MASS, 2dF, and SDSS galaxy redshift surveys (Sylos~Labini~\cite{sylos11}), problems with observations of baryon acoustic oscillations (Sylos~Labini~et~al.~\cite{sylos-bao09}), existence of structures with sizes $\sim$ 400 Mpc/h in the local Universe (Gott~et~al.~\cite{gott05}, Tully~et~al.~\cite{tully14}, and Pomarede~et~al.~\cite{pomarede2020}) 
and $\sim$1000 Mpc/h  structures in the spatial distribution of distant galaxies, quasars, and gamma-ray bursts
(Nabokov \& Baryshev~\cite{nabokov10}, Clowes~et~al.~\cite{clowes13}, Einasto~et~al.~\cite{einasto14}, and Horvath~et~al.~\cite{horvath15}), existence of old galaxies, and
supermassive black holes in quasars at redshift up to $\,z\sim7\,$ (Dolgov~\cite{dolgov2018}, Yang~et~al.~\cite{yang2020},
alternative interpretation of the shape of the CMBR correlation function
(Lopez-Corredoira \& Gabrielli~\cite{lopez-gab13}),
lack of CMBR power at angular scales larger 60 degrees and correlation of CMBR quadrupole with ecliptic plain (Copi~et~al.~\cite{cop. 10}).}}
\end{itemize}

These observational puzzles in the SCM interpretations of the astrophysical data give rise to the question:
Does the contemporary standard cosmological model present the ultimate physical picture of the Universe?
As was emphasized by Turner~\cite{turner02}, for making new cosmology one has to answer a new set of questions and the future world model will reveal deep connection between fundamental physics and cosmology. There may even be some big surprises, such as time variation of the constants or a new theory of gravity that eliminates the need for dark matter and dark energy~\cite{turner02}.
Intriguingly, besides~the above-mentioned observational puzzles,
there are several deep conceptual problems in the foundation of the SCM. Their solution could open the door  to construction future cosmology free of these~paradoxes.

The most intriguing conceptual puzzle of the SCM is the question about the physical sense of the mathematical space expansion $\,\,r(t)=S(t)\chi\,$.
Surprisingly,~almost one-hundred years
after Lemaıtre’s interpretation of cosmological redshifts as the effect of the space expansion, the~acute discussion again raised in professional cosmological literature about physical nature of the cosmological redshift and relation between mathematical geometrical concept of metric $\,g^{ik}(t)\,$ (dimensionless) and astronomical distance $\,r(t)\,$ 
(measured in physical unit  ``cm''):
Harrison 1993, 1995, 2000
~\cite{harrison93, harrison95, harrison00},
Francis~et~al.,~2007~\cite{francis07},
Baryshev  2008c, 2015~\cite{baryshev08c, baryshev15},
Kaiser 2014~\cite{kaizer2014},
Lopez-Corredoira 2014~\cite{lopez14},
Davis 2004, 2010~\cite{davis2004, davis2010}, 
Abramowicz 2007, 2009~\cite{abramowicz2007, abramowicz2009}, and
Peacock 1999, 2008~\cite{peacock1999, peacock2008}.

The fundamental cosmological observational fact, as discovered by Hubble 1929~\cite{hubble29}, is the linear (for the Local Universe) relation between observed redshift $z$ of the spectral lines and the distance $r\,$ (``cm'') to a galaxy, i.e.,~the observed Hubble Law is the linear redshift–distance relation.
To determine the distance $r$ to a galaxy, Hubble used the concept of the “standard candle”, i.e.,~an object with an apriori known luminosity. Note that Hubble called the observed redshift “an apparent velocity” \mbox{(in units of c)} because he measured the distance $r$ to a galaxy through the flux measurement of the standard candle. So, he did not measure the physical velocity of a galaxy as the change of distance with~time.
Hubble \& Tolman 1935~\cite{hubble35} suggested several observational tests for tesing the nature of the cosmological redshift.

The expanding space paradigm of the Standard Cosmological Model is the theoretical interpretation of the redshift as the Lemaıtre effect
in the expanding Friedmann homogeneous universe, where the space expansion velocity $V_{exp}(r)$ is identified with the spectroscopic-ally measured the spectral lines shift $cz = V_{app}(r)$. Note that Hubble did not discover the ``expansion of the Universe'', but found the linear  redshift-distance relation $\,z=Hr/c\,$
in the Local Universe~\cite{baryshev16}.

Below, we present several conceptual difficulties/paradoxes of the SCM, which already have been discussed in the~literature:
\begin{itemize}
\item[$\bullet$]{\emph{\textbf{Space expansion-Vacuum problem}: in the physical Universe there is no empty space, but~it is filled by the vacuum, so expansion of space must be accompanied by the creation of vacuum~\cite{baryshev08c,baryshev-t12}.}}
\item[$\bullet$]{\emph{\textbf{Vacuum energy paradox:} in the framework of the Einstein's geometrical gravity theory (GRT)  there is the paradox of too small value of the Lambda term, considered as the physical vacuum~\cite{weinberg89}.}}
\item[$\bullet$]{\emph{\textbf{1st Harrison's paradox (``energy-momentum non-conservation''):} physics of space expansion contains such puzzling phenomena as continuous creation of vacuum and violation of energy-momentum conservation for matter in any comoving volume, including photon gas of cosmic background \mbox{radiation~\cite{harrison95, harrison00,baryshev08c, baryshev15, baryshev-t12}.}}}
\item[$\bullet$]{\emph{\textbf{2nd Harrison's  paradox (``motion without motion''):}  a galaxy cosmological velocity is conceptually different from the galaxy peculiar velocity, in~particular, the~cosmological redshift in expanding space is not the Doppler effect, but~the Lemaitre effect is applicable to a receding galaxy, which can have velocity larger than the velocity of light (so cosmological redshift is a new physical phenomenon, which also includes the global gravitational cosmological redshift)
\cite{harrison93,harrison00,baryshev08c, baryshev15, baryshev-t12}.}}
\item[$\bullet$]{\emph{\textbf{Hubble-deVaucouleurs' paradox (``Hubble law is not a consequence of homogeneity''):} in the expanding space the linear Hubble law is the fundamental consequence of the assumed homogeneity, however modern observations reveal existence of strongly inhomogeneous (power-law correlated) large-scale galaxy distribution at interval of scales $1 \div  100$ Mpc, where the linear Hubble law is firmly established, i.e.,~just inside inhomogeneous spatial galaxy distribution of the Local Universe~\cite{baryshev94, baryshev98,baryshev16}
\cite{baryshev15, baryshev-t12} .}}
\end{itemize}

In the spirit of Sandage's practical cosmology, in order to~avoid the cosmological paradoxes, one must to find observational tests that could  confirm or reject 
the basic assumptions of the considered model.
For example, theoretical Hubble Diagram for different cosmological models
incorporates the directly observed fluxes, luminosity distances and redshifts for a particular class of
standard candles. Recently, the~observations of high-redshift gamma-ray bursts Hubble Diagram was suggested 
as a test of basic theoretical relations for alternative cosmological models
~\cite{shirokov20a, shirokov20b}.

As discussed in Baryshev 2015~\cite{baryshev15}, on~the verge of modern technology there are possibilities to perform  direct observational tests of the physical nature of the cosmological redshift. First crucial 
test of the reality of the space expansion was suggested by Sandage 1962~\cite{sandage62}, who noted
that the observed redshift of a distant object (e.g.,~quasar) in accelerated  expanding space must be changing with time, according to relation $dz/dt = (1+z)H_0 - H(z)$. In~terms of measured radial velocity, the~predicted change is about  $d\nu/dt \sim 1$ cm/s/yr, which is within the reach of the forthcoming ELT observations~\cite{probst2020,liske2008}.

Recently, Kopeikin 2012, 2015~\cite{kopeikin12, kopeikin15}
suggested a test for testing the space expansion in the Solar System. He noted that the equations of light propagation used  by Space Navigation Centers contain measurable terms which include  the Hubble constant $H_0$. 
Future space missions can measure this effect of the global cosmological expansion within the Solar System. The~predicted  frequency drift in the Doppler-tracking observations has
magnitude  
$d\nu/\nu = 2H_0\,t
\approx 4 \times 10^{-15} $
($H_0$/70km/s/Mpc)($\Delta t/10^3s$), where $H_0$ is the Hubble constant and
$\Delta t$ is the time interval of observations. For~the non-expanding Universe, the frequency drift equals~zero.

\subsection{Possible Fractal Cosmological Model in the Frame of~QFGT}
\label{sec5.3}

As we discussed above, the~gravity theory is the true basis of any cosmological models. Accordingly,~in~the frame of the quantum-field gravitation theory, there is the possibility to construct a cosmological model, which is based on the physical principles that are common with other fundamental quantum interactions. The~Poincare symmetry of the Minkowski spacetime, Noether's theorem on existence conceived energy-momentum tensor of the gravitational field, Wigner's classification of elementary particles,
and Feynman's path integral for gravity quantization are assumed to be applicable in QFGT cosmology. Especially, such principles of the quantum physics as quantum entanglement, non-locality, and irreversibility, are working in quantum cosmology that is based on~QFGT.

The Standard Cosmological Model (LCDM) has been developing for more than 30 years by many physicists before it gets the modern form with many important results. However, the possible cosmological models in the frame of QFGT have not been sufficiently developed yet. This is because of the absence of general interest to the Feynman's quantum-field approach to gravitation, existing misleading claims  on fatal internal contradictions of the field approach, and~ small astrophysical data that are relevant to~cosmology.

As we have demonstrated in preceding sections, the modern situation in theoretical physics and observational astrophysics is now dramatically changed, and~QFGT based cosmology started developing. 
The field gravity fractal (FGF) cosmological model now has preliminary qualitative character, but~it also contains several quantitative results, which can be tested . The~modern status of FGF cosmology allows one to formulate the really crucial observational tests of those basic interpretations of fundamental cosmological facts, such as the quantum nature and linearity of cosmological redshifts, together with the strong inhomogeneity of large scale spatial galaxy distribution at distances less than 400 Mpc ($z < 0.1$), where super-large galaxy clusters and filaments exist. Correlation analysis of the 2MRS all sky IR redshift survey
demonstrates that the observed spatial galaxy distribution has power-law complete correlation function, which is consistent with stochastic fractal structure having fractal dimension $D_F \approx 2$ in the scales interval $1 \div 100$ Mpc~\cite{tekhanovich16}. Hence, it is a good starting point for FGF~cosmology.

\subsubsection{Initial Assumptions of the FGF~Model}

A Field Gravity Fractal (FGF) cosmological model was suggested by Baryshev 1981
~\cite{baryshev81} and further developed in Baryshev 2008d~\cite{baryshev08d}
and Baryshev \& Teerikorpi 2012~\cite{baryshev-t12}. Crucial astrophysical tests of the FGF model  
by means of high-redshift gamma-ray bursts observations was recently considered in Shirokov~et~al.,~2020a, 2020b~\cite{shirokov20a, shirokov20b}.
The FGF model is based on the two~assumptions:
\begin{itemize}
\item[$\bullet$]{\emph{the  gravitational interaction  is described by the Poincare covariant Feynman's quantum-field gravitation theory in Minkowski spacetime; and,}}
\item[$\bullet$]{\emph{the total baryonic matter distribution (visible and dark) in the Local Universe is described by the stochastic fractal density law with critical fractal dimension $D_{crit} =2$.}}
\end{itemize}
Within the FGF framework, a new qualitative picture of the Universe could be developed, with~some quantitative results that may be tested by current and forthcoming observations.
The field gravity theory allows one to consider infinite matter distribution in  Minkowski space without the gravitational potential paradox. A~global evolution of matter is possible without space expansion and initial singularity. Cosmological redshift has global gravitational origin and relativistic quantum nature.
The global inertial rest frame is defined relative to the cosmic microwave background radiation by~measuring the isotropic distribution of the CMBR for a given reference frame in the Universe. The observed small galaxy velocity dispersion (around the linear Hubble law) corresponds the global quietness of the matter in the fractal Universe. The motion of galaxies (at different hierarchical levels of the large-scale structure) in many opposite directions compensate each other.
The energy-momentum tensor of the interaction plays the role of an effective cosmological~Lambda-term.

Instead of assumed in SCM homogeneous non-baryonic dark matter and dark energy,
it is assumed that the fractal distribution of $dark + luminous$ baryonic matter from the scales of galactic halos up to the radius of the Local Universe ($\sim$400 Mpc) has the fractal dimension of the total (luminous and dark) matter close to $D_F=2$ and  explains the observed linear Hubble law as the global gravitational redshift
in the Local~Universe.

\subsubsection{Universal Cosmological Solution and Global Gravitational~Redshift}

A specific feature of the 
QFGT is that the gravitational potential is restricted by the constant 
$|\phi| \leq c^2/2$  
(Section~\ref{sec-4-2-10}).
In particular, this is important in the cosmological problem, where we can obtain some quantitative results even using the generalized post-Newtonian equations.
In the case of a static spherically symmetric dust-like ($p. 0, e=0$) ball with finite radius $R_0$,
we got from the field Equations~(\ref{feq2-fgt}) the Equation~(\ref{delta-phi-sss}) for the   $\psi^{00} =\phi$ component in the form
\be
\Delta \phi 
 - \frac{8\pi G}{c^2}\varrho\phi -
 \frac{1}{c^2}
\left(\nabla \phi\right)^2 
= 4\pi G \varrho \,.
\label{cosmol-fgt-1}
\ee
This field equation can be used for both cases of finite (existing the center of mass) and infinite (\mbox{no center of mas}s) radius $R_0$.
The crucial ters in mthe left-hand side of the Equation~(\ref{cosmol-fgt-1}) appear due to  the
negative interaction mass density ($2\varrho\phi/c^2$) and positive  mass density 
$\,2\left(\nabla \phi\right)^2 
/(8\pi G c^2) $   of the gravitational field.
The main relativistic contribution to 
the Equation~(\ref{cosmol-fgt-1})
belongs to the negative interaction energy part, so, at the first step,
we can delete the non-linear term, and~consider the case of  large mass, when the gravity force goes to zero.
Hence, we obtain the simple equation
\be
\Delta \phi -
 \frac{8\pi G\varrho}{c^2}\phi =
 4\pi G \varrho \,.
\label{cosmol-fgt-2}
\ee

Note that Equation~(\ref{cosmol-fgt-2}) is similar to the Einstein's cosmological equation ~\cite{einstein17}, if~the Lmbda-term equals to $\Lambda = 8\pi G\varrho/c^2$.
For the infinite radius ($R_0=\infty$) of the mass distribution $\varrho$, the cosmological solution of Equation~(\ref{cosmol-fgt-2}) will be
\be
\phi = -\frac{c^2}{2}\,,
\,\,\, \qquad \,\,\,and\,\, \qquad
\,\,
\frac{d}{dr}\, \phi =0 \,,
\label{cosmol-fi-2}
\ee
which means that the net gravity force 
from the infinite mass distribution equals zero for any place in the Universe. This gives a natural solution of the infinite potential paradox of Newtonian gravity, which forbids pure Newtonian~cosmology. 

If we consider two mass-points (galaxies) at a finite distance $r= |\vec{r_1} - \vec{r_2}|$ between them, within~the constant mass  density $\varrho =\varrho_0=const$ infinitely distributed in space, then 
a new quantum gravity cosmological effect must be taken into account.
In this case, there is no preferred center of mass and around each point we can select a mass-ball with radius $R_0$ having center in the point under consideration. Due to isotropy of the mass distribution there is no global gravity force acting on the point $\vec{r_2}$ from the global mass-ball selected around point $\vec{r_1}$  having radius equals to the distance between these points ($r=R_0$). However in the quantum theory, there is exchange of virtual gravitons between the mass-point at the surface of the ball with all points in the ball. Hence, this is the difference between the local and global~gravity.

Ordinary local mass density fluctuations $\delta \varrho_{loc} = \varrho - \varrho_0$ 
produce local fluctuations of the gravitational potential 
$\,\delta \phi_{loc}(r)= \delta \phi(r_{12})\,$  for finite distances $r=r_{12}= |\vec{r_1} - \vec{r_2}|\,$.  However,~for large distances in a homogeneous mass distribution,  the  average value of the local gravitational potentials goes to zero
($\delta \phi_{loc}(r) \rightarrow 0 $ for $r \rightarrow \infty$).
However, in~the quantum theory of the gravitational interaction,  besides~the local gravitational potentials 
$\delta \phi_{loc}(r)$
(defined for preferred centers of mass fluctuations) 
one must consider also the difference of of global gravitational potential
$\delta \phi_{glob}(r) = \Delta \phi (r_{12}) = \phi(R_0) - \phi(0) )$ between the surface ($r=R_0$) and the center (r=0) of the mass-balls around each~mass-point.

For large distances (large radius of the ball), the global gravitational potential  goes to finite value 
($\delta \phi_{glob}(r) \rightarrow -c^2/2 $  for $r \rightarrow \infty$ ).
It means that, besides the ordinary local
gravitational frequency shift
($z_{loc} \approx \Delta\phi_{loc}/c^2$), due to local density fluctuations there is cosmological gravitational redshift ($z_{glob} \approx\Delta\phi_{glob} /c^2$)
due to global mass distribution around each points in the Universe. Importantly, the~local gravitational spectral shift can be both redshift and blueshift, corresponding to the motion from the center and to the center of the local density fluctuation.
While the global gravitational spectral shift is only the redshift, corresponding to the mass-point where photon was~emitted.

For the strictly homogeneous mass distribution $\varrho = \varrho_0$, the local density fluctuations equal zero
($\delta\varrho_{loc}=0$), so the local part of the gravitational spectral shift also equals zero. However, the global part of the gravitational redshift can be obtained from Equations~(\ref{z-grav-fg}) and  (\ref{phi-hom-ball}), though~the net gravity force equals zero (no preferred center). The~values of the global potential in the center ($r=0$) and in the surface ($r=R_0$) of the ball, we obtain
from Equation~(\ref{phi-hom-ball}):
\begin{equation}
\label{phi-hom-ball-3}
\phi(0) = -\frac{c^2}{2}
\left( 1 -
 \frac{ 1}{ ch(x_0)}
 \right)\,,\,\,\qquad \,\,\,
 \phi(x_0) = -\frac{c^2}{2}
\left( 1 -
 \frac{th (x_0)}{ x_0}
 \right)
\end{equation}
where $x_0 = R_0/a\,$, and~ $\,\,a=c/(8\pi G \varrho_0)^{1/2}$.
Parameter $a$ in this cosmological solution determines the Hubble radius and the Hubble time $a=R_H = c\,t_H$. 

Let us consider the event of the photon emission from the source at the center of the ball having radius equals to the distance between the source and detector/observer.  Accordingly, possible paths in the functional integral contain the whole ball of radius $r=R_0$. The~observed at the surface $R_0$ cosmological global gravitational redshift we get
from Equation~(\ref{z-grav-fg}) in the form: 
\begin{equation}
1+z_{cos}(R_0) = \frac{1 + z(0)}
{1+ z(R_0)} = 
\frac{\sqrt{1+2\phi(R_0)/c^2}}
{\sqrt{1+2\phi(0)/c^2}}
= \sqrt{\frac{sh(x_0)}{x_0}} 
\approx 1+ \frac{2\pi G \varrho_0}{3\,c^2}
\, R_0^2 \,\,, 
\label{cosmol-z-fgt-3}  
\end{equation}
where the last approximation is related to the small distances $r=R_0 << R_H$ . Note that according to Equation~(\ref{cosmol-z-fgt-3})  in homogeneous universe the gravitational cosmological redshift has quadratic law~$z \propto r^2$.

Why does the cosmological gravitational effect give the redshift? From the causality principle, the event of emission of a photon 
by the source marks the centre of
the ball, so it must precede the event of detection of the photon by an observer. The~latter event marks the spherical
edge where all potential observers are situated after the transition time $t = R_0/c$. Therefore, to calculate the
cosmological gravitational shift within the cosmologically distributed matter, 
one should consider the mass-ball
with the center in the source and with the radius $R_0$ of the ball equal to the distance $r$ between the source and an observer. Hence, the cosmological gravitational shift is the~redshift.

Note that, in textbooks by Peacock 1999~\cite{peacock1999}  and  Zeldovich \& Novikov 1984~\cite{zeldovich84}, the global cosmological gravitational shift was considered at classical level.
They put the observer to the center of the ball and so got the ``blueshift''. However in papers by  de Sitter 1916~\cite{desitter16a} and Bondi 1948~\cite{bondi1947} they put the light source to the center of the ball and hence got  the cosmological global gravitational ``redshift''.
In fact, the contradiction appears due to consideration of the classical GRT and Newtonian~gravity.

The quantum  interpretation of the cosmological gravitational redshift allows for one to solve the classical theory paradox of uncertain sign of the spectral shift. Indeed, 
the considered process start from the emission of photon in the center of
the ball, and~finished with the photon detection  on its surface. Accordingly, the quantum paths in the functional integral contain the whole ball of radius $r=R_0$.

Cosmological global gravitational redshift can be considered as a quantum non-local entanglement effect in the QFGT. The~collapse of the detected photon's wave function contains information regarding the total mass of the ball having radius $r=R_0$.
The non-locality and entanglement of the quantum mechanics is the experimentally verified fact and it is related to the basic concepts of modern quantum physics (see e.g.,~
Kadomtsev 2003~\cite{kadomtsev2003},
Rauch~et~al.,~2018~\cite{rauch2018}, and
Erhard, Krenn \& Zeilinger~2020~\cite{erhard2020}).

Let us consider the case of the fractal cosmologically distributed matter. The stochastic fractal galaxy distribution was introduced by Mandelbrot 1982
~\cite{mandelbrot82} as a model that describe the power-law slope ($\xi \sim r^{-\gamma}$) of Peebles's two point correlation functions. At~that time the slope was estimated as $\gamma = 1.8$, which corresponded to the fractal dimension of the observed luminous  galaxy distribution $D_F = 1.2$ at scales 
10 kpc $\div$ 10 Mpc. Modern data on the Local Universe (2MRS sample) demonstrate that the  visible galaxy distribution has fractal dimension $D_F \approx 2$ at scales  100 kpc $\div$ 100 Mpc~\cite{tekhanovich16,gabrielli05}.

Accordingly, we start from the stochastic fractal model of galaxy distribution  having fractal dimension $D_F =2$, hence   the fractal rest mass density law will be given by relation~\cite{mandelbrot82,gabrielli05}:
\be
\varrho(r) = \frac{\varrho_0 r_0}{r}\,
=\,\frac{\beta}{r}\,,
\label{cosmol-rho}
\ee
where conditional density $\varrho$ is defined as average mass within balls of  radius $r$ centered in each mass-point of the fractal structure, constant 
$\varrho_0 r_0 = \beta$ is related to the zero level of hierarchy.
For the case of the finite fractal ball having radius $r=R_0$,
the solution of Equation~(\ref{cosmol-fgt-2}) inside the ball ($r<R_0$) has the form
\be
\frac{\varphi(x)}{c^2} = -\frac{1}{2} +
\frac{I_1(4\sqrt{x})}{4\sqrt{x}
I_0(4\sqrt{x_0})}\,,
\label{cosmol-fi-d2}
\ee
where $x=r/R_H$ and~$R_H = c^2/(2\pi G\varrho_0 r_0$) is the Hubble radius
for the $D_F=2$ fractal universe, constant $\varrho_0 r_0 = \beta$ is the new
fundamental constant of the theory, and~$I_0\,,\,I_1\,$ are the modified Bessel functions.
The observed at the mass-ball surface $r=R_0$ cosmological global gravitational redshift: 
\begin{equation}
1 + z_{cos}(R_0) =
\left[
\frac{I_1(4\sqrt{x_0})}{2\sqrt{x_0}}
\right]^{\frac{1}{2}}
\approx 1 + \frac{2\pi G \varrho_0 r_0}{c^2}\,\,R_0\,\,,
\label{cosmol-z-fgt-o-r}   
\end{equation}
where $x_0 = R_0/R_H\,$, and~ $\,\,R_H=c^2/(2\pi G \beta)$ is the Hubble radius and the Hubble time is
$t_H = R_H/\,c$. The~last relation corresponds to the small distances 
$r=R_0<<R_H$.

From Equation~(\ref{cosmol-z-fgt-o-r}), in~the case of small distances between galaxies $R_0=r<<R_H$, we see that the global cosmological gravitational potential is a linear function of distance between a source and observer $\varphi (r) \propto r^1$; hence, the cosmological gravitational redshift
$z_{cos}(r) = \Delta\varphi/c^2 =( \varphi(r) - \varphi(0) ) /c^2$
will be
\be
z_{cos}(r) =
\frac{2\pi G \varrho_0 r_0}{c^2}\,\,r =
\frac{2\pi G \beta}{c^2}\,\,r
= \frac{H_g}{c}\,\,r \,\,.
\label{cosmol-z}
\ee
From Equation~(\ref{cosmol-z}), the~expression for the gravitational Hubble constant is given by relation:
\be
H_g =
2\pi  \varrho_0 r_0 \, \frac{G}{c}\,
=2\pi  \beta\, \frac{G}{c}\,\,.
\label{hubble-grav}
\ee
For a structure with fractal dimension 
$D_F = 2$ the constant
$\beta = \varrho_0 r_0$
may be actually viewed as a new
fundamental physical constant that determines the value of the gravitational Hubble constant. If~the value
of the fractal constant is 
$\beta = (1/2\pi )\,$ g/cm$^2$, as~it happens for an ordinary galaxy, where~e.g., one~can take
$\varrho_0 = 5.2 \times 10^{-24}$(g/cm$^3$  and $r_0 = 10\,$kpc, then
$H_g = 2\pi \beta G/c = 68.7$ (km/s)/Mpc.

\subsubsection{The Structure and Evolution of the Field-Gravity Fractal~Universe}

From complete correlation function analysis, the~first observational evidence on the fractal structure of the galaxy Universe was presented by Pietronero 1987~\cite{pietronero1987} and developed by his team in~\cite{sylos11,sylos98,pietronero1986,pietronero1995,pietronero1997}.
Especially, the crucial role of  different selection effects on the correlation function analysis  was discussed 
\footnote{
For the history of the ``fractal debate'' see~\cite{baryshev-t02,baryshev-t06}}. 
Modern observational data on the Local Universe ($z<0.1$)  demonstrate that the galaxy spatial distribution has the power-law complete correlation function with slope $\gamma \approx 1$ at scales of up to 100 Mpc, which corresponds to the stochastic fractal distribution with fractal dimension
$D_F = 3 - \gamma = 2$
\cite{gabrielli05,baryshev-t12,tekhanovich16}. There are also observations of super-large structures
(galaxies, gamma-ray bursts, quasars) having sizes up to 1000~Mpc, so  inhomogeneity continues to very large scales
\cite{sylos11,sylos-bao09,gott05,tully14,pomarede2020,nabokov10,clowes13,einasto14,horvath15}.

According to Equation~(\ref{cosmol-z})
for the global gravitational potential, 
the mass $M_F(r)$ of the fractal structure within distances $r<<R_H$  is given by the relation
\be
M_F(r) =
2\pi  \varrho_0 r_0 \, r^2
= 4.8 \times 10^{11} M_{\odot} \left(\frac{r}{10\,kpc}\right)^2\,\,,
\label{cosmol-mass-grav}
\ee
where the fractal constant equals
$\beta=\varrho_0 r_0 =(1/2\pi)\, g/cm^2$. Equation~(\ref{cosmol-mass-grav})  gives a  reasonable values for the mass that are close to a total galaxy mass (including dark matter) within the radius r about 10 kpc, and~also to the value of the total mass of the Universe within the Hubble radius $r=R_H$.

However, a problem appears from the estimation of the gravitating mass using Equation~(\ref{cosmol-mass-grav}). 
\mbox{To produce the gravitational} Hubble law on scales of about 10 Mpc the total mass within such balls should be $M(r=10\,Mpc)=4.8\,10^{16}\,M_{\odot}$.
Such values much exceed the mass of the luminous matter and this is why the FGF model is compelled to assume that a sufficient amount of dark matter has the fractal distribution with $D_F = 2$. Additionally, to~have sufficiently small fluctuations
in the Hubble law in different directions around an observer, the fractal should be a special class: isotropic with small~lacunarity.

The observed distribution of luminous matter (galaxies) on scales from 10 kpc up to 100 Mpc is well approximated by a fractal distribution with $D_F\approx 2$
(\cite{sylos11,baryshev-t12,tekhanovich16}). This means that, within the FGF model, both dark and luminous matter is similarly distributed on these scales. The~nature of the fractal dark matter has to be determined from future observations.
The current restrictions on possible dark matter candidates include
dead stars, neutron and quark stars, Jupiters, planet size objects, asteroids and comets, Pfeniger's hydrogen cloudlets, and~also macroscopic quark dust
~\cite{jacobs15,delpopolo14}.

For large distances ($r>>R_H$), the total gravitating mass is
$M(r) = (c^2/2G)\,r$
for both $D_F = 2$ and $D_F = 3$ fractal structures. For~scales that are close to  $R_H$
the fractal dimension of dark matter may become $D_F=3$, corresponding to a homogeneous distribution.
Hence, it may be a transition zone from scales about 100 Mpc (where fractal dimension $D_F \approx 2$) to scales about 1000 Mpc, where mass distribution becomes homogeneous. 

Intriguingly, for~the general fractal matter distribution with fractal dimension $D_F=2$, the~mass density---radius relation Equation~(\ref{cosmol-mass-grav}) gives
$\varrho r \sim 1\,\,$g/cm$^2$ and looks universal, starting
from the elementary particles scales
\mbox{($\varrho \sim 10^{13}$ g/cm$^3;\,\,r \sim 10^{-13}$ cm)} and continues at galactic scale\mbox{ ($\varrho \sim 10^{-24}$ g/cm$^3;\,\,r \sim 10^{24}$ cm)} and holds up to
the Hubble radius ($\varrho \sim 10^{-28}$ g/cm$^3;\,\,r \sim 10^{28}$ cm).
Accordingly, the universal
linear gravitational redshift law within the fractal structure with $D_F=2$ would have deep roots in the fundamental
physics and $H_g$ can be expressed as a combination of fundamental constants
of microphysics via expressions $\varrho_0$ and $ r_0$ for nuclear matter
$(h, c, m_p, m_e)$
(\cite{baryshev-r88,baryshev94b}).
\mbox{To understand the global gravitational} cosmological redshift, one needs to develop Ghc gravitation theory on the common
basis with modern quantum~physics.


The evolution of the fractal Universe can be considered to be a process of dynamical and chemical evolution in static Minkowski spacetime. In~the space,
filled by matter and cosmic background radiation, there is a special frame of reference, namely the one where the matter is at rest on average relative to the CMBR. This frame of reference also allows
for one to speak about a universal time and the arrow of time is determined by the growth of the local entropy. Initial fluctuations in the homogeneous gas of primordial hydrogen exponentially grow into large scale structures according to the classical scenarios by Jeans 1929~\cite{jeans29}
and Hoyle 1953~\cite{hoyle53}.
The fractal structure of matter distribution with $D_F = 2$ could naturally originate as the result of the evolution
of the initial fluctuations within the explosion scenario
(Schulman \& Seiden 1986~\cite{schulman86}).
The fractal structure with critical dimension $D_F^{crit} = 2$ is also preferred in the dynamical evolution of self-gravitating N-body system (\mbox{Perdang 1990~\cite{perdang90}; de
Vega~et~al., 1996~\cite{devega96}; 1998~\cite{devega98}).
}

The time-scale of the largest structures evolution is determined by the characteristic Hubble time: $t_H \approx R_H/c \propto (\varrho_H)^{-1/2} \approx 10^{10}$ years.
The total evolution time of the Universe may be several orders of magnitude larger, which could be tested by
observations at high redshifts and numerical simulations of the large-scale structure and galaxy formation
in static space, but dynamically evolving~matter.

According to the classical argument by Hoyle (1982~\cite{hoyle82},
1991~\cite{hoyle91}),
the cosmic microwave background radiation
could be a remnant of the evolution of stars because
the CMBR  energy density equals to the energy released by the nuclear reactions in stars of all generations during the Hubble time.
The optical photons that are radiated by stars could be thermalized by scattering and gravitational deflections by structures of different masses and scales.
The fractal dark matter is also a product of the process of stellar evolution and large scale structure formation. Hence, in the frame of the FGF cosmological model all three phenomena---the cosmic background radiation, the~fractal
large scale structure, and~the Hubble law---could be consequences of a unique evolution process of the initially homogeneous matter (e.g. cold hydrogen~gas).

\subsubsection{Crucial Cosmological Tests of the Fractal~Model}

The philosophical, methodological, and sociological aspects of the development of the science on the whole Universe was recently analyzed by
Lopez-Corredoira~\cite{lopez14}, who emphasized the important role of alternative ideas in modern cosmology.  The~mathematical and physical basis for the construction of alternative
cosmological models was discussed by Baryshev \& Teerikorpi~\cite{baryshev-t12}.
An alternative cosmological model must be consistent with the firmly established results of  theoretical and experimental physics,  and~with contemporary astrophysical observations. Moreover, it should be sufficiently developed to predict 
the crucial observational tests, which 
can distinct between the standard and proposed~model.

The field gravity fractal cosmological model satisfies the above demands. FGF is based on modern physics and existing cosmologically relevant observations.
The fractality of large-scale galaxy distribution  is consistent with the correlation analysis of large  galaxy redshift surveys, such as 2MASS, 2dF and SDSS (e.g.,~\cite{sylos98,sylos11,sylos14,tekhanovich16,baryshev16}).
Accordingly, at~least for the interval of  scales $1 \div 100$ Mpc, there is a power law relation between the average galaxy number density $n(R)$ and the radius $R$ of test spheres $R$, in~the form $n(R) \propto R^{-\gamma}$ (see reviews by Sylos Labini~\cite{sylos11},
Baryshev \& Teerikorpi~\cite{baryshev-t12}, Baryshev~\cite{baryshev16}).
Such~power law behavior of the galaxy clustering is known as the \textbf{de Vaucouleurs law} \cite{devac70,devac71}.
Note that the power law correlation function is the characteristic feature of the discrete stochastic fractal structures in general physics (phase transitions, strange attractors, dynamical chaos, structure growth) and it has clear mathematical presentation (e.g.,~Gabrielli~et~al.~\cite{gabrielli05}).

\begin{figure}
\begin{center}
\includegraphics[width=11cm]
{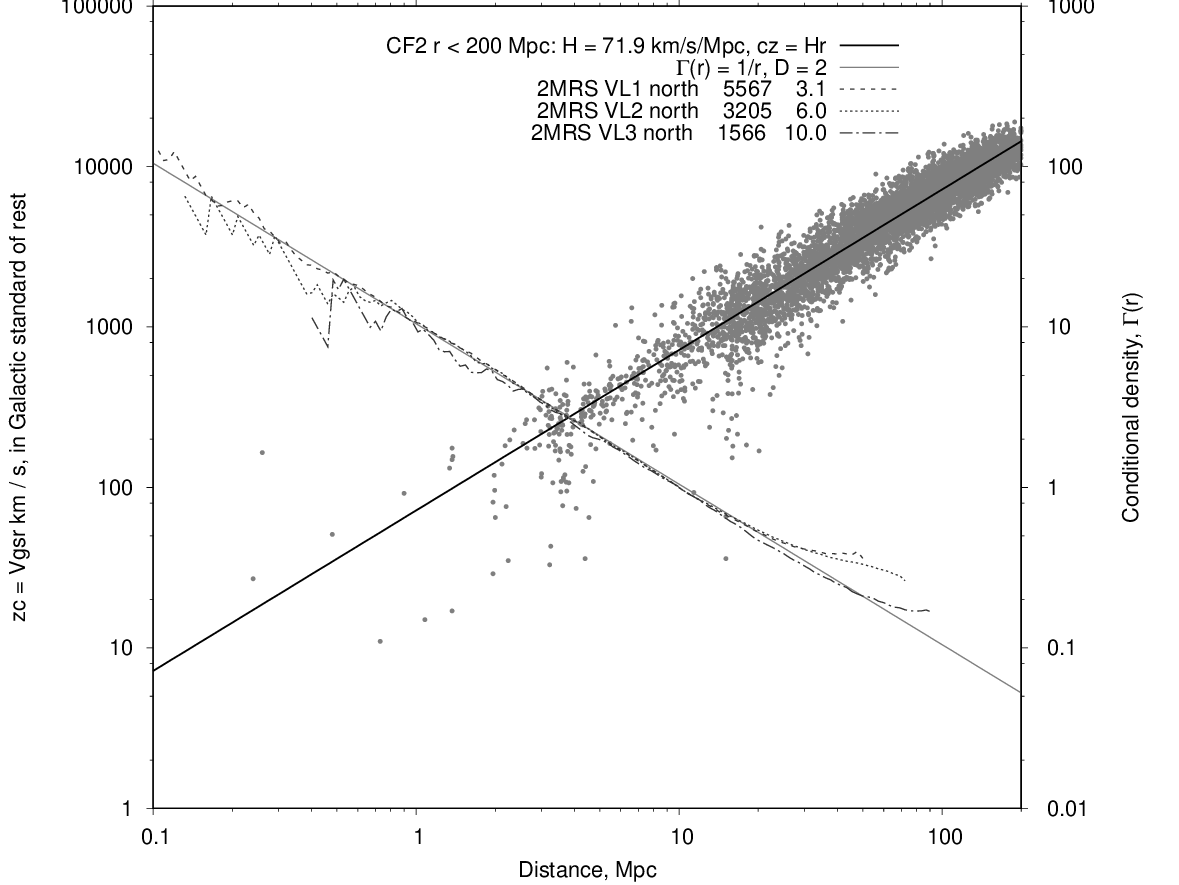}
\caption{\label{fig:1}
Presentation of the Hubble-deVaucouleurs (HdeV) paradox in the Local Universe~\cite{baryshev16}. The~Hubble linear redshift law (\mbox{$cz = H\times R$}) \cite{tully2016} and the fractal de Vaucouleurs density law \mbox{$\Gamma(r) = k\,r^{-\gamma}$} 
with $\gamma \approx 1$ 
\cite{tekhanovich16},
coexist at the common length-scale interval 1$\div$100 Mpc. While, in~the frame of the SCM, the~linear redshift-distance relation is the strict consequence of homogeneity~\cite{peebles91}. 
}
\end{center}
\end{figure}



The observed linearity of the redshift-distance relation, i.e. the \textbf{Hubble law} \cite{hubble29} in the Local Universe
(see Figure~\ref{fig:1}),
was confirmed by modern studies that were based on Cepheid distances to local galaxies, supernova distances, Tully-Fisher distances, and other distance indicators. 
It was demonstrated that at the very small distances around our Galaxy  ($1 \div 10$) Mpc  the linear Hubble law is well established (Ekholm~et~al.~\cite{ekholm01}, Karachentsev~et~al.~\cite{kara03}). It is remarkable, that the  dispersion  of the galaxy peculiar velocities relative to the linear Hubble law is about 30 km/s. 
Within the whole Local Universe, from 1 Mpc up to 200~Mpc, the linear Hubble law 
was firmly established by 
Sandage~\cite{sandage10},
Tully, Courtois \& Sorce~\cite{tully2016}.

A puzzling conclusion is that the Hubble law, i.e.,~the strictly linear redshift-distance relation, \mbox{is observed just inside strongly} inhomogeneous galaxy distribution, i.e.,~deeply inside fractal structure at scales
$1 \div 100$ Mpc 
(see Figure~\ref{fig:1}).
This empirical fact, called \textbf{``Hubble-deVaucouleurs paradox''} (Baryshev \& Teerikorpi~\cite{baryshev-t12}, Baryshev~\cite{baryshev16}), presents a profound challenge to the standard model where the homogeneity is the basic explanation of the Hubble law, and ``the connection between homogeneity and Hubble's law was the first success of the expanding world model'' (Peebles~et~al.~\cite{peebles91}).

However, contrary to this expectation, modern data show a good linear Hubble law within very inhomogeneous spatial distribution of the Local Universe galaxies (Figure~\ref{fig:1}) in common  interval of scales ($1 \div 100$ Mpc). It leads to the  conceptual problem for the standard model (based on homogeneity), because~observations demonstrate  that the linear Hubble law is not a consequence of the homogeneity of spatial galaxy distribution (as it was believed during long time). Actually, this is a crucial test for the FGF model, which predicts the linear Hubble law within fractal matter distribution having fractal dimension two, as~it is observed for luminous matter.
The problem of the dark matter still exists for FGF model. 
As we emphasized above, according to Equations~(\ref{cosmol-z}) and  \ref{cosmol-mass-grav}), giving the value of the gravitational Hubble constant and
the corresponding gravitating mass within the fractal ball,  there should be sufficient amount of dark matter, which have the same  fractal dimension as the luminous galaxies.

The second crucial test of the FGF cosmological model at high redshifts was recently suggested by Shirokov~et~al.~\cite{shirokov20a}. They analyzed
the prospects of using the high-redshift long gamma-ray bursts  Hubble diagram as a test of the basic cosmological principles. Analysis of the Hubble
diagram allows for one to test several fundamental cosmological principles while using the directly observed flux–distance–redshift relation.
In the frame of the FGF model the redshift–distance relation is given by the Equation~(\ref{cosmol-z-fgt-o-r}), where the radius $R_0 = r$  is the distance to a galaxy. 
Future THESEUS space observations of GRBs~\cite{amati2018}
and accompanying
multimessenger ground-based studies will be a powerful tool for testing the basic cosmological~principles.



\section{Conclusions}

The whole observable Universe is the cosmic laboratory where fundamental physical laws are tested using modern multimessenger astronomy facilities.
Especially deep problems of the theory of gravitational interaction can be tested by modern astronomical observations.
Modern theoretical physics delivers two main approaches for description of gravity phenomena---Einstein's general relativity theory (GRT) and Feynman's quantum field gravitation theory (QFGT). Both of the approaches have important successes and also open problems, which were reviewed~above. 

The Quantum-Field Gravitation Theory can be constructed on the common basis with modern quantum physics, which includes such fundamental quantum physics principles as: particle-wave duality, uncertainty principle, probability amplitude, superposition of quantum states, quantum entanglement, Poincare symmetry of Minkowski spacetime, and Lagrangian formalism of the quantum field~theory.

The quantum nature of the gravitational interaction requires different approach to the gravity physics than the Riemannian geometry of the General Relativity Theory. Although~GRT can be reconciled with the pure spin-2  quantum field theory, there are clear conceptual tensions between geometrical and quantum~approaches.

First, the~common bases for all fundamental physical interactions is the principal  role of the total symmetry of the Minkowski spacetime (Poincare group), which must be accepted in the quantum-field approach to gravitation. 
It is the necessary condition for unification of gravitational interaction with other physical fundamental forces. Hence, the Minkowski spacetime is not an ``arbitrary apriori geometry'', but, according to Noether theorem, it is the necessary and sufficient condition for the existence and conservation of the energy-momentum tensor of the gravitational field. In~fact, the~Minkowski spacetime guarantee the positive localizable energy density of the gravitational~waves. 

Second, the~Wigner classification of the elementary particles/fields is also based on the Poincare group. In~particular, the~source of the gravity field---Energy-Momentum Tensor of matter---in~accordance with irreducible representations of the Poincare group, generates two kinds of quantum particles/fields having spin-2 and spin-0.
The conservation of the source EMT and the gauge invariance of the field equation allows to delete four (from 10) degrees of freedom of the symmetric second rank tensor potential. Hence, the remaining six dofs correspond to spin-2 (five dofs) plus spin-0 (1 dof) real dynamical fields. 
The trace of the source EMT ($T=\eta_{ik}T^{ik}$) is the source of the scalar spin-0 field, in~addition to the traceless spin-2~field.

The third, all of the classical relativistic gravity effects are explained by the QFGT, when one takes the composite structure (spin-2 plus spin-0 fields) of the symmetric second rank tensor potential, generated by the matter EMT, into account.
Moreover, the~QFGT opens new ways for performing the experiments/observations in gravity physics for testing the existence of internal repulsive scalar field, generated by the trace of the matter~EMT.

In cosmology, GRT is the basis of the standard cosmological model together with cosmological principle of matter homogeneity and space expanding paradigm.
In the frame of QFGT, there are new possibilities for constructing cosmological models on the basis of non-expanding Minkowski space with dynamical and chemical matter evolution. The field gravity fractal cosmological model delivers new interpretation of the observed linear redshift-distance relation in the Local Universe, as~the global quantum gravitational redshift~effect.

Importantly, the~quantum understanding of the gravity physics can be tested by laboratory and solar system experiments, by~observations of relativistic compact objects, gravitational waves, large-scale structure, and high redshift objects while using modern multi-messenger~astronomy.


\acknowledgments{I thank for many discussions and collaborative works 
V.~V.~Sokolov, S.~A.~Oschepkov, P.~Teerikorpi, L.~Pietronero, F.~Sylos~Labini, G.~Paturel, S.~I.~Shirokov,
D.~I.~Nagirner, N.~A.~Topchilo, A.~A.~Raikov,
and A.~A.~Tron'. Also I grateful to 
M.~K.~Babadzhanyntz, V.~V.~Ivanov, Yu.~N.~Parijskij, V.~V.~Vitkovskij and V.~L.~Gorokhov for support during performing this~work.}

\centering
\subsection*{Funding}
This research received no external funding.

\subsection*{Conflict of interest}
The authors declare no conflicts of interest.

\subsection*{The following abbreviations are used in this manuscript:}
\begin{tabular}{@{}lcl}
GRT & ~~ & general relativity~theory \\
QFGT & ~~ & quantum-field gravitation~theory\\
EMT & ~~ & energy momentum~tensor\\
SCM & ~~ & standard cosmological~model \\
CMBR & ~~ & cosmic microwave background~radiation\\
LCDM & ~~ & lambda cold dark matter (model)\\
UGI & ~~ & The principle of universality of gravitational~interaction \\
RCO & ~~ & relativistic compact~objects\\
BH & ~~ & black~hole\\
FGF & ~~ & field gravity fractal (model)\\
HdeV & ~~ &  Hubble$-$de Vaucouleurs (paradox)\\
vDVZ & ~~ & van Dam-Veltman-Zakharov (paradox)
\end{tabular}

\end{document}